\documentclass[fleqn,usenatbib]{mnras}

\usepackage{newtxtext,newtxmath}

\usepackage[T1]{fontenc}

\DeclareRobustCommand{\VAN}[3]{#2}
\let\VANthebibliography\thebibliography
\def\thebibliography{\DeclareRobustCommand{\VAN}[3]{##3}\VANthebibliography}

\usepackage{graphicx}	
\usepackage{amsmath}	
\usepackage{orcidlink}
\usepackage{lineno}
\usepackage{xspace}
\usepackage{soul}
\usepackage{color}
\usepackage{csquotes}
\usepackage{enumitem}
\usepackage{longtable}
\usepackage{gensymb}
\usepackage{natbib}
\usepackage[table]{xcolor}
\usepackage{booktabs}
\usepackage{array}

\definecolor{blazeorange}{rgb}{1.0, 0.6, 0.2}
\definecolor{seagreen}{rgb}{0.18, 0.55, 0.34}
\definecolor{rufous}{rgb}{0.66, 0.11, 0.03}
\definecolor{royalfuchsia}{rgb}{0.79, 0.17, 0.57}
\definecolor{scarlet}{rgb}{1.0, 0.13, 0.0}
\definecolor{royalpurple}{rgb}{0.47, 0.32, 0.66}
\definecolor{darkblue}{rgb}{0, 0, 0.66}
\definecolor{violet}{rgb}{0.5,0.,0.5}

\title[GRB~250702B]{GRB~250702B: Discovery of a Gamma-Ray Burst from a Black Hole Falling into a Star}

\author[E.~Neights, E.~Burns et al.]{Eliza~Neights\orcidlink{0009-0005-0762-4507},$^{1,2}$\thanks{These authors contributed equally to this work.}\thanks{Email: eliza.neights@gmail.com}
Eric~Burns\orcidlink{0000-0002-2942-3379},$^{3}$\footnotemark[1]\thanks{Email: ericburns@lsu.edu}
\newauthor
Chris~L.~Fryer\orcidlink{0000-0003-2624-0056},$^{4}$
Dmitry~Svinkin\orcidlink{0000-0002-2208-2196},$^{5}$
Suman~Bala\orcidlink{0000-0002-6657-9022},$^{6}$
Rachel~Hamburg\orcidlink{0000-0003-0761-6388},$^{6}$
Ramandeep~Gill\orcidlink{0000-0003-0516-2968},$^{7,8}$
\newauthor
Michela~Negro\orcidlink{0000-0002-6548-5622},$^{3}$
Megan~Masterson\orcidlink{0000-0003-4127-0739},$^{9}$
James~DeLaunay\orcidlink{0000-0001-5229-1995},$^{10,11}$
David~J.~Lawrence\orcidlink{0000-0002-7696-6667},$^{12}$
\newauthor
Sophie~E.~D.~Abrahams\orcidlink{0000-0003-0482-1984},$^{13}$
Yuta~Kawakubo\orcidlink{0000-0002-2064-3164},$^{14}$
Paz~Beniamini\orcidlink{0000-0001-7833-1043},$^{15,8,1}$
Christian~Aa.~Diget\orcidlink{0000-0002-9778-8759},$^{13}$
\newauthor
Dmitry~Frederiks\orcidlink{0000-0002-1153-6340},$^{5}$
John~Goldsten\orcidlink{0000-0001-7009-6288},$^{12}$
Adam~Goldstein\orcidlink{0000-0002-0587-7042},$^{6}$
Alexander~D.~Hall-Smith\orcidlink{0000-0002-6734-6204},$^{13}$
Erin~Kara\orcidlink{0000-0003-0172-0854},$^{9}$
\newauthor
Alison~M.~Laird\orcidlink{0000-0003-0423-363X},$^{13}$
Gavin~P.~Lamb\orcidlink{0000-0001-5169-4143},$^{16}$
Oliver~J.~Roberts\orcidlink{0000-0002-7150-9061},$^{6}$
Ryan~Seeb\orcidlink{0009-0005-4094-3278},$^{17,18}$
V.~Ashley~Villar\orcidlink{0000-0002-5814-4061},$^{19,20}$
\newauthor
Aldana~Holzmann~Airasca\orcidlink{0009-0007-8169-4719},$^{21,22}$
Joseph~R.~Barber,$^{23}$
P.~Narayana~Bhat\orcidlink{0000-0001-7916-2923},$^{24}$
Elisabetta~Bissaldi\orcidlink{0000-0001-9935-8106},$^{25,21}$
\newauthor
Michael~S.~Briggs\orcidlink{0000-0003-2105-7711},$^{26,24}$
William~H~Cleveland\orcidlink{0009-0003-3480-8251},$^{6}$
Sarah~Dalessi\orcidlink{0000-0003-1835-570X},$^{26,24}$
Davide~Depalo\orcidlink{0000-0001-6690-7789},$^{25,21}$
Misty~M.~Giles,$^{23}$
\newauthor
Jonathan~Granot\orcidlink{0000-0001-8530-8941},$^{15,8,1}$
Boyan~A.~Hristov\orcidlink{0000-0001-9556-7576},$^{24}$
C.~Michelle~Hui\orcidlink{0000-0002-0468-6025},$^{17}$
Andreas~von~Kienlin\orcidlink{0000-0002-0221-5916},$^{27}$
\newauthor
Carolyn~Kierans\orcidlink{0000-0001-6677-914X},$^{2}$
Daniel~Kocevski\orcidlink{0000-0001-9201-4706},$^{17}$
Stephen~Lesage\orcidlink{0000-0001-8058-9684},$^{26,24}$
Alexandra~L.~Lysenko\orcidlink{0000-0002-3942-8341},$^{5}$
Bagrat~Mailyan\orcidlink{0000-0002-2531-3703},$^{28}$
\newauthor
Christian~Malacaria,$^{29}$
Tyler~Parsotan\orcidlink{0000-0002-4299-2517},$^{2}$
Anna~Ridnaia\orcidlink{0000-0001-9477-5437},$^{5}$
Samuele~Ronchini\orcidlink{0000-0003-0020-687X},$^{30,10,11}$
Lorenzo~Scotton\orcidlink{0000-0002-0602-0235},$^{24}$
\newauthor
Aaron~C.~Trigg\orcidlink{0009-0006-8598-728X},$^{31}$
Anastasia~Tsvetkova\orcidlink{0000-0003-0292-6221},$^{5}$
Mikhail~Ulanov\orcidlink{0000-0002-0076-5228},$^{5}$
P\'eter~Veres\orcidlink{0000-0002-2149-9846},$^{26,24}$
Maia~Williams\orcidlink{0000-0002-0025-3601},$^{32,33}$
\newauthor
Colleen~A.~Wilson-Hodge\orcidlink{0000-0002-8585-0084},$^{17}$
and Joshua~Wood\orcidlink{0000-0001-9012-2463}$^{17}$
\\
$^{1}$Department of Physics, The George Washington University, 725 21st St NW, Washington, DC 20052, USA\\
$^{2}$Astrophysics Science Division, NASA Goddard Space Flight Center, 8800 Greenbelt Road, Greenbelt, MD 20771, USA\\
$^{3}$Department of Physics \& Astronomy, Louisiana State University, Baton Rouge, LA 70803, USA\\
$^{4}$Center for Theoretical Astrophysics, Los Alamos National Laboratory, Los Alamos, NM 87545, USA\\
$^{5}$Ioffe Institute, Polytekhnicheskaya, 26, St. Petersburg, 194021- Russian Federation\\
$^{6}$Science and Technology Institute, Universities Space Research Association, Huntsville, AL 35805, USA\\
$^{7}$Instituto de Radioastronom\'ia y Astrof\'isica, Universidad Nacional Aut\'onoma de M\'exico, Antigua Carretera a P\'atzcuaro $\#$ 8701, Ex-Hda. San Jos\'e de la Huerta, \\Morelia, Michoac\'an, C.P. 58089, M\'exico\\
$^{8}$Astrophysics Research Center of the Open University (ARCO), The Open University of Israel, P.O Box 808, Ra’anana 4353701, Israel\\
$^{9}$MIT Kavli Institute for Astrophysics and Space Research, Massachusetts Institute of Technology, Cambridge, MA 02139, USA\\
$^{10}$Department of Astronomy and Astrophysics, The Pennsylvania State University, 525 Davey Lab, University Park, PA 16802, USA\\
$^{11}$Institute for Gravitation and the Cosmos, The Pennsylvania State University, University Park, PA 16802, USA\\
$^{12}$Johns Hopkins University Applied Physics Laboratory, Laural, MD 20723, USA\\
$^{13}$School of Physics, Engineering and Technology, University of York, Heslington, York YO10 5DD, UK\\
$^{14}$Department of Physical Science, Aoyama Gakuin University, 5-10-1 Fuchinobe, Chuo-ku, Sagamihara, Kanagawa 252-5258, Japan\\
$^{15}$Department of Natural Sciences, The Open University of Israel, P.O Box 808, Ra'anana 4353701, Israel\\
$^{16}$Astrophysics Research Institute, Liverpool John Moores University, IC2 Liverpool Science Park, 146 Brownlow Hill, Liverpool, L3 5RF, UK\\
$^{17}$ST12 Astrophysics Branch, NASA Marshall Space Flight Center, Huntsville, AL 35812, USA\\
$^{18}$Department of Physics, Brown University, Providence, RI 02912, USA\\
$^{19}$Center for Astrophysics \textbar{} Harvard \& Smithsonian, 60 Garden Street, Cambridge, MA 02138-1516, USA\\
$^{20}$The NSF AI Institute for Artificial Intelligence and Fundamental Interactions\\
$^{21}$Istituto Nazionale di Fisica Nucleare, Sezione di Bari, I-70126 Bari, Italy\\
$^{22}$Universit\`a degli studi di Trento, via Calepina 14, 38122 Trento, Italy\\
$^{23}$Amentum Space Exploration Division, Huntsville, AL 35806, USA\\
$^{24}$Center for Space Plasma and Aeronomic Research, University of Alabama in Huntsville, Huntsville, AL 35899, USA\\
$^{25}$Dipartimento Interateneo di Fisica, Politecnico di Bari, Bari, Italy\\
$^{26}$Department of Space Science, University of Alabama in Huntsville, 320 Sparkman Drive, Huntsville, AL 35899, USA\\
$^{27}$Max-Planck-Institut f\"{u}r extraterrestrische Physik, Giessenbachstrasse 1, D-85748 Garching, Germany\\
$^{28}$Department of Aerospace, Physics and Space Sciences, Florida Institute of Technology, Melbourne, FL 32901, USA\\
$^{29}$INAF-Osservatorio Astronomico di Roma, Via Frascati 33, I-00078, Monteporzio Catone (RM), Italy\\
$^{30}$Gran Sasso Science Institute (GSSI), 67100, L'Aquila, Italy\\
$^{31}$NASA Postdoctoral Program Fellow, NASA Marshall Space Flight Center, Huntsville, AL, 35812, USA\\
$^{32}$Department of Physics and Astronomy, Northwestern University, Evanston, IL 60208, USA\\
$^{33}$Center for Interdisciplinary Exploration and Research in Astronomy (CIERA), Northwestern University, 1800 Sherman Avenue, Evanston, IL 60201, USA\\
}

\date{Accepted XXX. Received YYY; in original form ZZZ}

\pubyear{\the\year{}}

\begin{document}
\label{firstpage}
\pagerange{\pageref{firstpage}--\pageref{lastpage}}
\maketitle

\begin{abstract}
Gamma-ray bursts are the most luminous electromagnetic events in the universe. Their prompt gamma-ray emission has typical durations between a fraction of a second and several minutes. A rare subset of these events have durations in excess of a thousand seconds, referred to as ultra-long gamma-ray bursts. Here, we report the discovery of the longest gamma-ray burst ever seen with a $\sim$25,000~s gamma-ray duration, GRB~250702B, and characterize this event using data from four instruments in the InterPlanetary Network and the Monitor of All-sky X-ray Image. We find a hard spectrum, subsecond variability, and high total energy, which are only known to arise from ultrarelativistic jets powered by a rapidly-spinning stellar-mass central engine. These properties and the extreme duration are together incompatible with all confirmed gamma-ray burst progenitors and nearly all models in the literature. This burst is naturally explained with the helium merger model, where a field binary ends when a black hole falls into a stripped star and proceeds to consume and explode it from within. Under this paradigm, GRB~250702B adds to the growing evidence that helium stars expand and that some ultra-long GRBs have similar evolutionary pathways as collapsars, stellar-mass gravitational wave sources, and potentially rare types of supernovae.
\end{abstract}

\begin{keywords}
gamma-ray burst: individual: GRB~250702B -- gamma-rays: general -- methods: observational
\end{keywords}



\section{Introduction}\label{sec:intro}
Gamma-ray bursts (GRBs) are brief flashes of gamma rays which are traditionally separated into classes based on their prompt duration, referred to as short and long GRBs separated by a fiducial 2~s threshold. Using \enquote{gamma-ray burst} as a phenomenological term, the shortest class of GRBs are magnetar giant flares \citep{2008ApJ...680..545M,2021ApJ...907L..28B,2025A&A...694A.323T,rastinejad2021probing}. However, the majority of short GRBs arise from neutron star mergers, as confirmed with associated kilonovae and gravitational waves \citep{1989Natur.340..126E,2017ApJ...850L..40A}. Most long GRBs arise from rapidly-rotating massive stars known as collapsars, as verified with associations to broad-line type Ic supernovae \citep{1998Natur.395..670G,1999ApJ...524..262M,cano2017observer}. Neutron star merger and collapsar GRBs are powered by collimated, ultrarelativistic outflows called jets. In these progenitor cases, the prompt emission is followed by broadband synchrotron radiation observed across the electromagnetic spectrum, which is referred to as afterglow. Lastly, a small number of GRBs have later been identified to originate from the tidal disruption of stars by supermassive black holes at the center of galaxies, with prompt durations of a few days \citep{2011Sci...333..203B,2012ApJ...753...77C,2015MNRAS.452.4297B}. 

Ultra-long GRBs are a rare class with prompt durations of $\gtrsim$1,000~s \citep{2014ApJ...781...13L}. When excluding tidal disruption events (TDEs), the longest gamma-ray duration of such an event is $\sim$15,000~s for GRB~111209A. This event was associated with SN~2011kl, the most luminous supernova seen following a GRB \citep{2013ApJ...779...66S,2016ApJ...817L...8B,2019A&A...624A.143K}. GRB~101225A, another ultra-long GRB with prompt duration $\gtrsim$2000~s, also has some evidence for supernova emission \citep{2011Natur.480...72T}. Theoretically, the longest accretion time possible with a collapsar GRB is a few thousand seconds, physically limited by the angular momentum in a single star spinning at break-up velocity \citep{fryer2025explaining}. As we generally expect the prompt emission timescale to be of similar order to the accretion time, collapsars struggle to explain the longest ultra-long GRBs. 

GRB~250702B was first identified when the \textit{Fermi} Gamma-ray Burst Monitor (\textit{Fermi}-GBM) triggered multiple times over a $\sim$3 hour period on July 2, 2025 in response to impulsive signals consistent with originating from the same position \citep{2025GCN.40891....1N,2025GCN.40931....1N}. The prompt emission of GRB~250702B was observed by numerous X-ray and gamma-ray monitors \citep{2025GCN.40903....1D,2025GCN.40910....1K,2025GCN.40914....1F,2025GCN.40923....1S}, with Konus-\textit{Wind} identifying the duration of this event as at least comparable to GRB~111209A. The \textit{Einstein Probe} (EP) Wide-field X-ray Telescope observed X-rays from GRB~250702B in individual exposures over a 17 hour period on July 2, and a stacking analysis found a signal beginning a day earlier on July 1 \citep{2025GCN.40906....1C}. The first precise localization of GRB~250702B is $(\rm RA,~Dec) = (284.6901 \degree, -7.8741 \degree)$ with an uncertainty of 2~arcseconds, measured by the \textit{Swift} X-Ray Telescope \citep[XRT][]{2025GCN.40919....1K}, which is the position used for our analysis.

Follow-up observations were performed across the electromagnetic spectrum, with detections in near-infrared, X-ray, and radio \citep{2025GCN.40924....1M,2025GCN.40961....1L,2025GCN.40985....1B,2025GCN.41014....1O,2025GCN.41053....1S,2025GCN.41054....1A,2025GCN.41059....1A,2025GCN.41096....1G,2025GCN.41147....1G}, and non-detections in optical and very high-energy gamma-rays \citep{2025GCN.40908....1K,2025GCN.40918....1B,2025GCN.40929....1P,2025GCN.40949....1B,2025GCN.40952....1S,2025GCN.40986....1L,2025GCN.41067....1P,2025GCN.41095....1N}. Observations by the Very Large Telescope and \textit{Hubble Space Telescope} resolved the host galaxy, proving an extragalactic origin of GRB~250702B and showing the event as offset from the host galaxy center, disfavoring a supermassive black hole TDE origin \citep{levan2025day,carney2025grb250702b,oconnor2025grb250702b}. The multiwavelength follow-up observations can be modeled as synchrotron radiation from a fairly typical forward and reverse shock \citep{levan2025day,oconnor2025grb250702b,carney2025grb250702b}. 
\citet{gompertz2025grb250702b} utilize data from the \textit{James Webb Space Telescope} to measure a redshift of 1.036, show the position of GRB~250702B to be in a star-forming region, and show no obvious transient at the position of GRB~250702B at 25.5~days after the GRB in the rest frame of the source.

Here, we present the gamma-ray analysis of GRB~250702B. The extreme duration and unusual properties of this event require combining observations from multiple monitors to gain a complete understanding. We describe our analysis in Section~\ref{sec:analysis}, and the physical inferences which follow in Section~\ref{sec:inferences}. In Section~\ref{sec:exclusion-progenitor}, we discuss the progenitor options that can be excluded based on the gamma-ray observations. The combination of the duration and rapid variability of GRB~250702B require an alternative origin, which we propose in Section~\ref{sec:helium-merger} is a helium star merger. All times in this work are referenced against midnight UTC on July 2, 2025, i.e. T0=2025-07-02T00:00:00. Throughout this paper, we use the final cosmological parameters measured from the \textit{Planck} mission \citep{2020A&A...641A...6P}: $H_0=67.4$~km/s/Mpc and $\Omega_m=0.315$ for a flat universe.

\section{Prompt Gamma-Ray Burst Analysis}\label{sec:analysis}
\begin{figure*}
    \centering
    \includegraphics[width=\textwidth]{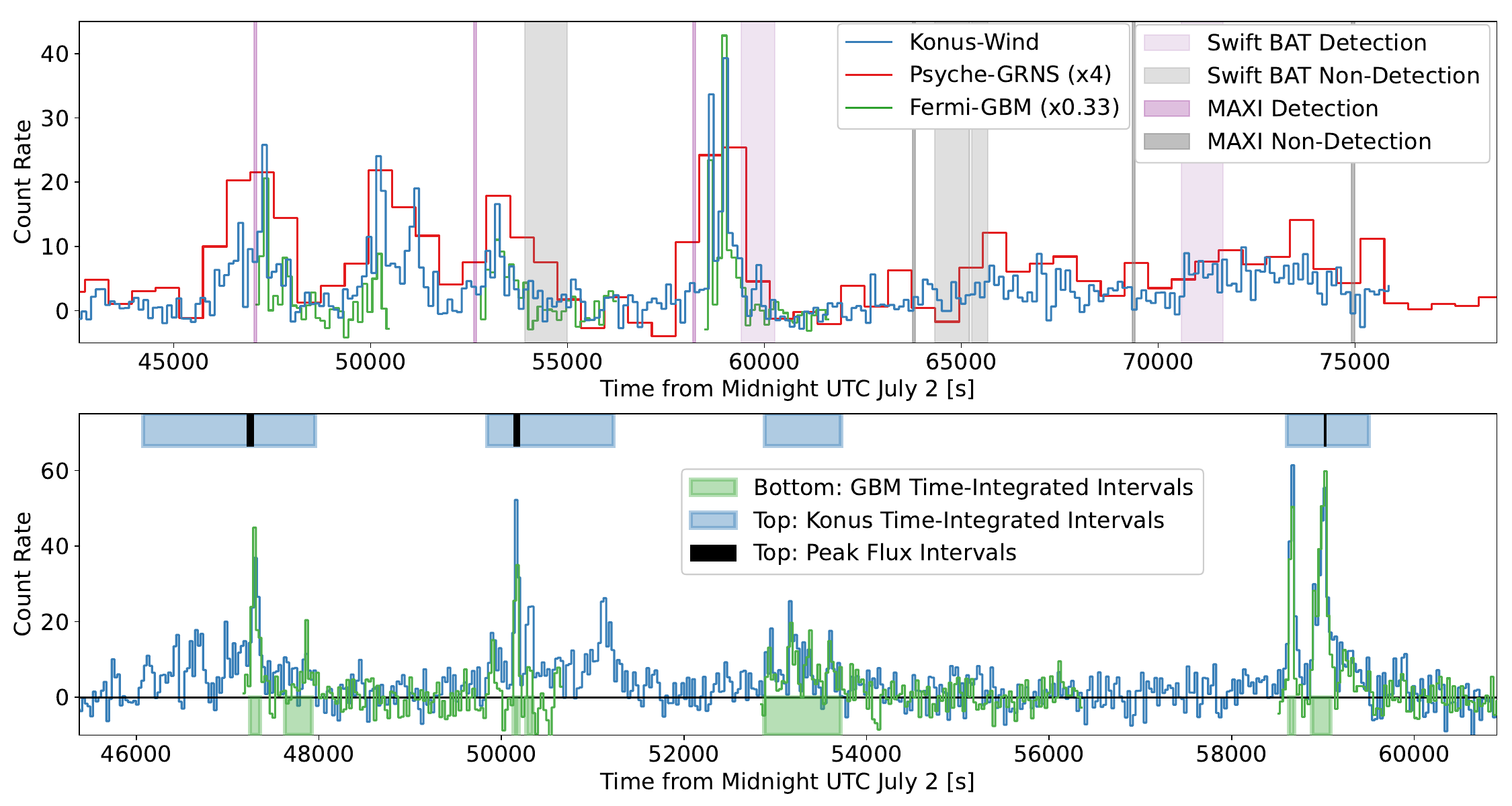}
    \caption{The combined, background-subtracted gamma-ray lightcurve of GRB~250702B. For Konus-\textit{Wind}, we show counts in the 75--315~keV energy range, for \textit{Psyche}-GRNS the 30--230~keV energy range, and for \textit{Fermi}-GBM 50-500~keV for NaI and 400--2,000~keV for BGO detectors. Detector lightcurves are scaled and \textit{Psyche} is shifted by 1,000~s (to account for light travel time) for visualization. Intervals where the data for a given instrument are not useful are removed (i.e. times with unrelated GRBs, Earth-occluded times, etc.). \textit{Top}: Lightcurves for \textit{Psyche}-GRNS with 600~s temporal resolution and Konus-\textit{Wind} and \textit{Fermi}-GBM both at 120~s temporal resolution. The prompt gamma-ray emission from GRB~250702B begins at least by T0+46,074~s based on the rapid rise and lasts at least until T0+71,600~s based on the last significant flare, as confirmed by \textit{Swift}-BAT emission from the source. The BAT non-detections show the burst has quiescent intervals. MAXI information confirms these results. \textit{Bottom}: A view of the brightest region of gamma-ray emission with \textit{Fermi}-GBM and Konus-\textit{Wind} data shown at 30~s resolution. Also shown are the selection intervals for detailed analyses.}
    \label{fig:lc}
\end{figure*}

GRB~250702B was detected by several GRB monitors, five of which are used in this paper: \textit{Fermi}-GBM \citep{2009ApJ...702..791M}, Konus-\textit{Wind} \citep{1995SSRv...71..265A}, the Burst Alert Telescope onboard the \textit{Neil Gehrels Swift Observatory} \citep[\textit{Swift}-BAT;][]{Barthelmy05}, the \textit{Psyche} Gamma-Ray and Neutron Spectrometer \citep[GRNS;][]{psyche}, and the Monitor of All-sky X-ray Image (MAXI) \citep{matsuoka2009maxi}. Individually, each instrument only contributes partial information necessary to fully characterize this event. As such, we use the appropriate instruments for each analysis, measuring each quantity with more than one instrument when possible as summarized in Table~\ref{tab:instruments}.

\textit{Fermi}-GBM is in low Earth orbit and therefore has only partial coverage of GRB~250702B due to Earth occluding the source and detector downtime as a result of orbital regions of high particle activity. \textit{Fermi}-GBM provides photon-by-photon continuous data with high spectral and precise temporal resolution across the 8~keV to 40~MeV band using two Bismuth Germanate (BGO) detectors and twelve Sodium Iodide (NaI) detectors \citep{2009ApJ...702..791M}. However, its background stability is on the order of a minute. 

\textit{Swift}-BAT is also in low Earth orbit with partial sky coverage. It is a coded aperture mask allowing for arcminute level spatial information. It has an instantaneous field of view of $\sim$15\% of the sky and serendipitously observes $\sim$80\% of the sky each day. \textit{Swift} points at fixed positions, allowing for characterization of signals on long timescales. The detectors are composed of Cadmium Zinc Telluride, which measure energies in the $\sim$15--350~keV range \citep{Barthelmy05}.

Konus-\textit{Wind} is far from Earth and thus has total coverage of the event. It is a transient monitor in a Lissajous orbit around the first Lagrange point of the Sun–Earth system at a distance of $\sim$5 light seconds. Konus-\textit{Wind} consists of two identical thallium-doped NaI spectrometers S1 and S2, pointing to the ecliptic poles. As it did not trigger on GRB~250702B, we utilize the continuous (waiting mode) data, which has 2.944~s temporal resolution. We use the S2 detector for most analyses, in which GRB~250702B is at an incident angle of $75.2\degree$, which has three energy bands: 18--76~keV, 76--316~keV, and 316--1250~keV. We also use the S1 detector in one analysis, with energy bands 23--96~keV, 96--398~keV, and 398--1628~keV.

The \textit{Psyche}-GRNS, also far from Earth with full coverage of the event, contains a Germanium detector built to determine the composition of the asteroid 16~Psyche. The GRNS is surrounded on most sides by a plastic scintillator anticoincidence shield built to separate cosmic rays from photons. It is this anticoincidence shield which we use as a GRB instrument within the InterPlanetary Network, which will be detailed in a future publication. These data have the most stable backgrounds and complete coverage, but the temporal resolution for continuous data is 600~s and the photon response has not yet been calculated. Lightcurves of GRB~250702B use the low-energy channels covering $\sim$30--230~keV, while the maximum photon energy analysis uses the high-energy channels covering $\gtrsim$350~keV.

The Monitor of All-sky X-ray Image (MAXI) is an X-ray monitoring instrument mounted on the Japanese Experiment Module–Exposed Facility aboard the International Space Station. Operating since 2009, it conducts nearly continuous all-sky surveys in the 0.2--30~keV energy range using Gas Slit Cameras (GSC) and Solid-state Slit Cameras (SSC). MAXI scans most of the sky every 92 minutes, allowing for serendipitous coverage of the position of GRB~250702B.

\begin{table*}
\caption{The instruments utilized for each analysis. Primary instruments are those which are used in the reported values. Secondary instruments provide confirmation.}\label{tab:instruments}
\centering
\begin{tabular}{rccccc}
\hline
& \textit{Fermi}-GBM & Konus-\textit{Wind} & \textit{Psyche}-GRNS & \textit{Swift}-BAT & MAXI\\ \hline
Duration                      & -         & Primary       & Primary        & Primary   & Primary   \\
Minimum Variability Timescale & Primary      & -       & -           & Secondary    & -  \\
Maximum Photon Energy         & Primary      & -          & Secondary        & -        & - \\
Spectra                       & Primary      & Primary       & -           & -       &  - \\
Quasiperiodic Oscillations    & -         & Primary       & -        & -       & -  \\
\hline
\end{tabular}
\end{table*}

\subsection{Gamma-Ray Lightcurve and Duration}
\begin{figure*}
    \centering
    \includegraphics[width=0.48\textwidth]{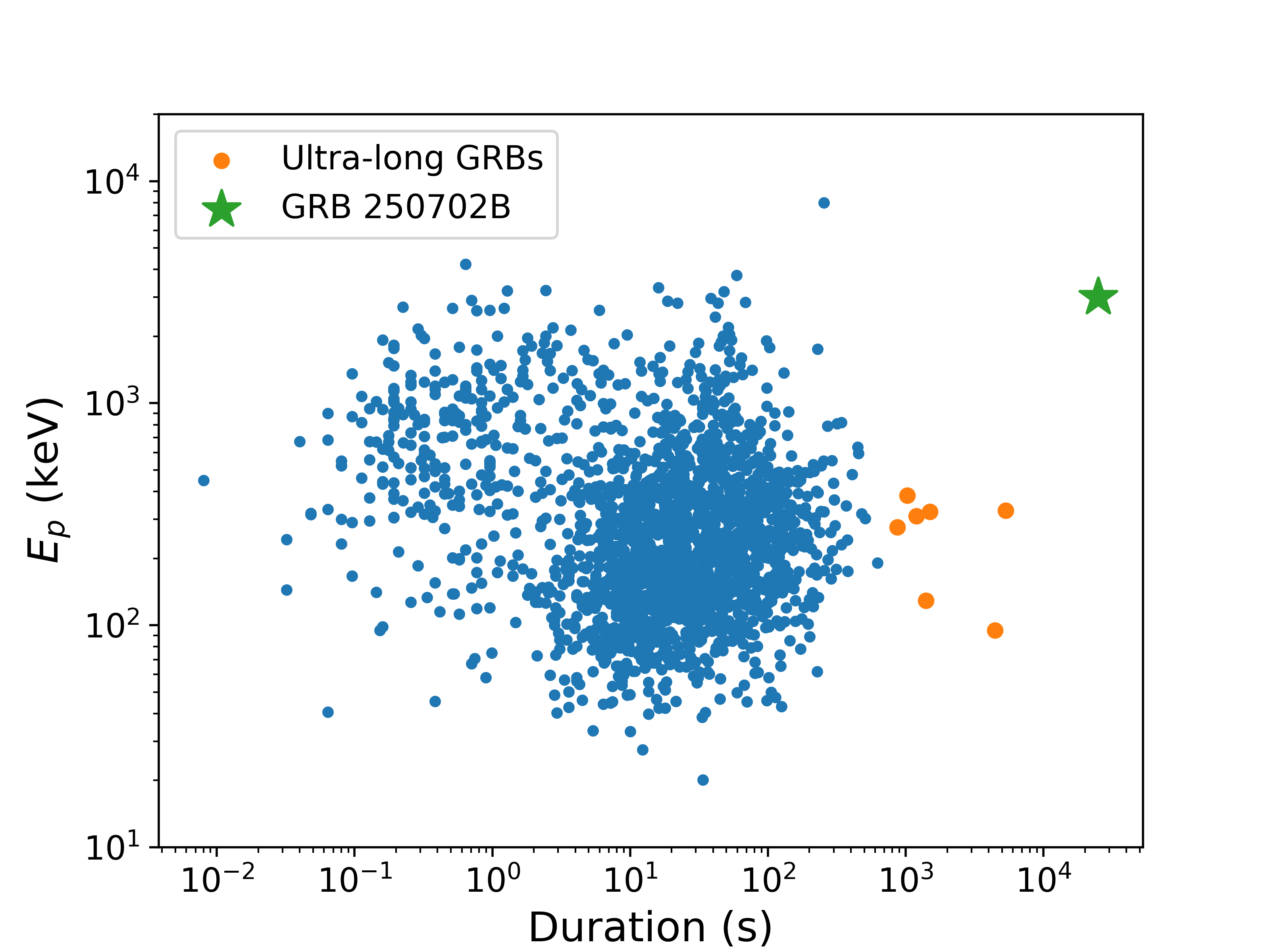}
    \includegraphics[width=0.48\textwidth]{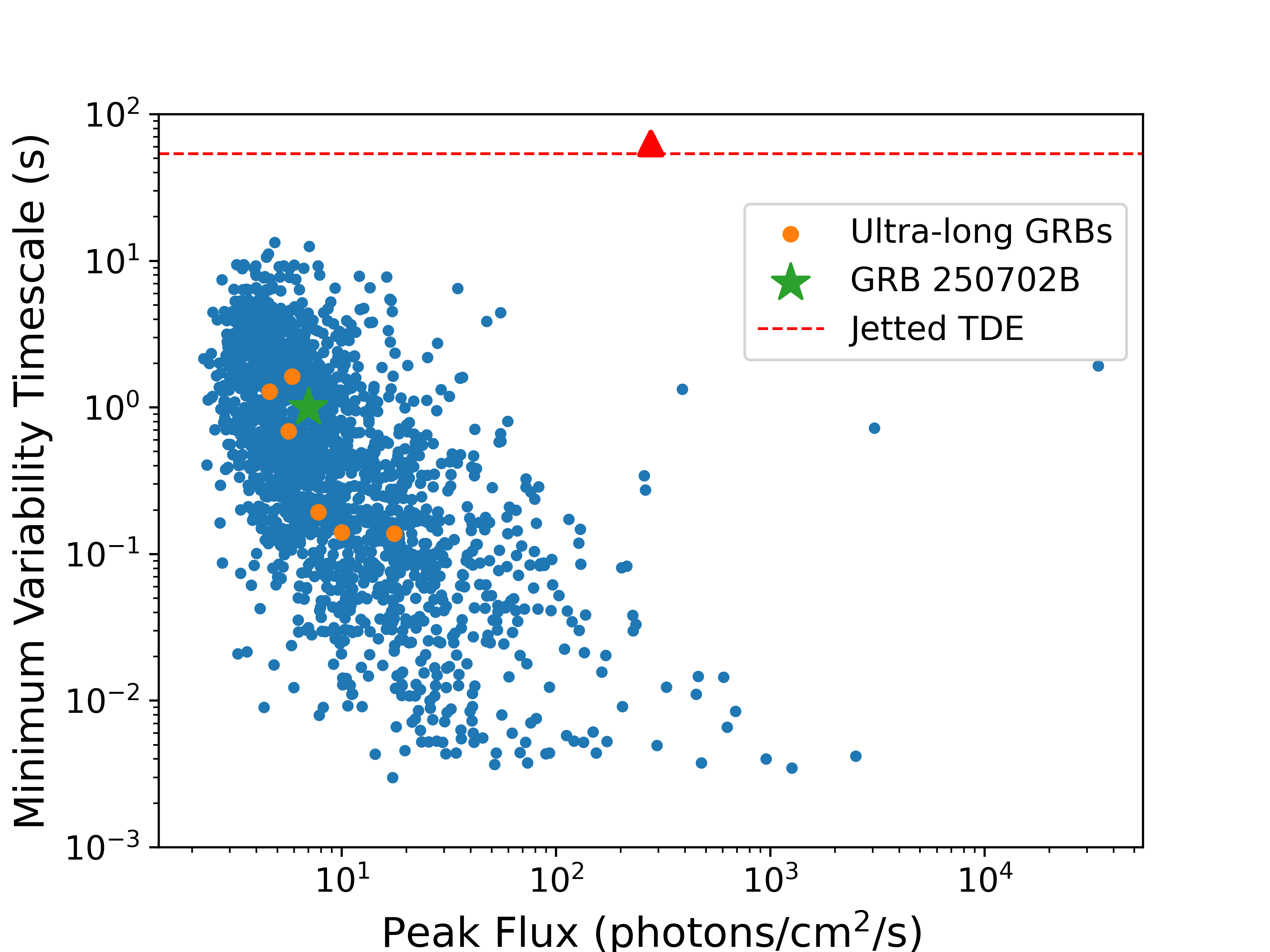}
    \caption{GRB~250702B in the context of \textit{Fermi}-GBM GRBs. \textit{Left}: The duration and peak energy in the peak flux interval. \textit{Right}: The observed MVT as a function of 64~ms peak photon flux. Also shown is an approximate lower limit on the known MVT from TDEs, based on Swift~J1644+57, detailed in Appendix~\ref{app:tde}. While the duration of GRB~250702B is an extreme outlier and the $E_p$ is unusually high, the MVT and peak flux are typical values.}
    \label{fig:t90-epeak-mvt}
\end{figure*}

We combine information from a number of instruments to build the lightcurve shown in Fig.~\ref{fig:lc} and determine the duration of GRB~250702B with the approach detailed in Appendix~\ref{app:lightcurve}.  We find that the prompt gamma-ray duration begins by T0+46,074~s from the onset of Konus-\textit{Wind} emission and lasts at least until T0+71,600~s from the end of the last \textit{Swift}-BAT significant detection. This gives an observed gamma-ray duration of $\gtrsim$25,000~s and up to $\sim$30,000~s based on emission whose association is ambiguous. This corresponds to a rest-frame duration of $\gtrsim$12,500~s. For comparison, the previous record holder was GRB~111209A, measured by Konus-\textit{Wind} to have a prompt duration of $\sim$15,000~s and a rest-frame duration of $\sim$9,000~s \citep{golenetskii2011konus,vreeswijk2011grb}. GRB~250702B is unambiguously the longest GRB ever identified, shown in Fig.~\ref{fig:t90-epeak-mvt}.

Appendix~\ref{app:lightcurve} also shows a search for gamma-ray emission over wider intervals, identifying no confident detections. Because of the extended X-ray emission beginning on July~1, we calculate upper limits on the gamma-ray emission at these times. Using Konus-\textit{Wind} data we calculate an approximate upper limit for a soft spectrum on the 2.944~s timescale of $\sim$1.5$\times10^{-7}$~erg/cm$^2$/s. This is the deepest limit available from a mission with nearly complete coverage of the event.

\subsection{Minimum Variability Timescale}
The minimum variability timescale (MVT) is the shortest timescale over which statistically significant variability is detected in the lightcurve. This is linked to the size and dynamics of the emitting region, with shorter MVTs suggesting a smaller emission region or faster central engine variability due to the finite speed of light. We compute the MVT following the procedure in \citet{golkhou2014}, in which a sliding wavelet is used to compare the measured variability power to that from random statistical fluctuations. 

We analyze the \textit{Fermi}-GBM triggers listed in Table~\ref{table:gbm-triggers} in Appendix~\ref{app:interval-selections} using 0.1~ms bins, as well as the bright GBM analysis intervals described in Table~\ref{table:gbm-bright-intervals-raw} and Table~\ref{table:gbm-bright-intervals}, and shown in Fig.~\ref{fig:lc}, using 1~ms bins. In all cases, we use NaI detectors with a viewing angle within 60$\degree$ of the source. The shortest MVT values arise from T0+47,285--47,300~s, T0+50,068--50,359~s, and T0+58,975--58,990~s, with values of $1.2\pm0.5$~s, $1.0\pm0.4$~s, and $1.4\pm0.7$~s, respectively. We perform a similar analysis using the \textit{Swift}-BAT event data, available around the \textit{Fermi}-GBM triggers via the Gamma-ray Urgent Archiver for Novel Opportunities \citep[GUANO;][]{GUANO} pipeline, finding weaker constraints (i.e., larger MVT). The figures showing these fits and the BAT table are shown in Appendix~\ref{app:mvt}. As the MVT is an upper limit, and the signal-to-noise ratio (SNR) is low, these results are consistent. We therefore measure an MVT of $\sim$1~s for GRB~250702B. This is well within the normal distribution for GRBs from stellar-mass central engines, as shown in Fig.~\ref{fig:t90-epeak-mvt}. In the rest frame, the MVT is $\sim$0.5, giving subsecond variability.

\subsection{Spectral Lags}

The temporal difference between the lightcurves of a GRB in different energy bands is known as spectral lag. The spectral lag is defined as positive when high-energy photons precede low-energy photons. In general, long GRBs show a positive spectral lag \citep{1995_Cheng, 1997_band, Norris+00lag, Ukwatta+10lag}, while short GRBs are characterized by a spectral lag consistent with zero \citep{Norris+00lag, Ukwatta+10lag}. Spectral lag may therefore help in categorizing GRBs \citep{gehrels06, Zhang+06ag, norris06}, though the measured difference between short and long bursts may be due to photon statistics.

We measure the spectral lag between the 25--50~keV and 100--300~keV lightcurves using the cross-correlation method on the GBM analysis intervals in Table~\ref{table:gbm-bright-intervals-raw}. We use the NaI detectors with a viewing angle within 60$\degree$ of the source. To identify the characteristic timescale of the correlation, we test four different temporal bin widths for each interval: 0.064~s, 0.128~s, 0.256~s, and 0.512~s. A statistically significant correlation, defined by a cross-correlation function $\text{SNR}>3$, was detected in several intervals. In all of these, the measured spectral lag is consistent with zero, with significant and reliable measurements summarized in Table~\ref{tab:lag_results}.

\begin{table}
\caption{Summary of significant spectral lag measurements, where the cross-correlation SNR exceeds the detection threshold of 3 and the fit is numerically stable. The measured spectral lag is consistent with zero for all intervals.}
\centering
\begin{tabular}{cccc}
\hline
GBM Interval & Bin Width (s) & Spectral Lag (s) & SNR \\
\hline
1a & 0.256 & $0.02 \pm 0.14$ & 3.81 \\
 & 0.512 & $-0.1 \pm 0.4$ & 3.25 \\
2 & 0.256 & $0 \pm 30$ & 3.18 \\
3a & 0.064 & $0.4 \pm 0.3$ & 3.05 \\
6 & 0.512 & $0.0 \pm 0.3$ & 3.1 \\
7a & 0.256 & $0.0 \pm 0.2$ & 3.83 \\
7b & 0.512 & $0.3 \pm 0.6$ & 3.0 \\
7d & 0.512 & $0.2 \pm 0.4$ & 3.09 \\
\hline
\end{tabular}
\label{tab:lag_results}
\end{table}

\subsection{Spectrum}
We perform time-resolved spectral analysis of bright intervals and peak flux intervals in both \textit{Fermi}-GBM and Konus-\textit{Wind}. The details of the background and source selections are described in Appendix~\ref{app:lightcurve}. The specifics of the spectral analysis and complete results are detailed in Appendix~\ref{app:spectral-analysis}. The best-fit spectra are summarized in Table~\ref{table:spectra-summary}. The results for bright intervals with robust background estimates are included here, summarizing the measurements we consider to be reliable. 

Given the respective data limitations, the \textit{Fermi}-GBM and Konus-\textit{Wind} spectral results are in broad agreement. The indices generally agree, as do the fluence and peak flux values (once accounting for the different temporal selections and errors). All intervals are best fit with a cutoff power-law model with photon index between $\sim$-1.3 and $\sim$-0.6. Both instruments report peak energy ($E_{p}$) values which are high for long GRBs, though the values disagree. Because we are limited to continuous data from Konus, the available energy channels preclude measuring $E_{p}$ above $\sim$1.5~MeV. GBM measures multiple intervals with $E_{p}\gtrsim3$~MeV. As these values are measured with two background approaches and we find maximum photon energies of at least 5~MeV (Section~\ref{sec:maxphoton}), GRB~250702B has an unusually high peak energy for a long GRB \citep[e.g.][]{2021ApJ...913...60P}.

\begin{table*}
\caption{The spectral results of \textit{Fermi}-GBM and Konus-\textit{Wind}. All intervals are best fit with a cutoff power law, where $\alpha$ is the photon index and $E_p$ is the peak energy. The analysis intervals are shaded on the lightcurve in Fig.~\ref{fig:lc}. The \textit{Fermi}-GBM results are reported for the preferred spectral fits for analysis intervals selected using Bayesian blocks, as well as the peak flux measured over a 1~s period within Interval~7. The total fluence values reported are lower limits as the source intervals considered do not include all intervals in which emission is observed. GBM uncertainties are given at the 90\% confidence level, and fluence and flux are reported over the 1~keV to 10~MeV energy range. The Konus-\textit{Wind} results are reported for four intervals with the total being the combined spectral fit over all intervals, as well as during peak flux intervals. Konus uncertainties are given at the $1\sigma$ confidence level, and fluence and flux are reported over the 10~keV to 10~MeV energy range.} 
\centering
\begin{tabular}{cccccccccc|}
    \hline
    Instrument & Interval & Time Range - T0 & $\alpha$ & $E_p$ & Fluence \\
     & & (s) & & (MeV) & (erg/cm$^2$) \\
    \hline
    \rule{0pt}{2.5ex}
     GBM & 1 & 47,245--47,355 & $-1.22${\raisebox{0.5ex}{\tiny$^{+0.08}_{-0.08}$}} & $3.5${\raisebox{0.5ex}{\tiny$^{+0.2}_{-2}$}} & $6.1${\raisebox{0.5ex}{\tiny$^{+0.6}_{-0.5}$}}$\times10^{-5}$ \\ [0.5ex]
     & 6 & 58,625--58,685 & $-1.14${\raisebox{0.5ex}{\tiny$^{+0.06}_{-0.06}$}} & $3.4${\raisebox{0.5ex}{\tiny$^{+0.9}_{-0.7}$}} & $7.2${\raisebox{0.5ex}{\tiny$^{+0.3}_{-0.4}$}}$\times10^{-5}$ \\ [0.5ex]
     & 7 & 58,880--59,085 & $-1.24${\raisebox{0.5ex}{\tiny$^{+0.04}_{-0.04}$}} & $4.3${\raisebox{0.5ex}{\tiny$^{+1}_{-0.8}$}} & $1.79${\raisebox{0.5ex}{\tiny$^{+0.07}_{-0.07}$}}$\times10^{-4}$ \\ [0.5ex]
     & Total & & & & $>3.1${\raisebox{0.5ex}{\tiny$^{+0.1}_{-0.1}$}}$\times10^{-4}$ \\ [0.5ex]
    Konus & 1 & 46,075--47,959 & $-1.1${\raisebox{0.5ex}{\tiny$^{+0.5}_{-0.3}$}} & $0.8${\raisebox{0.5ex}{\tiny$^{+1.3}_{-0.2}$}} & $1.3${\raisebox{0.5ex}{\tiny$^{+0.7}_{-0.2}$}}$\times10^{-4}$ \\ [0.5ex]
     & 2 & 49,843--51,227 & $-1.2${\raisebox{0.5ex}{\tiny$^{+0.3}_{-0.2}$}} & $0.8${\raisebox{0.5ex}{\tiny$^{+0.7}_{-0.2}$}} & $1.2${\raisebox{0.5ex}{\tiny$^{+0.4}_{-0.2}$}}$\times10^{-4}$ \\ [0.5ex]
     & 3 & 52,884--53,720 & $-1.3${\raisebox{0.5ex}{\tiny$^{+0.5}_{-0.2}$}} & $1.2${\raisebox{0.5ex}{\tiny$^{+9}_{-0.6}$}} & $7${\raisebox{0.5ex}{\tiny$^{+6}_{-2}$}}$\times10^{-5}$ \\ [0.5ex]
     & 4 & 58,607--59,505 & $-0.6${\raisebox{0.5ex}{\tiny$^{+0.4}_{-0.3}$}} & $0.7${\raisebox{0.5ex}{\tiny$^{+0.3}_{-0.2}$}} & $1.4${\raisebox{0.5ex}{\tiny$^{+0.3}_{-0.2}$}}$\times10^{-4}$ \\ [0.5ex]
     & Total & & $-1.0${\raisebox{0.5ex}{\tiny$^{+0.3}_{-0.2}$}} & $0.8${\raisebox{0.5ex}{\tiny$^{+0.6}_{-0.2}$}} & $4.6${\raisebox{0.5ex}{\tiny$^{+1.5}_{-0.7}$}}$\times10^{-4}$ \\ [0.5ex]
 \end{tabular}
 \begin{tabular}{cccccccccc|}
    \hline
     Instrument & Interval & Time Range - T0 & $\alpha$ & $E_p$ & Flux  \\
     & & (s) & & (MeV) & (erg/cm$^2$/s) \\
    \hline
    \rule{0pt}{2.5ex}
    GBM & 7 & 59,024.082--59,025.106 & $-1.01${\raisebox{0.5ex}{\tiny$^{+0.13}_{-0.2}$}} & $3${\raisebox{0.5ex}{\tiny$^{+2}_{-2}$}} & $2.8${\raisebox{0.5ex}{\tiny$^{+0.3}_{-0.4}$}}$\times10^{-6}$ \\ [0.5ex]
    Konus & 1 & 47,284.533--47,308.085 & $-0.8${\raisebox{0.5ex}{\tiny$^{+0.5}_{-0.3}$}} & $0.8${\raisebox{0.5ex}{\tiny$^{+1}_{-0.2}$}} & $7${\raisebox{0.5ex}{\tiny$^{+4}_{-2}$}}$\times10^{-7}$ \\ [0.5ex]
     & 2 & 50,137.269--50,193.205 & $-0.9${\raisebox{0.5ex}{\tiny$^{+0.2}_{-0.2}$}} & $1${\raisebox{0.5ex}{\tiny$^{+0.8}_{-0.3}$}} & $9${\raisebox{0.5ex}{\tiny$^{+4}_{-2}$}}$\times10^{-7}$ \\ [0.5ex]
     & 4 & 59,019.316--59,034.036 & $-0.7${\raisebox{0.5ex}{\tiny$^{+0.4}_{-0.3}$}} & $0.8${\raisebox{0.5ex}{\tiny$^{+0.6}_{-0.2}$}} & $8${\raisebox{0.5ex}{\tiny$^{+3}_{-2}$}}$\times10^{-7}$ \\
    \hline
\end{tabular}
\label{table:spectra-summary}
\end{table*}

\subsection{Intrinsic Energetics}
The observed peak flux and fluence at Earth can be converted into isotropic-equivalent total intrinsic peak luminosity, $L_{\rm iso}$, and total energetics, $E_{\rm iso}$, by accounting for the inverse square law and converting to a rest-frame bolometric energy range, i.e. 1--10,000~keV, after accounting for cosmological expansion and redshift \citep{bloom2001prompt}. The Konus time-integrated spectral fit gives $E_{\rm iso} \gtrsim 1.4${\raisebox{0.5ex}{\tiny$^{+0.4}_{-0.2}$}}$\times10^{54}$~erg and rest-frame peak energy $E_{p,i,z}=1.7${\raisebox{0.5ex}{\tiny$^{+1.3}_{-0.4}$}}~MeV. Adding the sum of the GBM measurements and accounting for the partial coverage gives reasonably consistent results. Konus data gives $L_{\rm iso}=4.8${\raisebox{0.5ex}{\tiny$^{+2.0}_{-1.0}$}}$\times10^{51}$~erg/s with a peak flux peak energy $E_{p,p,z}=1.6${\raisebox{0.5ex}{\tiny$^{+1.3}_{-0.4}$}}~MeV while GBM data gives $L_{\rm iso}\sim4.0\times10^{51}$~erg/s and $E_{p,p,z}=6${\raisebox{0.5ex}{\tiny$^{+4}_{-4}$}}~MeV. 

\begin{figure*}
    \centering
    \includegraphics[width=0.32\textwidth]{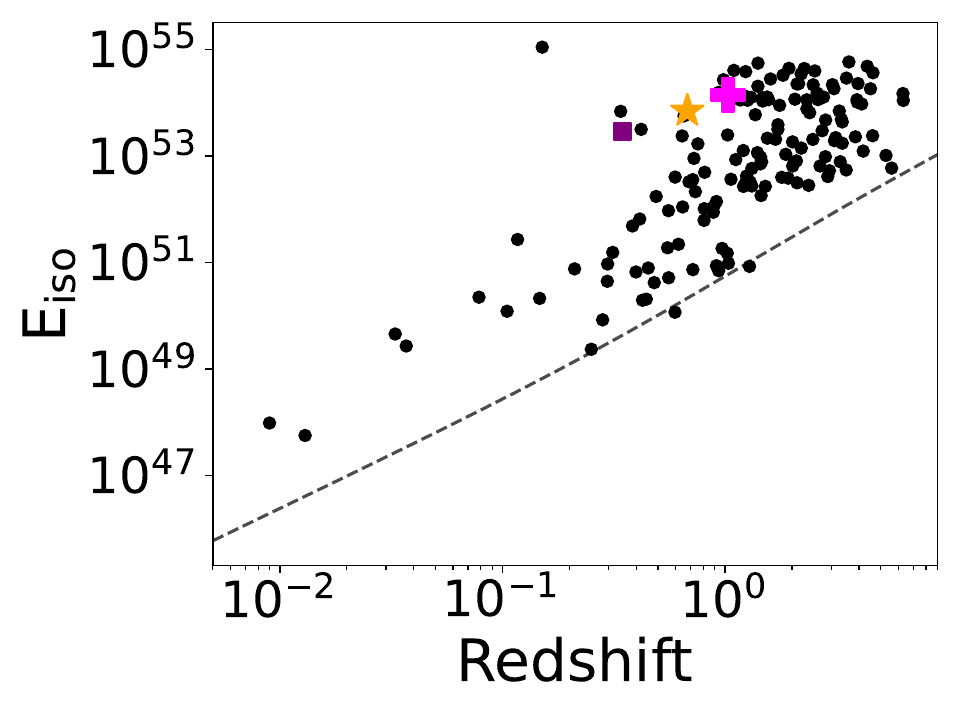}
    \includegraphics[width=0.32\textwidth]{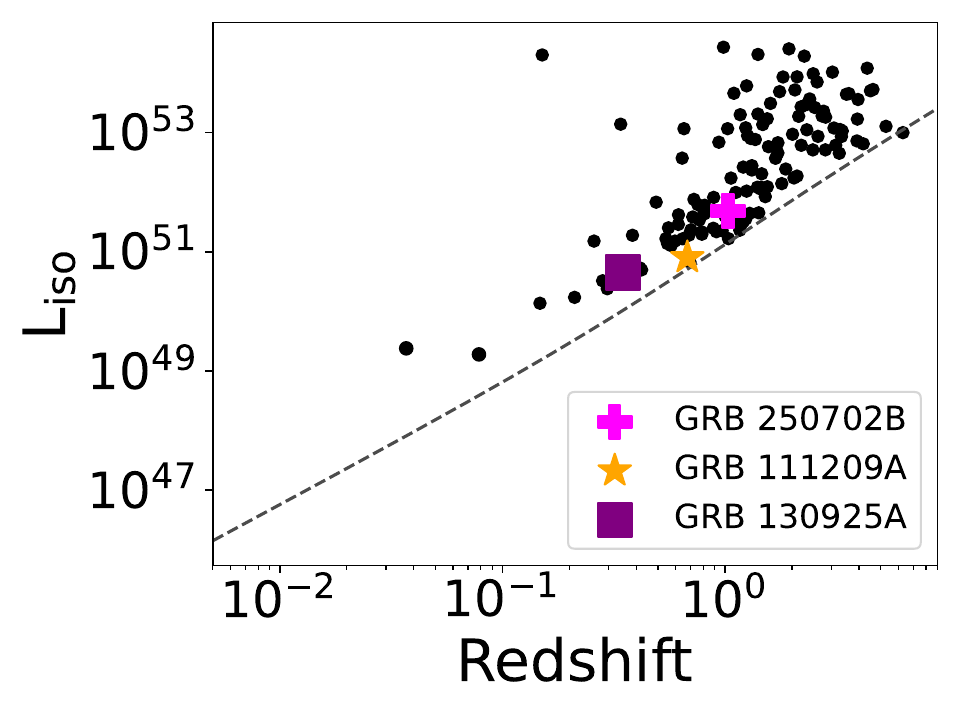}
    \includegraphics[width=0.32\textwidth]{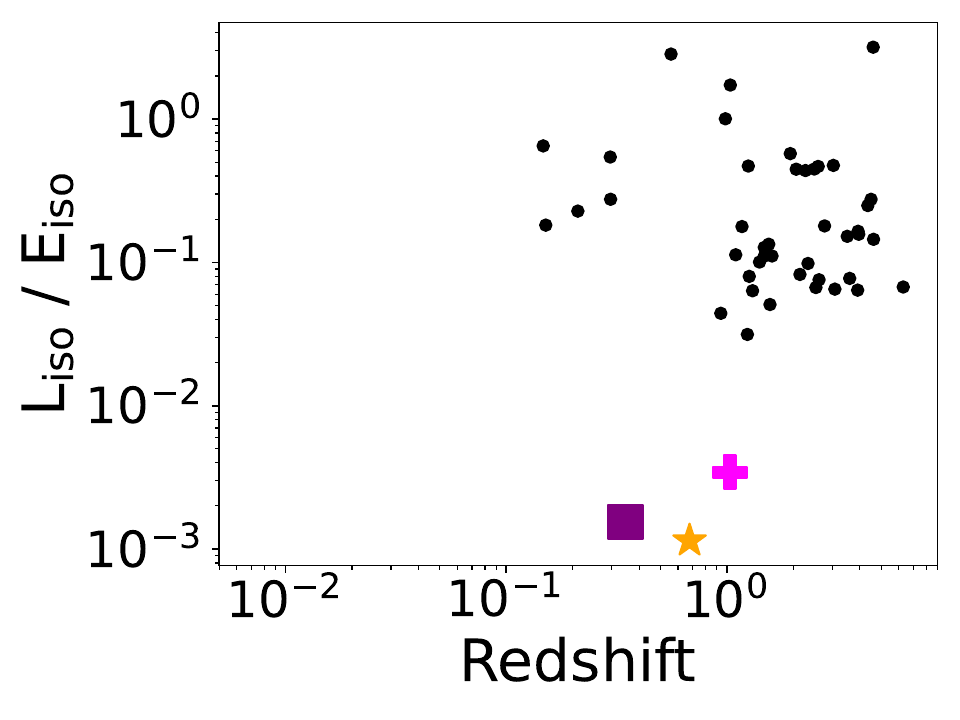}
    \caption{GRB~250702B intrinsic energetics compared to the broader population of GRBs from \citet{burns2023grb}. We also highlight the second and third longest bursts. These events have high $E_{\rm iso}$ and low $L_{\rm iso}$ values that fall within the known distributions, but have extreme ratios of these two values incompatible with the broader population. The dashed line in the left two plots correspond to the approximate \textit{Fermi}-GBM detection threshold.}
    \label{fig:energetics}
\end{figure*}

\begin{figure*}
    \centering
    \includegraphics[width=0.7\textwidth]{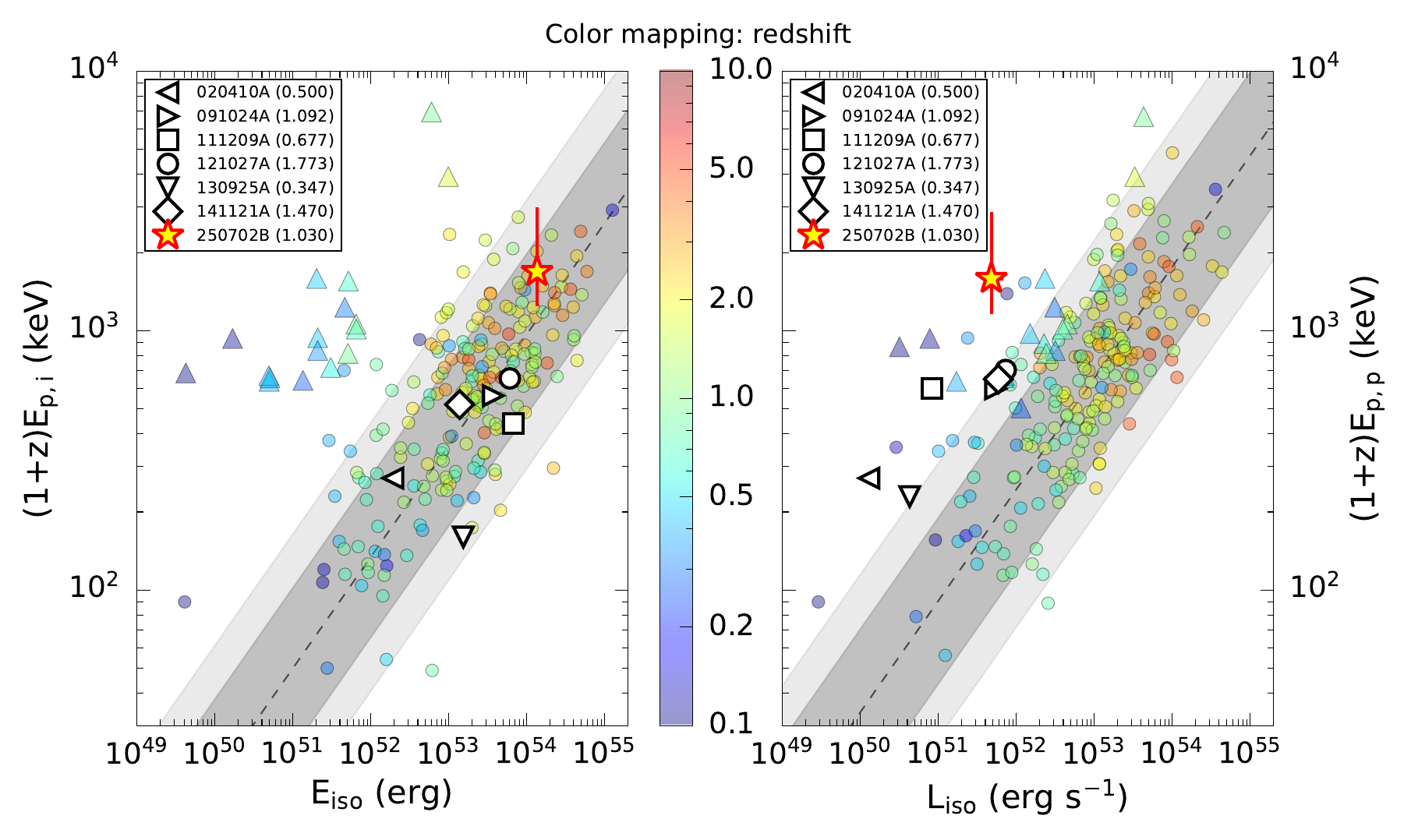}
    \caption{The Konus-\textit{Wind} Amati (left) and Yonetoku (right) relations with 68\% and 90\% confidence intervals, using data from \citet{tsvetkova2017konus,tsvetkova2021konus}. $E_{p,i}$ is the peak energy from the time-integrated interval, while $E_{p,p}$ is the peak energy from the peak flux interval. GRB~250702B is marked with a star, while other ultra-long GRBs (defined as durations above 1,000~s) with measured redshift are shown with white shapes.}
    \label{fig:relations}
\end{figure*}

These measures allow us to place GRB~250702B in the context of other bursts, shown in Fig.~\ref{fig:energetics}, showing fairly typical $E_{\rm iso}$ and $L_{\rm iso}$, but the ratio between these values is a clear outlier to the broader distribution shown. The Amati and Yonetoku relations are shown in Fig.~\ref{fig:relations}. Intriguingly, GRB~250702B is harder than expected from the broader long GRB Yonetoku relation, with the GBM $E_p$ resulting in even greater inconsistency than the Konus value. Further, other ultra-long GRBs also exceed the expected $E_p$ value for their luminosity at the $\gtrsim$90\% level, suggesting a distinct population. 

\citet{levan2025day} and \citet{oconnor2025grb250702b} utilize afterglow measurements to infer properties of the jet, including the isotropic-equivalent kinetic energy, $E_{k,\rm iso}$ and an exceptionally narrow half-jet opening angle, $\theta_j\lesssim1\degree$. \citet{oconnor2025grb250702b} find the collimation-corrected kinetic energy to be $E_{k,\rm coll}=3.49${\raisebox{0.5ex}{\tiny$^{+1.36}_{-1.02}$}}$\times10^{50}$~erg. With this, we infer a collimation-corrected gamma-ray energy release $E_{\gamma,\rm coll}=5.82${\raisebox{0.5ex}{\tiny$^{+8.75}_{-5.6}$}}$\times10^{49}$~erg. This together gives a total jet energy of $E_{\rm jet,coll}=4.08${\raisebox{0.5ex}{\tiny$^{+1.62}_{-1.16}$}}$\times10^{50}$~erg, and the gamma-ray efficiency is $\sim$14\%{\raisebox{0.5ex}{\tiny$^{+22\%}_{-13\%}$}}. Using the Konus $L_{\rm iso}$ value, the collimation-corrected peak luminosity is $L_{\gamma,\rm coll}=2.0${\raisebox{0.5ex}{\tiny$^{+3.06}_{-1.94}$}}$\times10^{47}$~erg/s. These collimation corrected energetics are all within the normal distribution for typical long GRBs \citep{tsvetkova2017konus,o2023structured,levan2025day,oconnor2025grb250702b}. However, GRB~250702B has a higher $E_p$ for its $E_{\gamma,\rm coll}$ value compared to expectations from the broader population, while its peak luminosity $L_{p}$ for its $L_{\gamma,\rm coll}$ falls within the normal range (see fig. 13 \citealt{tsvetkova2021konus}). This is the reverse of the isotropic-equivalent relations.

\subsection{Maximum Photon Energy}\label{sec:maxphoton}
Photons in excess of 1~MeV are often detected in GRB prompt emission and are used to put limits on the bulk Lorentz factor from pair opacity arguments. Thus, we seek to understand the highest energy photons observed from GRB~250702B. To quantify this, we search \textit{Psyche}-GRNS and \textit{Fermi}-GBM data for the highest energy channel numbers which show significant emission using a basic SNR calculation. The \textit{Psyche}-GRNS high-energy data channels 30-40 contain a 3$\sigma$ excess from T0+58,100~s to T0+59,400~s (referenced to time at Earth), corresponding to a deposited energy of 2.1--2.8~MeV. In \textit{Fermi}-GBM BGO data, analysis Interval 6 has a 3.4$\sigma$ excess over 5.1--6.3 MeV and Interval 7 a 3.4$\sigma$ excess over 4.9--5.5 MeV. The energy ranges here refer to the energy deposited into the detector which is a (probabilistic) lower bound on the incident photon energies. Thus, we have observer frame photons above $\gtrsim$5~MeV, and rest-frame photons above $\gtrsim$10~MeV. 

\subsection{Quasi-periodic Oscillations}
We use Konus-\textit{Wind} data to search for quasi-periodic oscillations, taking advantage of the complete coverage of the $\sim$25,000~s gamma-ray emission interval and 2.944~s temporal resolution. We use two approaches, as detailed in Appendix~\ref{app:qpo}. One method is a standard power spectrum analysis, leveraging the full photon statistics. The other is a cross-spectrum analysis, using both Konus detectors to isolate white noise and some instrumental effects. There is some excess at $\sim$3~mHz but neither approach rejects the power-law only model at the 99\% confidence level (each reporting a p-value of $\sim$2\%). Thus, we find no significant quasi-periodic oscillation in GRB~250702B, but we encourage similar searches in other ultra-long GRBs. We find no excess power at low frequencies, excluding the possibility raised in  \citet{levan2025day} of $\sim$2,825~s periodicity.

\section{Prompt Gamma-Ray Burst Inferences}\label{sec:inferences}
The measurement of the gamma-ray properties in the previous section enable a number of key inferences on this event and the central engine which created it. These are crucial measurements for understanding the progenitor system. We briefly note the intrinsic energetics suggest a comparable total energy reservoir available to ultra-long GRBs compared to collapsars, with the only clear distinction being that this power dissipated over a far longer duration. 

\subsection{Central Engine Duration}\label{sec:engine-duration}
The extreme observed duration requires a long-lived central engine. The 12,500~s rest-frame gamma-ray duration relates to the timescale of the most rapid accretion (see Section~\ref{sec:bhsize}). However, the central engine activity in GRB~250702B is longer than this time. First, early \textit{Swift}-XRT observations beginning at $\sim$T0+92,000~s show rapid fading, indicative of the end of flaring behavior \citep{2025GCN.40919....1K}. Second, EP reports detections of this event from stacked observations on July 1 \citep{2025GCN.40906....1C}. At these distances, an X-ray signal must be produced by accretion and is a good indicator of central engine activity \citep{parsotan2024photospheric}. If we take the EP signal start time as $\sim$noon on July 1 and have uncertainty of half a day, the overall central engine time (rest-frame) is $\sim$90,000 $\pm$ 20,000~s. The bright and impulsive gamma-ray signal occurs $\sim$45,000-60,000~s after the EP detection begins (with $\sim$20,000~s uncertainty from the lack of precise start time). This delay to peak power is unusual.

\subsection{Bulk Lorentz Factor}\label{sec:bulk-Gamma}
The coasting bulk Lorentz factor ($\Gamma_0$) of the relativistic jet, before it is decelerated by its interaction with the external medium, can be obtained from compactness arguments \citep{Piran-99}. Since GRBs are intense sources of gamma-rays, an ultra-relativistic jet ($\Gamma_0\gg1$) is required so that the photons near and above the $\nu F_\nu$ peak energy, $E_p$, are not absorbed due to $\gamma\gamma$-annihilation ($\gamma\gamma\to e^-e^+$). In this case, the spectral cutoff due to $\gamma\gamma$-annihilation occurs at the energy $E_{\rm cut}>E_p$ where the optical depth to $\gamma\gamma$-annihilation is $\tau_{\gamma\gamma}(\Gamma_0,E_{\rm cut})=1$. Here we model $\tau_{\gamma\gamma}(\Gamma_0,E)$ using the standard approach from \citet{Granot+08} to infer $\Gamma_0$ when a spectral cutoff is seen or obtain an estimate of the minimum Lorentz factor ($\Gamma_{0,\min}$) when the observed spectrum only extends to some maximum energy ($E_{\max}$). More details are provided in Appendix~\ref{app:gamma}. 

For GBM Interval 7 of GRB~250702B, the MVT is $\sim$1.4~s and the maximum observed photon energy is $E_{\max}\sim5$~MeV, with both parameters listed in the observer frame. The best-fit cutoff power-law model shows a spectral peak at $E_p\sim4$\,MeV, with a hard photon index $\alpha=-1.24$ below this energy. When interpreting the sharp spectral break as the high-energy spectral cutoff, with $E_{\rm cut}=E_p$, this model yields an estimate of the true bulk Lorentz factor of $\Gamma_0\simeq81$ for the source redshift of 1.036. A power-law spectral model that also yields an acceptable fit to the data and extends to $E_{\max}$ is used to obtain $\Gamma_{0,\min}=56$. Both estimates compare favorably with $\Gamma_0$ inferences from afterglow studies (see Appendix~\ref{app:gamma} as well as \citealt{levan2024heavy} and \citealt{oconnor2025grb250702b}) that also require an ultra-relativistic jet in this event. In both spectral models, we find that the emission region is highly optically thick to Thomson scattering by the created $e^\pm$-pairs, in which case the observed spectrum is significantly modified and explains the hard photon index of $\alpha\gtrsim-2$ at energies below the spectral cutoff energy \citep{Gill-Granot-18}.

\subsection{Black Hole Size}\label{sec:bhsize}
The shortest variation which can be produced by an emitting region is determined by its physical size and the finite speed of light, i.e., a one light-second diameter object will not have variability at subsecond timescales, though it can be longer. Thus, minimum variability timescales in emission from accretion disks give insight into the physical size of the central engine. In the case of black holes, the Schwarzschild crossing time scales linearly with black hole mass, so constraints on physical size are also constraints on mass.

However, as detailed in Section~\ref{sec:bulk-Gamma}, our analysis requires the emitting region to be moving towards us at relativistic velocities. The emission from relativistic jets typically occurs at a radius much larger than that of the central object. For example, a shell with a large Lorentz factor emitted with an interval $\Delta t$ after an initial shell with a smaller Lorentz factor $\Gamma$ collides with the latter at $R\approx 2c\Gamma^2 \Delta t$. The Doppler contraction of the observed pulses means that the observed variability is much shorter than $R/c$ and is instead of order $R/(2c\Gamma^2)\sim \Delta t$ \citep{1997ApJ...490...92K}. In other words, the two factors cancel out and the observed variability is on the order of the variability that is produced by the central engine. In practice, the observed values are substantially above this lower limit.

For example, for a stellar-mass black hole, this timescale is $<$100~$\mu$s, whereas the variability timescales observed in GRB lightcurves, which are thought to host such stellar-mass black hole central engines, are orders of magnitude larger with typical values being $\sim$0.01--10~s \citep{golkhou2014}. The gamma-ray emission from GRB~250702B has a rest-frame MVT of $\sim$0.5~s and is produced in a relativistic jet. Based on the comparable MVTs seen in collapsar GRBs, which arise from $\sim$3~M$_\odot$ black holes, we favor a stellar-mass central engine for GRB~250702B. The majority of collapsar GRBs have MVTs above $\sim$15~ms \citep{veres2023extreme}. A fiducial scaling of our observed rest frame MVT against this value, i.e. 3~M$_\odot \times0.5$~s/$0.015$~s, gives a nominal upper limit on the black hole mass of GRB~250702B of $\lesssim$100~M$_\odot$. 

\section{Excluded Progenitors}\label{sec:exclusion-progenitor}
A key question is the physical progenitor system which created this transient. We here describe numerous options from the literature which are sufficiently advanced to be testable, largely selected from \citet{fryer2019understanding} and from models invoked to explain previous ultra-long GRBs. Each model is compared with our results directly as summarized in Table~\ref{tab:progenitor}. A detailed discussion of each individual progenitor scenario, including scenarios where a normal collapsar produces such a long signal due to effects induced in propagation, is provided in Appendix~\ref{app:progenitor}. We utilize the engine duration, power, and evolution discussion in Section~\ref{sec:helium-merger} in our consideration of whether a given model can explain GRB~250702B.

\begin{table*}
\caption{Known or theoretical progenitors for gamma-ray burst signatures, listed with their gamma-ray properties. The progenitor parameter values are only order of magnitude, reflecting significant theoretical or observational uncertainty, but are sufficient for our purposes. Viability is marked in the last column, sometimes relying on additional information detailed in the text, with the only viable options involving a stellar-mass compact object consuming a star and involving more angular momentum than can be contained within a star.}\label{tab:progenitor}
\centering
\begin{tabular}{rccccc}
\hline
 & Engine & Minimum Variability & Maximum Photon & Power \& & Viable \\
 & Duration (s) &  Timescale (s) &  Energy (MeV) & Profile & (Y/N)\\
\hline
X-ray Binaries & 1,000,000 & - & 0.5 & Y & N \\
Magnetar Giant Flare & 0.01--0.1 & 0.001--0.1 & 5 & N & N \\
Neutron Star Mergers & 0.01--10 & 0.001--1 & 10 & N & N \\
White Dwarf Mergers & 100--10,000 & - & - & N & N \\
Tidal Disruption Event & 250,000 & $\gtrsim$40 & 1 & Y & N \\
IMBH Tidal Disruption Event & $\sim$10,000 & $\gtrsim$10 & 1 & N & N \\
Micro Tidal Disruption Event & 10,000-100,000 & 0.01--10 & 10 & Y & Y \\
Carbon-Oxygen Collapsar & 1--1,000 & 0.01--10 & 10 & N & N \\
Helium Collapsar & 1--1,000 & 0.01--10 & 10 & N & N \\
Binary Helium Star Merger & 1--1,000 & 0.01--10 & 10 & N & N \\
Helium Merger & $\sim$100,000 & 0.01--10 & 10 & Y & Y \\
\textbf{GRB 250702B} & $\sim$100,000 & 0.5 & 10 & - & -\\
\hline
\end{tabular}
\end{table*}

Most long GRBs arise from collapsars, which are an ideal scenario to describe GRB~250702B except for the extreme duration. As detailed in the Section~\ref{sec:helium-merger}, extreme angular momentum can arrest material in the disk, extending accretion time. \citet{fryer2025explaining} explore the maximal durations that can occur when a star is spinning at breakup, where faster velocities would spin the star apart, finding a maximum value of a few thousand seconds. Thus, these events are inconsistent with the measured duration of GRB~250702B at two orders of magnitude, and we are motivated to explore other scenarios. 

X-ray binaries and other Galactic sources are excluded by our $\sim$10~MeV rest-frame photons and the identification of the host galaxy in \citet{levan2025day}. Magnetar giant flares and neutron star mergers are excluded because of insufficient durations by orders of magnitude. White dwarf mergers, carbon-oxygen collapsars, helium collapsars, and binary helium star mergers are excluded because their durations cannot reproduce the total central engine time by $\sim$two orders of magnitude and because each would predict a peak power at early times, in contrast to the significant delay to peak power observed in GRB~250702B. 

Traditional TDEs from supermassive black hole mergers are excluded because of their long MVTs. For a direct comparison we repeat our MVT analysis for the \textit{Swift}-BAT observation of the TDE Swift~J1644+57 and find a rest-frame value of $\sim$40~s (shown in Fig.~\ref{fig:t90-epeak-mvt} and detailed in Appendix~\ref{app:tde}). As this is the shortest MVT ever seen from a TDE, the MVT of known TDEs are $\gtrsim$2 orders of magnitude greater than GRB~250702B. Follow-up observations also disfavor this origin due to the non-nuclear position of the transient with respect to the host and lack of late-time transient light \citep{levan2025day,gompertz2025grb250702b}.

\citet{levan2025day} consider the possibility of a white dwarf tidal disruption by an intermediate mass black hole (IMBH) to explain GRB~250702B. This model faces several issues. With respect to gamma-rays, such a model is still inconsistent with our MVT based on the scaling of our jetted emitting region suggesting the central engine mass to be $\lesssim$100~M$_\odot$. Indeed, a white dwarf disrupting around an IMBH was already invoked to explain $\sim$100~s variability \citep{krolik2011swift} which is $>$100 times larger than our value. Further, the duration of a white dwarf merger with an IMBH will not be longer than typical white dwarf mergers where the peak signal is $\sim$150~s and the longest duration is less than 15,000~s, supported both by our subsequent central engine modeling as well as hydrodynamical simulations in \citet{oconnor2025grb250702b}.

More generally, GRB~250702B appears to have fairly typical collapsar jet energetics except it is extreme in having a high peak energy and particularly narrow jet \citep{oconnor2025grb250702b}. Powerful jets are only known to arise from rapidly spinning black holes, with collapsar GRB stellar-mass black holes spun up during accretion of the star and active galactic nuclei spun up by accretion over enormous timescales. The power from a Blandford-Znajek jet is proportional to both the accretion rate and the square of the black hole spin (equation~\ref{eq:BZform}); most engine models have similar relations. There is no obvious reason for an IMBH to be rapidly spinning in the prograde direction to the orbit of a white dwarf it disrupts. Even if this occurs, a white dwarf is a substantially smaller energy reservoir than available in collapsars, even before accounting for losses to thermonuclear explosions \citep{2009ApJ...695..404R}. The accretion power from these systems will be lower than models invoking stellar-mass black holes. 

Narrow jets are thought to be collimated by their central engine and surrounding material. As an IMBH engine should be less efficient than a typical collapsar, and also have less material in the polar regions, there is no obvious explanation for the narrow jet. One may ameliorate these problems and the required energy reservoir by invoking disruption of a massive star around an IMBH, but this would exacerbate the MVT problems \citep{rees1988tidal,krolik2011swift}. 

In either IMBH case, there is also no obvious explanation for the significant delay to peak power output. Additionally, follow-up observations show a temporal decay more typical of GRB afterglow rather than the fiducial -5/3 power-law decay expected in TDEs \citep{levan2025day,oconnor2025grb250702b}. Thus, we do not consider this model viable as an IMBH TDE origin would require black hole mass to MVT scalings which differ from other GRBs, expect different durations, power profiles, and temporal decays, and is unlikely to produce the jet properties seen in GRB~250702B.

\section{Helium Star Mergers}\label{sec:helium-merger}\label{sec:COM}
The only models which naturally explain GRB~250702B involve a rapidly-spinning stellar-mass black hole and a total energy reservoir of similar order to collapsar GRBs, given our measured intrinsic energetics. Since the duration cannot be achieved with the angular momentum possible in a single star (or two), orbital angular momentum must be tapped. Further, the orbital angular momentum must be added as the engine is starting, so that the star does not fully spin itself apart before accretion can occur. Thus, viable progenitor scenarios require the infall of a black hole into a star, with variations including evolution of field binaries, dynamical capture, or a hybrid option. We briefly introduce these options here. If helium stars expand, and there is reason to believe this occurs \citep[e.g.][]{van1976structure,linden2012rarity,fryer2025explaining}, then helium star mergers (i.e., the merger of a compact object with a stripped star) should occur fairly often in the universe when compared with other GRB progenitors. When helium star mergers do occur, the extreme angular momentum and stellar-mass engine should power a jet. Thus, we focus on the field binary helium merger model below with Section~\ref{sec:he-jet} and Section~\ref{sec:he-sn} focusing on our expectations for jetted and supernova emission in GRB~250702B, and Section~\ref{sec:he-pop} discussing population-level expectations.

Massive stars go through a series of expansion phases that, in binary systems, can lead to a situation where the binary companion is immersed in the expanding stellar envelope.  The loss of orbital angular momentum in this common envelope scenario (through friction from tidal forces or bow shocks) cause the binary orbit to shrink~\citep{2013A&ARv..21...59I}.  Mass transfer phases in binaries are roughly classified into 3 categories: expansion during hydrogen burning, expansion after hydrogen exhaustion, and expansion after helium depletion (there can be multiple stellar expansions after helium depletion). The helium core merger scenario covers cases where a compact object falls into the secondary star in the system. Initial work on this progenitor focused on the scenario where the common envelope occurs in a massive star/black hole binary after hydrogen exhaustion~\citep{1999ApJ...520..650F,2001ApJ...550..357Z}. For a fraction of these common envelope scenarios, the orbit tightens so much in the hydrogen common envelope that it ultimately merges into the helium core of the star. The angular momentum lost from the orbit goes into the helium star and when the black hole reaches the center of the core, this high angular momentum will cause the helium core to accrete through a disk.  This disk can produce the magnetic fields required to drive jets and viscosity in the disk will drive strong winds. This will explode the star and produce a supernova, similarly to the supernova engine in collapsars.

Variants of this system exist. The compact object can be a neutron star instead of a black hole.  Because the neutron star will accrete rapidly due to neutrino cooling~\citep{1996ApJ...460..801F}, it will quickly collapse to a black hole.  The system will also occur in more evolved cores (e.g. after helium or even CO depletion). Thus, the phrase helium star merger generically refers to the merger of a compact object, which is or will become a black hole, with a stripped star \citep{1999ApJ...520..650F}. Such a model has been invoked to explain previous ultra-long GRBs including GRB~101225A \citep{2011Natur.480...72T}. We note that a helium star merger will occur only in regions with active star formation, as is seen in GRB~250702B \citep{gompertz2025grb250702b}.

A related possibility is tidal capture of a star by stellar-mass compact object due to dynamical interactions in a dense stellar environment. The compact object can begin as a neutron star or a black hole, but a neutron star would accrete into a black hole. This results in a micro-TDE \citep{Perets2016}. Alternatively, micro-TDEs may also arise in field binaries, where the natal-kick due to the collapse of one star sends it into an eccentric orbit that results in disruption of the companion on the newly formed compact object. Beniamini (in preparation) shows that the association with a stellar-mass compact object naturally explains the rapid gamma-ray variability, hard gamma-ray spectrum, and off-center galactic location. At the same time, the fallback time naturally explains the very long duration. The rates of partial/repeating disruptions via this channel are comparable to those of full disruptions. The former can provide a natural explanation for the observed X-ray precursor.

\subsection{Jets}\label{sec:he-jet}
\begin{figure*}
    \centering
    \includegraphics[width=0.49\textwidth]{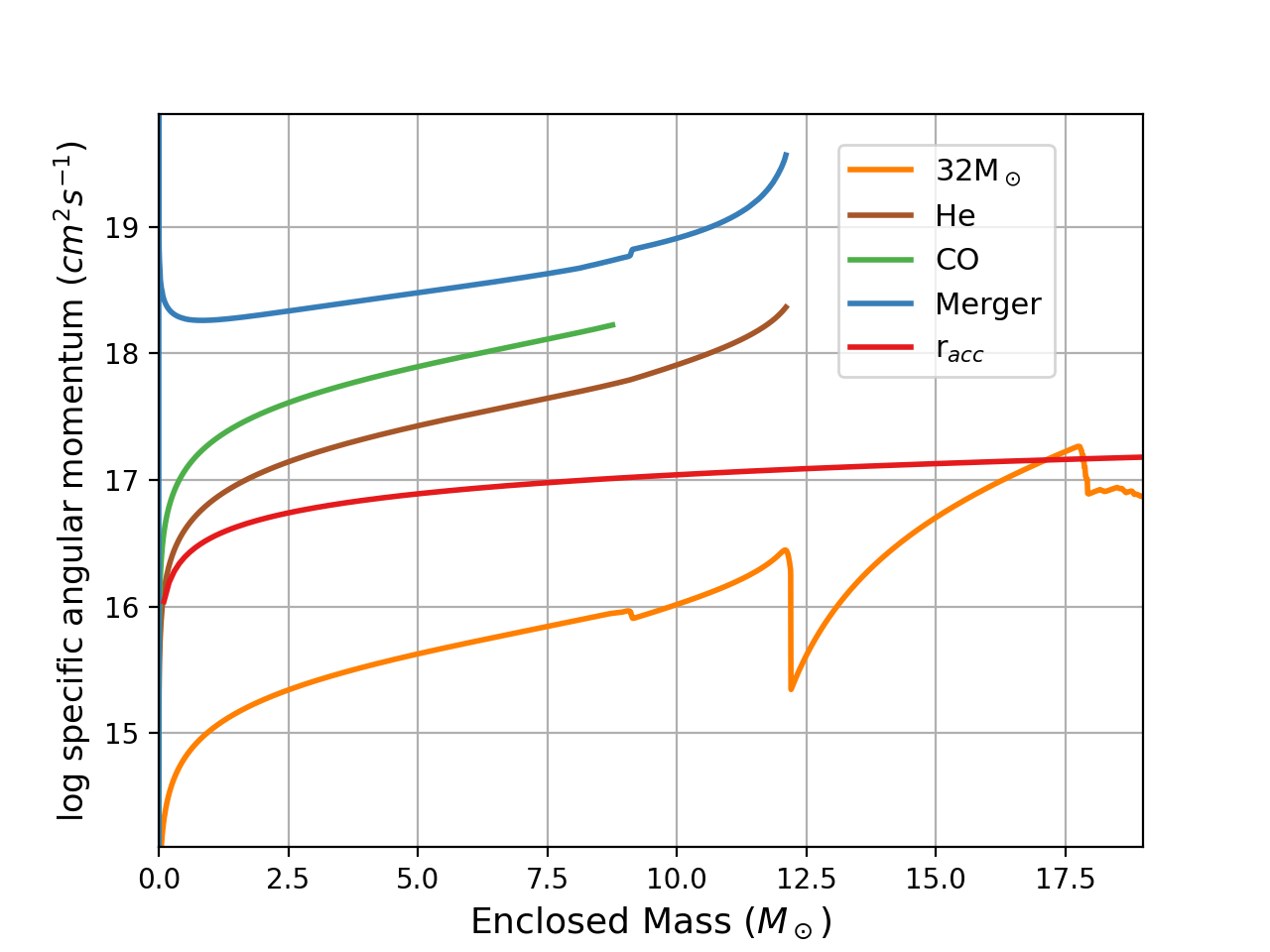}
    \includegraphics[width=0.49\textwidth]{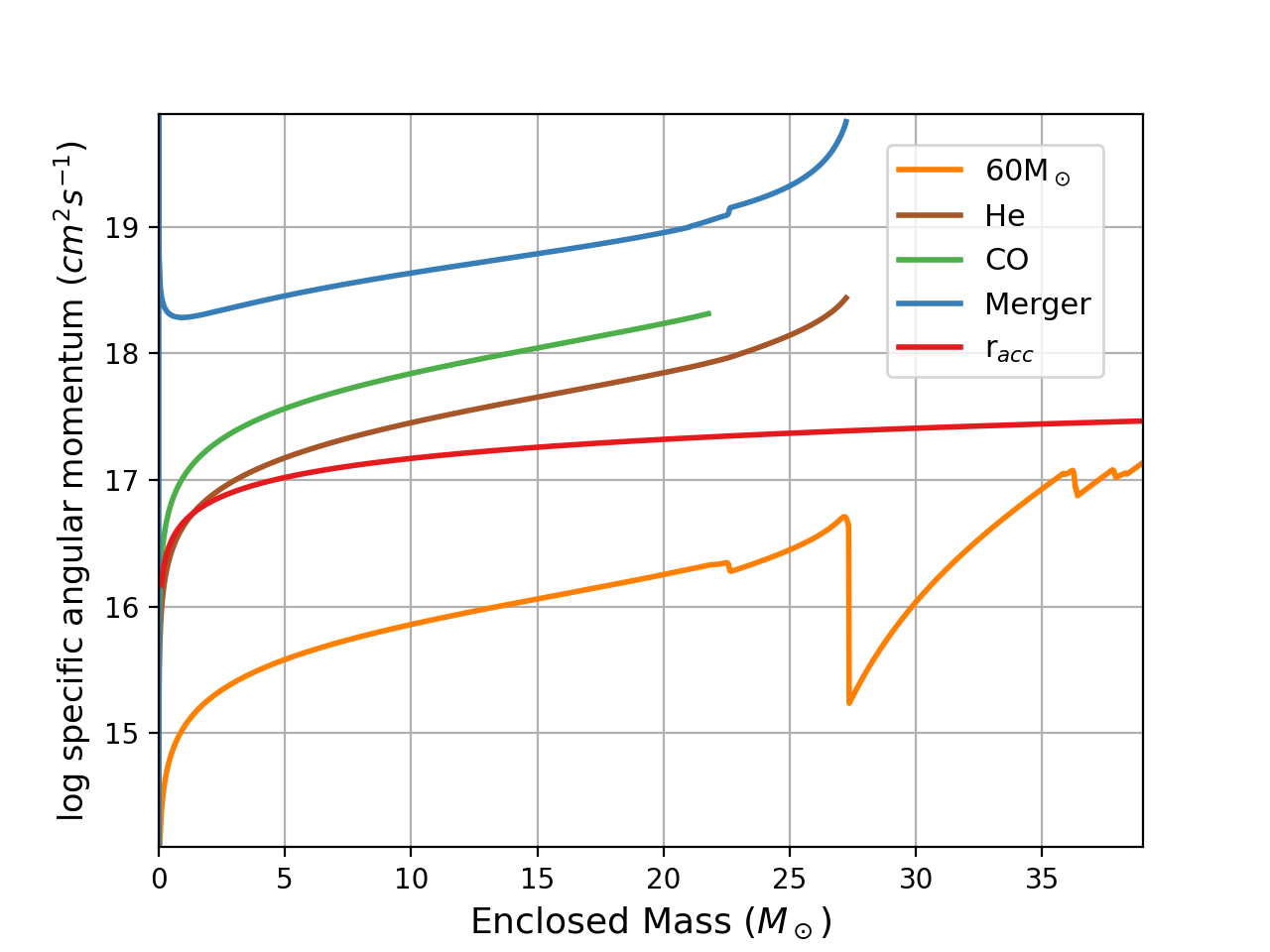}
    \caption{Specific angular momentum versus mass coordinate for a single star with moderate dynamo models locking different burning layers, tidally-locked Helium and CO star collapsars \citep{fryer2019understanding,fryer2025explaining}, and our helium star merger model.  The two plots show the results for the cores produced by two different 1/10th solar, stellar models (32 and 60\,M$_\odot$ zero-age main sequence stars) at CO depletion. A powerful jet requires an accretion disk at $\sim$100~km, i.e. viable GRB models must exceed the red line above.}
    \label{fig:angularmom}
\end{figure*}

We seek to understand the properties of the jets which can be produced in helium mergers. Estimating the duration of jets produced in these mergers requires a set of assumptions about accretion timescale and the jet power. We assume that the rotational angular momentum in the star post-merger is set by the orbital angular momentum lost as the compact remnant inspirals.  The angular momentum profile of the star post merger, compared to the angular momentum profiles of tidally-locked binaries \citep{fryer2019understanding} and single stars with moderate dynamo models locking different burning layers, is shown in Fig.~\ref{fig:angularmom}.  

\begin{figure*}
    \centering
    \includegraphics[scale=0.5]{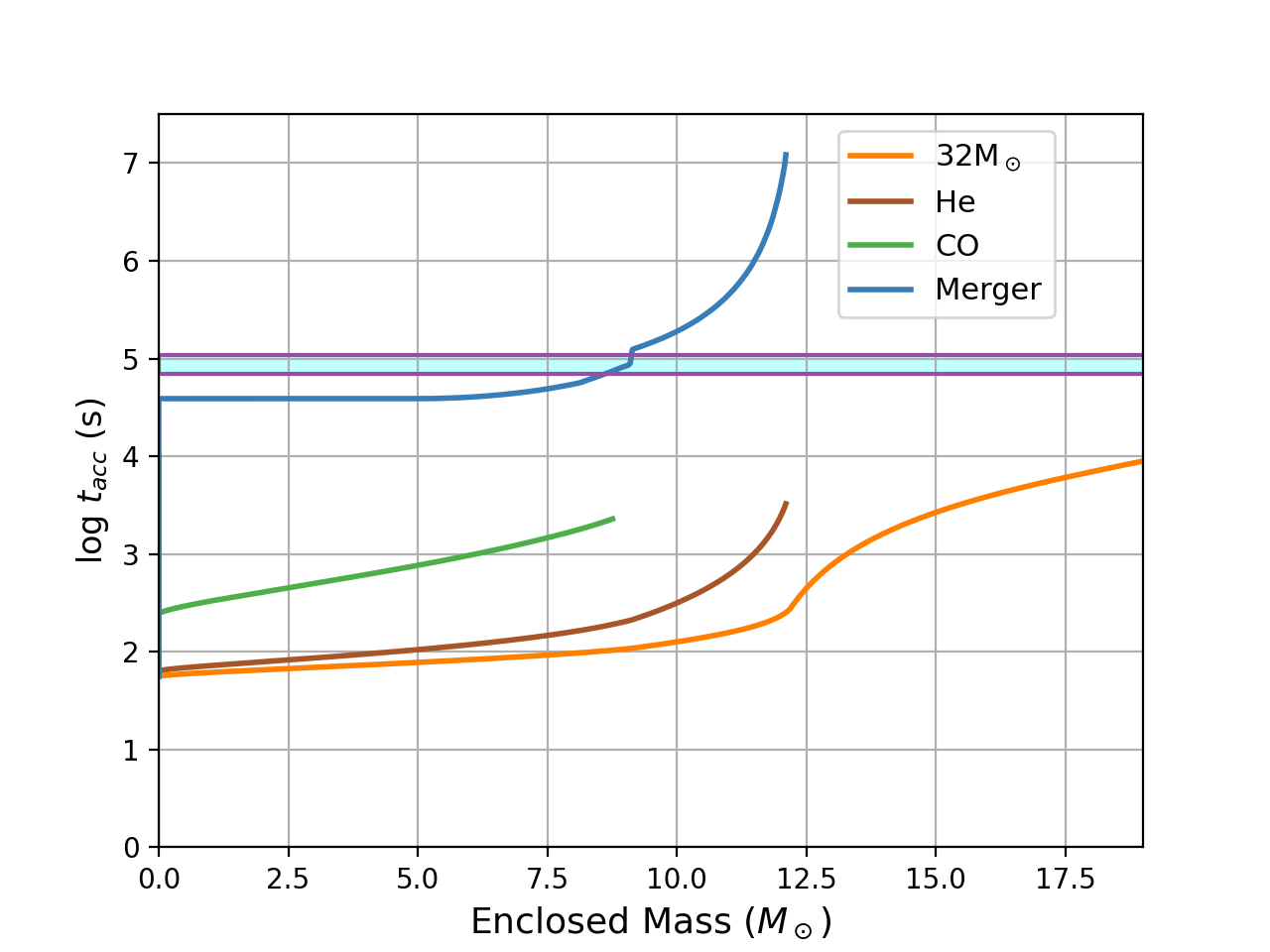}
    \includegraphics[scale=0.5]{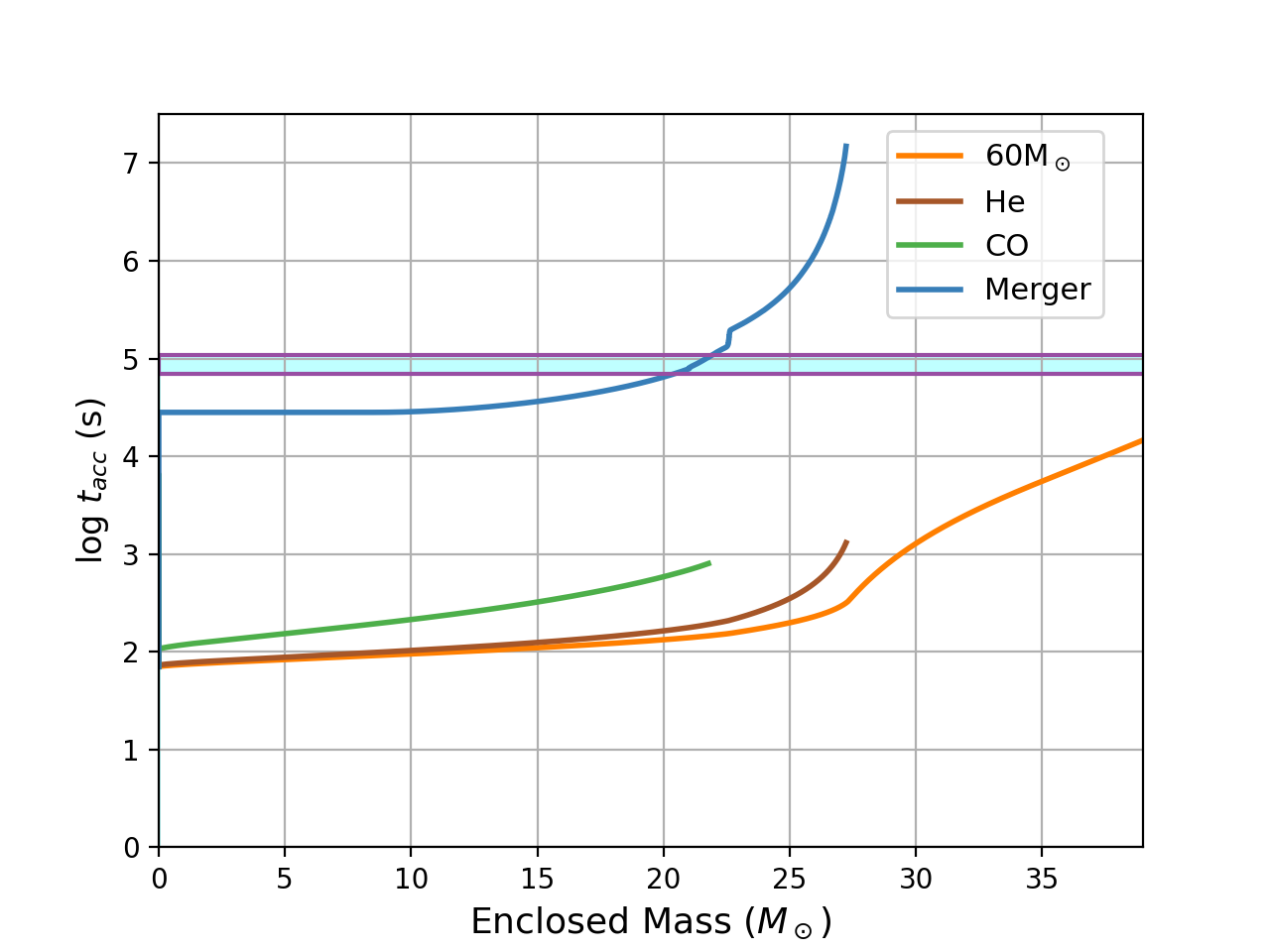}
    \caption{Accretion timescales versus mass coordinate for the models in Fig.~\ref{fig:angularmom}.  The timescales for single stars are dominated by the free-fall time, but the accretion timescale for helium-mergers is entirely set by the disk accretion timescale. Our inferred central engine duration is shown as a horizontal line, where only the helium star merger is remotely viable.}
    \label{fig:tacc}
\end{figure*}

For general exploration of GRB accretion times, the accretion rate is set by the sum of the free-fall time and disk accretion time for these mergers.  The free-fall time is given by:
\begin{equation}
    t_{\rm ff} = \frac{\pi r^{3/2}}{2 \sqrt{2GM}}
    \label{eq:freefall}
\end{equation}
where $r$ is the position of the material, $G$ is the gravitational constant and $M$ is the enclosed mass. The corresponding timescale for accretion through an $\alpha-$disk is:
\begin{equation}
    t_{\rm disk} = \frac{2 \pi r^{3/2}_{\rm disk}}{\alpha \sqrt{GM}} = \frac{2 \pi j_{\rm rot}^3}{\alpha G^2 M^2}
\end{equation}
where $r_{\rm disk}$ is the radius where the material hangs up in the disk set by the specific angular momentum, $j_{\rm rot}$, and $\alpha$ is the effective viscosity assuming a standard $alpha$-disk model~\citep{1973A&A....24..337S}. Fig.~\ref{fig:tacc} shows the accretion timescales for our single star, tidally spun-up, and merger models.  Typically, the free-fall time is rapid. If the angular momentum is not extremely high, the disk is compact (e.g. 100-1,000~km) and the free-fall time can dominate the accretion timescale.  This will be true for most models. For helium star mergers the extreme angular momentum sets the accretion timescale. The helium star merger accretion timescale can be orders of magnitude higher than any collapsar case and is the only scenario which can explain GRB~250702B. Other models are insufficient by more than an order of magnitude.

\begin{figure}
    \centering
    \includegraphics[scale=0.5]{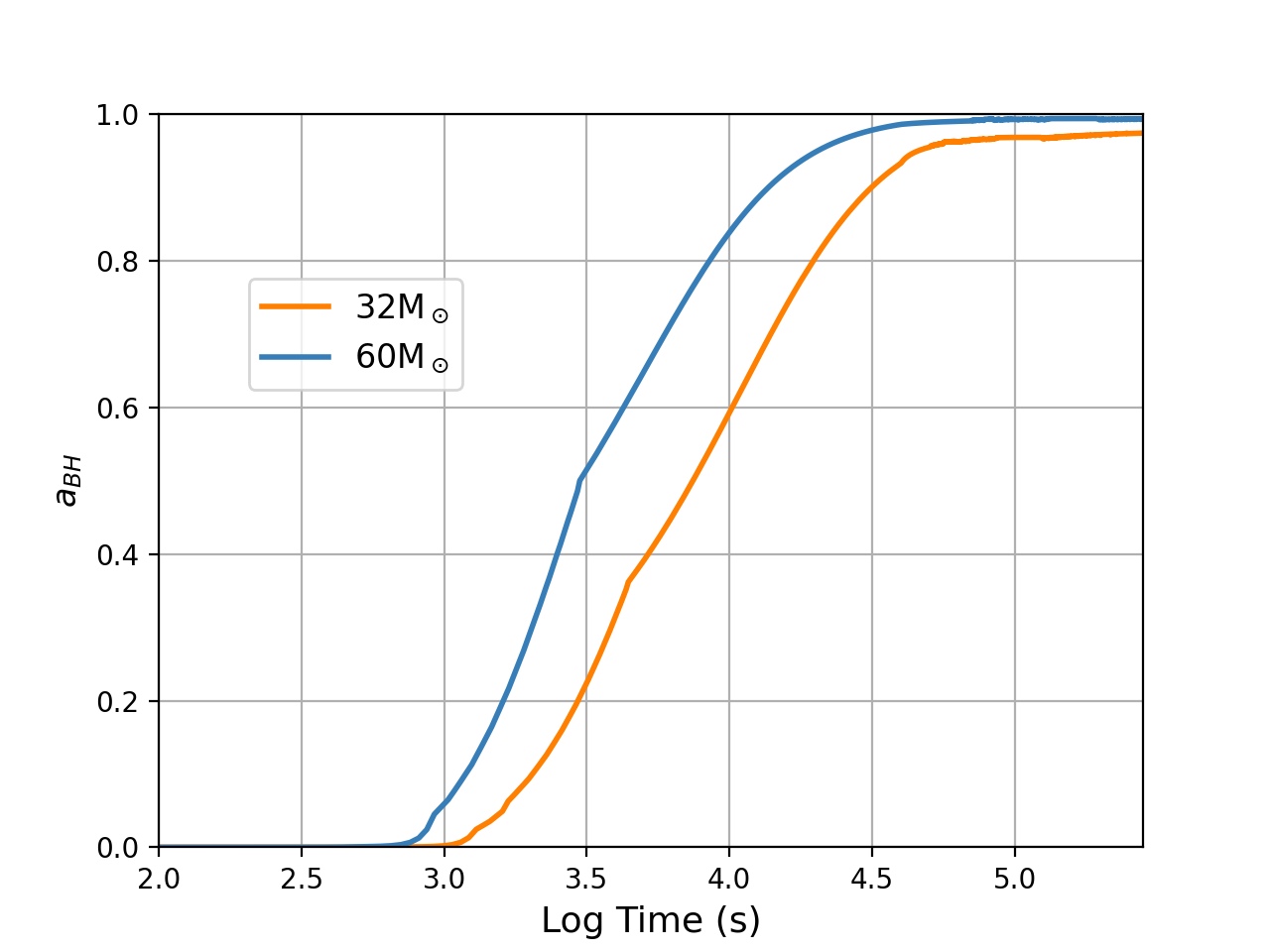}
    \includegraphics[scale=0.5]{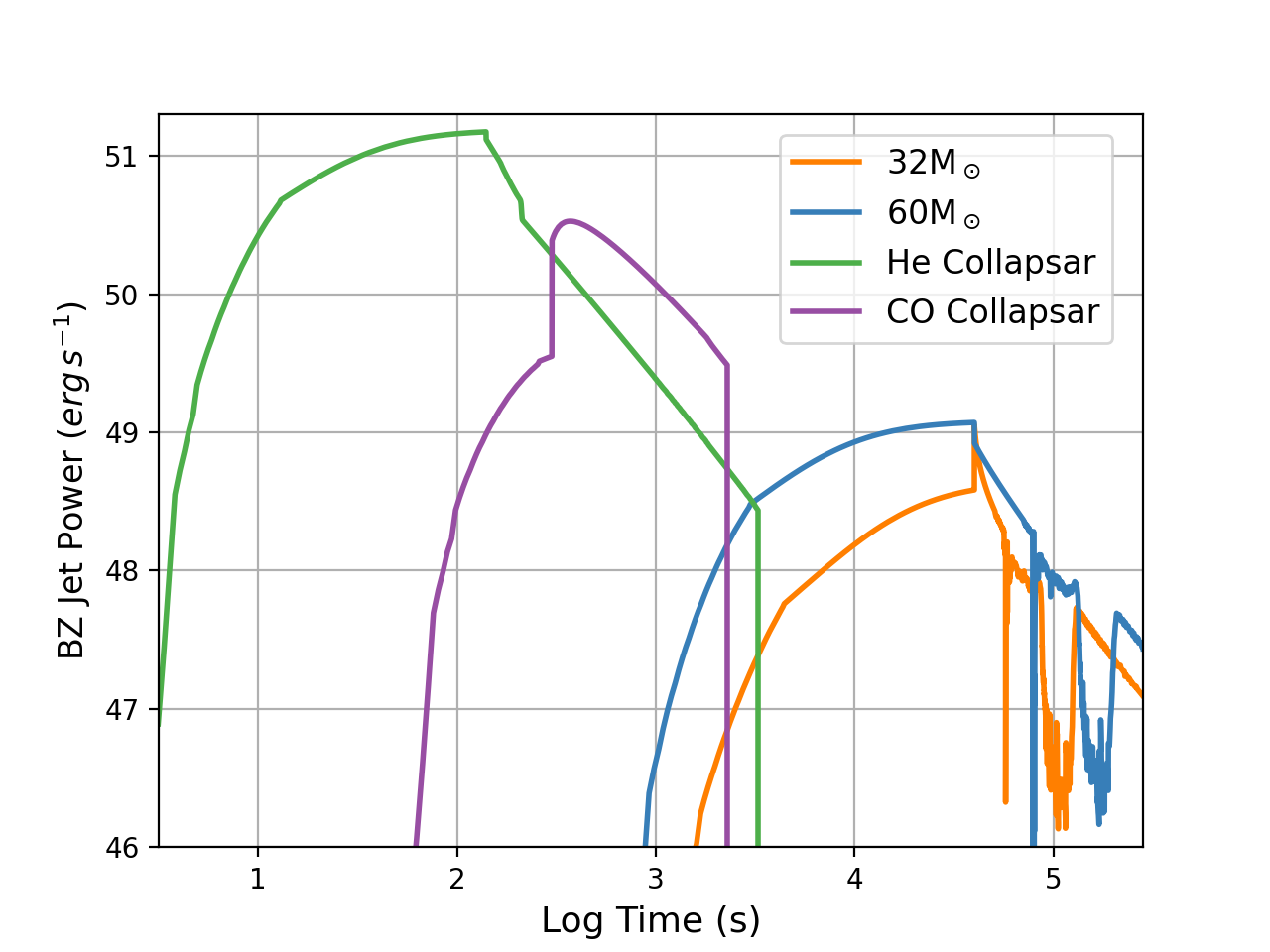}
    \caption{\textit{Top}: The black hole spin as a function of time for our 32 and 60~M$_\odot$ helium star merger progenitors. For these calculations, we assume that the compact remnant spirals into the center without accreting, starting the compact remnant with a mass of 2\,M$_\odot$ and a spin ($a_{\rm BH}=0$).  In reality, the compact remnant will accrete mass and spin up during the inspiral phase~\citep{2001ApJ...550..357Z}. \textit{Bottom}: The idealized Blandford-Znajek jet power from the combination of black hole spin and accretion rate in our systems of interest using equation~\ref{eq:BZform}. We include two tidally-spun up collapsar models using the CO and He cores of a 30\,M$_\odot$ zero-age main sequence star assuming they are in a tight binary with an orbital separation just beyond the Roche Radius~\citep[for more details, see][]{fryer2025explaining}.}
    \label{fig:helium-star-power}
\end{figure}

The accretion timescale is not necessarily the same as the timescale of the jet.  The dependence of the jet power on the black hole spin ($a_{\rm BH}$) and accretion rate through the disk ($\dot{M}_{\rm disk}$) is still being studied.  Here we use the \cite{1977MNRAS.179..433B} formula to determine the possible available power: 
\begin{equation}
    P_{\rm jet} = 3 \times 10^{51} a^2_{\rm BH} \frac{\dot{M}_{\rm disk}}{0.1\,{\rm M}_\odot/\rm s} {\rm erg/s}
    \label{eq:BZform}
\end{equation}
From our accretion timescale as a function of enclosed mass, we estimate the evolution of the black hole spin and the power in the Blandford-Znajek jet, as shown in Fig.~\ref{fig:helium-star-power}.  The luminosity is sensitive to the stellar models.  For example, the large drop in the power above $10^5$~s is due to the density jump at the boundary between the CO and He layers of the star.  

The total energy deposited into the jet is distributed into the prompt electromagnetic release, the kinetic energy of the jet, and contributes to the explosion of the supernova. As no supernova is observed following GRB~250702B \citep{gompertz2025grb250702b} we cannot quantify the latter. Thus, the sum of our measured gamma-ray and kinetic energies in the jet are a lower limit on the total jet power, which can be compared with our theoretical expectations. However, we must also account for accretion-to-jet efficiencies, which are at most a few percent \citep{2025arXiv250803689M,2025ApJ...980L..28W}, giving our theoretical expectation on the order of 1\% of the Blandford-Znajek prediction.

Fig.~\ref{fig:helium-star-power} gives the peak accretion power as $\sim$(0.3--1)$\times10^{49}$~erg/s for our two examples, predicting an approximate jet power of $\sim$(0.3--1)$\times10^{47}$~erg/s. Our measurement of the GRB~250702B collimation-corrected peak luminosity is $\sim$2$\times10^{47}$~erg/s. These values thus compare favorably.

Similarly, the sum of the total measured gamma-ray and kinetic energies is $\sim$(2-6)$\times10^{50}$~erg \citep{oconnor2025grb250702b}.  The integrated energy available in the 32~M$_\odot$ helium star merger model is 3$\times10^{53}$~erg, and for the 60~M$_\odot$ it is 7$\times10^{53}$~erg, giving a scaled expectation of $\sim$(3--7)$\times10^{51}$~erg. We measure $\sim$4$\times10^{50}$~erg, giving favorable agreement when accounting for the energy required for the predicted supernova. As a sanity check, we note the total energy of our helium merger model is comparable to the expected total energy from collapsar models ($\sim$(2--3)$\times10^{53}$~erg), which with our 1\% efficiency expectation recover the observed collapsar total jet energies \citep[e.g.][]{o2023structured}.

We emphasize that our helium merger model also explains the delay from central engine onset to duration, while nearly all other progenitor scenarios do not. The power will increase over a~few~$\times10^4$~s before reaching the peak output for a similar timescale, and is then followed by a $\sim$monotonic decrease in accretion over the next $\sim$day. This matches the profile inferred from EP and gamma-ray observations described in Section~\ref{sec:engine-duration}.


\subsection{Supernova}\label{sec:he-sn}
The other major observable from these events will be supernovae. The JWST observations of GRB~250702B find no transient at the position at a rest-frame time of T0+25.5~days, excluding a typical Ic broad-lined supernovae seen following collapsar GRBs \citep{gompertz2025grb250702b}. This does allow for less luminous supernovae to remain undetectable, due to the red galaxy and the high amounts of extinction. We thus seek to understand how a supernova following GRB~250702B may compare to those following classic collapsars.

Supernovae are generally powered at least partially by the heat from the radioactive decay of $^{56}$Ni. Core-collapse supernovae are also powered by shock heating as the explosion front hits the circumstellar material. In collapsar GRBs it is the combination of the shock from the jet and the disk winds which explode the star at high velocities, with some $^{56}$Ni created in the accretion disk which power the Ic-broad lined supernovae. We expect the same general picture in helium merger GRBs. However, supernovae following helium mergers could be any type of stripped envelope supernova (as detailed below). We study these two power sources in the context of GRB~250702B.

\begin{figure}
    \centering
    \includegraphics[scale=0.45]{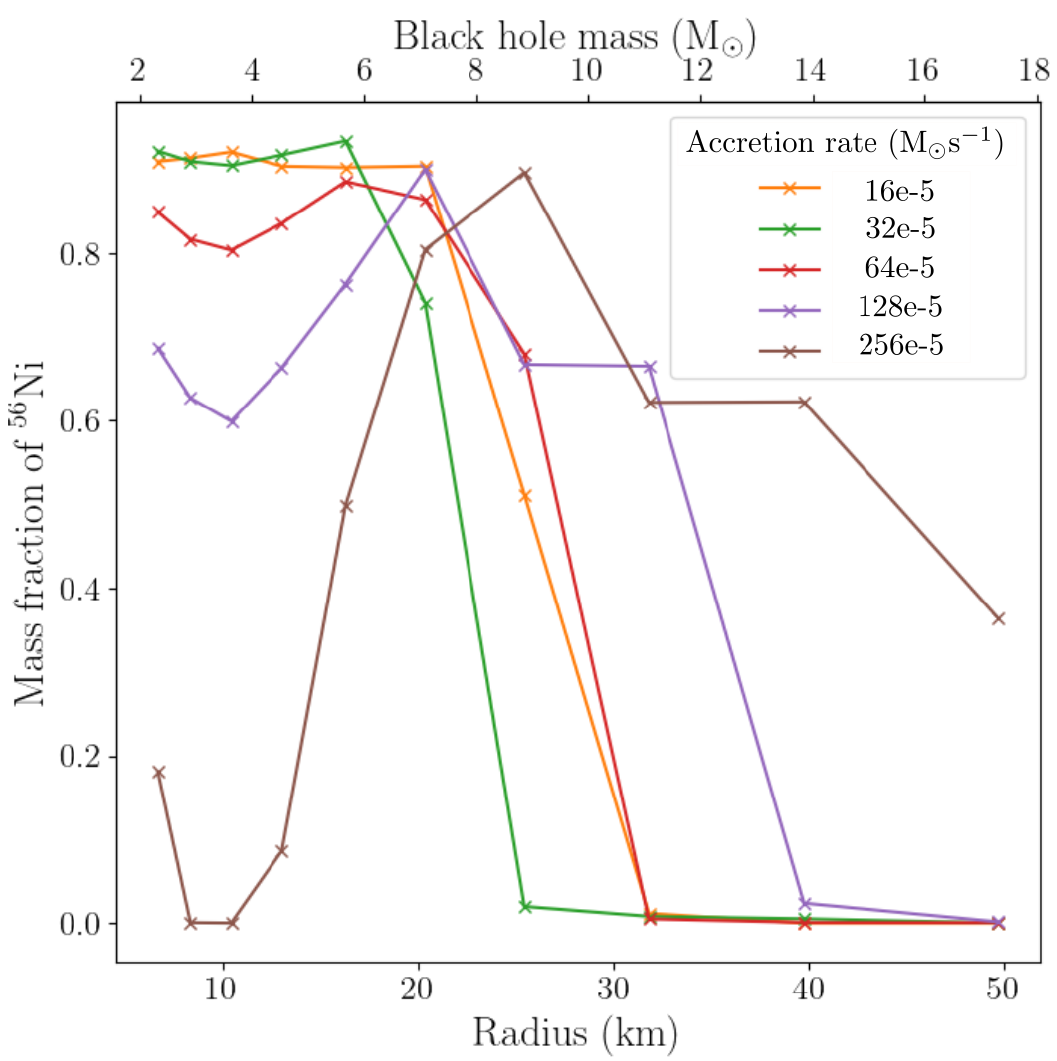}
    \caption{$^{56}$Ni mass fraction as a function of the position in the accretion disk.  The innermost radius depends upon the mass of the black hole.  The innermost stable circular orbit for a maximally rotating black hole dictates the innermost radius of the disk. The upper x-axis shows the position of the innermost disk radius for a set of black hole masses at maximal spin.  The $^{56}$Ni in the disk drops precipitously when the black hole exceeds 5\,M$_\odot$ for all accretion rates expected for helium mergers. Lower-mass black holes will produce some $^{56}$Ni, but we expect the total yield to be low \citep{Abrahams2025ni}.}
    \label{fig:niprod}
\end{figure}

$^{56}$Ni is expected to be produced in accretion disks, but the amount of $^{56}$Ni production depends sensitively on the accretion rate and the black hole mass (which dictates the innermost radius of the disk).  Fig.~\ref{fig:niprod} shows the $^{56}$Ni mass fraction as a function of the position in the disk for a range of accretion rates.  We assume that material is ejected across the entire disk (with perhaps some bias toward the inner region) as viscous heating drives a disk wind~\citep[for example,][]{2023ApJ...956...71K}.  For the expected accretion rates and black hole masses in our helium merger scenario, we expect much of the material to be ejected from regions of the disk that do not produce $^{56}$Ni, particularly for black hole masses above 5~M$_\odot$.  The $^{56}$Ni yield from helium mergers will be much lower than typical collapsars. Assuming the maximal scenarios for $^{56}$Ni in the models we consider, the highest yield would be only $\sim0.075$~M$_\odot$. This is consistent with the JWST limit on a $^{56}$Ni yield of $\lesssim$0.22~M$_\odot$.

The lower accretion rates required to match the duration of GRB~250702B will produce weaker disk outflows. Further, in the case of a common envelope the region along the angular momentum axis could be relatively clean and the jet shock will deposit less power into the surrounding material. Thus, the stellar outflows will be weaker, and the shock heating in helium merger systems less powerful. All together, we expect the supernovae from these mergers to be much dimmer than typical collapsar models, being consistent with the JWST supernova non-detection \citep{gompertz2025grb250702b}.

\subsection{Population Considerations}\label{sec:he-pop}
Population level expectations are key to exploring GRB progenitors. We here discuss whether past observations of GRBs and supernovae are consistent with our expectations of this model, and the population-level predictions which may be tested in future events. 

We note the helium star merger model has been invoked in past events. While GRB~101225A has a duration which can be explained through other means, the faint supernova is consistent with the helium merger model expectation \citep{2011Natur.480...72T}. In contrast, GRB~111209A has the second longest duration but has the most luminous supernova seen following a GRB \citep{2019A&A...624A.143K}, possibly tied to the much wider jet half-opening angle of $\sim$20$^\degree$ \citep{2013ApJ...779...66S} depositing more jet shock energy into the stellar envelope. Thus, if that is also a helium merger, the expected supernova brightness may have a large range of luminosities.

For GRB emission, the extreme angular momentum in helium mergers forces a long accretion timescale, inconsistent with the duration of typical long GRBs. This explains the lack of Ib supernovae following typical long GRBs, though these may be observed in related transients \citep[e.g.][]{rastinejad2025ep}. The lower intrinsic luminosities of ultra-long GRBs limit detection distances, and the ultra-long durations make identification more difficult. Thus, we find no reason to disbelieve this progenitor channel from the archival GRB sample.

In the helium star merger scenario the associated supernova must arise from a stripped star. Thus, we may expect Ib and Ic supernovae. These may be broad-lined supernovae if enough energy is deposited into the stellar envelope (though supernova seem from these events off-axis may have lower observed velocities due to ejecta asymmetry). If significant hydrogen from the common envelope is swept up in the ejecta the associated supernova could be IIb. Similar physical models have been invoked to explain narrow emission-line supernovae. For example, \citet{2022ApJ...932...84M} suggest this progenitor for Ibn and Icn supernovae, and, if so, could also explain (the currently unobserved) Idn, Ien \citep{schulze2025extremely}, and IIn supernovae \citep{gagliano2025evidence}. Here the supernovae is surrounded by circumstellar material which emits lines when heated by the supernova. This material would be ejected by the compact object interactions with the stripped star partner in the years before merging. These events do show lower $^{56}$Ni yields than normal core-collapse supernovae \citep{maeda2022properties, Farias2025}.

Additionally, if the merger is produced during a hydrogen common envelope event where the compact object continues to inspiral into the core of the star, we would expect an extensive circumstellar medium (with hydrogen).  This would enhance the shock heating and likely produce hydrogen features.  The debris above the angular momentum axis for these hydrogen common envelope systems may contaminate the jet, preventing it from producing GRBs, producing supernovae (or fast blue optical transients) instead \citep{2022ApJ...932...84M,2025ApJ...988...30H}. Similarly, if the secondary star explodes before the black hole is engulfed, this matches the tight-binary scenario invoked to explain the angular momentum required for typical collapsar GRBs \citep{fryer2025explaining}, which relates to future stellar-mass gravitational wave sources.

Lastly, the host galaxy properties and offsets are key to probing other GRB progenitor channels. In the helium merger scenario, during the formation of the compact object, mass ejection and a potential compact remnant kick can cause the binary to gain a net momentum.  The velocity distribution of these massive star binaries tends to peak at $20$~km/s and extend, depending on the compact remnant kick up to 100--200~km/s.  Only in these extreme cases would the merger occur far from the birthplace of these binary stars. As massive stars live only for a short time, the majority of helium mergers should occur close to where they are born. They will thus track active star formation in individual galaxies, occurring within the stellar field, and the cosmological star formation history as a population. Because the compact object can begin as a neutron star, these events should track host galaxies and metallicity dependencies more like standard core-collapse supernovae rather than the low metallicity preference of collapsars. 

With future proof that this merger scenario occurs, we can start to gain insight into the merger process itself and the subsequent mass ejection.  Coupled with radiation-hydrodynamics calculations of the shock interactions, we can use observations and observational limits of the associated supernova to probe both the properties of the explosion and the mass ejection during the last common envelope phase.  Upper limits on the peak supernova emission place limits on the $^{56}$Ni yield and shock heating.  An observation, particularly with spectra determining the presence of hydrogen or helium lines, will provide crucial clues into the extent at which helium stars expand. This is already strongly suggested via GRB observations by this event and population-level inferences on collapsar GRBs \citep{fryer2025explaining}.

\section{Conclusion}\label{sec:conclusion}
Gamma-ray bursts have been enigmatic objects since their discovery more than half a century ago. After detections of $\sim$15,000 GRBs, GRB~250702B is still unique. It has subsecond variability, typical intrinsic energetics, high bulk Lorentz factor, and no spectral lag, all of which are fairly typical in GRBs. However, it has record duration, is inconsistent with the peak energy for its luminosity in the collapsar Yonetoku relation (as are other ultra-long GRBs), and has an exceptionally narrow jet. 

While we considered numerous GRB models, the only one which naturally explains the properties observed in GRB~250702B is the fall of a stellar-mass black hole into a star. We focus on the field binary evolution to a helium merger as our preferred explanation. This model makes a number of testable predictions, even with current knowledge. Ultra-long GRBs from helium mergers should track star formation, with individual events arising from star-forming regions and the population tracking the cosmic star formation rate evolution. They can arise from higher metallicity regions than collapsar GRBs. Lastly, helium merger GRBs should be followed by stripped envelope supernovae.

There are a number of opportunities where theory and simulation investment are warranted. The unusual behavior of an idealized engine in the Blandford-Znajek scenario may allow for unique predictions and tests of accretion and jets. The types of stripped enveloped supernovae, such as broad-lined or narrow emission line supernovae, can likely be narrowed. Predictions on the minimal $^{56}$Ni forged in accretion disks can be refined and form a floor for the minimal supernova luminosity. And modeling from long before merger may allow for constraints on amounts and distributions of viable circumstellar material. 

Lastly, these events are difficult to identify. They are intrinsically lower luminosity than collapsar GRBs, which most instruments were designed to detect. The most sensitive GRB monitors are limited to low Earth orbit, where they lack the continuous viewing timescales necessary to probe ultra-long GRBs. Certainly \textit{Swift}, \textit{Einstein Probe}, and \textit{SVOM} provide opportunity to identify these events, but the broad characterization requires instruments like Konus-\textit{Wind} and \textit{Psyche}-GRNS. For the first time since the identification of ultra-long GRBs, we have two distant monitors with stable backgrounds on the required timescales in these instruments, allowing for spatial information from the InterPlanetary Network. When paired with the new Legacy Survey of Space and Time by the Vera Rubin Observatory, we may expect more regular identification of ultra-long GRBs. Further, the \textit{Compton Spectrometer and Imager} will be able to individually identify the brighter ultra-long events \citep{2024icrc.confE.745T}. Thus, we strongly encourage investment in theory, simulation, and prioritized follow-up of these events.

\section*{Author Contributions}
Eliza Neights helped with organizing the meetings and writing the paper, performed the complex \textit{Fermi}-GBM background fitting and source selections used in her and other analyses, performed the \textit{Fermi}-GBM spectral analysis and contributed related plots and text, and contributed general text to the paper. Eric Burns helped with organizing the meetings, bringing together the relevant teams, and the helium merger model scenario, contributed major text to all sections in the paper, performed the maximum photon energy analysis, calculated intrinsic energetics and placed the bursts in context, determined duration, performed the analysis of the high level \textit{Psyche} data, and is co-lead of the InterPlanetary Network. Chris Fryer performed the quantitative work on the helium merger model, contributed key analysis on excluding other progenitor scenarios, and contributed much of the relevant text and figures. Dmitry Svinkin performed much of the Konus-\textit{Wind} analysis including the spectral analysis, the measures in intrinsic energetics, the Amati and Yonetoku relations and the analysis of the other ultra-long GRBs for their inclusion, the Konus-\textit{Wind} gamma-ray upper limits over \textit{Einstein Probe} intervals, the background fitting, and the relevant text. Suman Bala contributed the MVT and spectral lag analyses. Rachel Hamburg performed the search for extended emission in Fermi-GBM. Ramandeep Gill provided the discussion on the required bulk Lorentz factor. Michela Negro contributed the cross-spectrum QPO analysis. Megan Masterson contributed the power-spectrum QPO analysis. Jimmy DeLaunay contributed the \textit{Swift}-BAT analysis and relevant text and figures. David J. Lawrence processed the \textit{Psyche}-GRNS data and provided instrument and mission expertise to enable proper handling of the data. Sophie E. D. Abrahams performed simulations leading to the $^{56}$Ni production predictions. Yuta Kawakubo provided the MAXI detection and non-detection information. Paz Beniamini provided the text on tying minimum variability timescale to central engine size, key arguments excluding other progenitors, and contributed the text on the micro TDE scenario.

Christian Aa. Diget contributed to the $^{56}$Ni production predictions. Dmitry Frederiks provided the Konus-\textit{Wind} data in his role as PI and contributed to the analysis and interpretation of this and other data. John Goldsten provided the channel to energy mappings for \textit{Psyche}-GRNS to allow for proper data selection including the calculation of the maximum photon energy. Adam Goldstein provided input on the novel background approach and general use of the GDT. Alexander Hall-Smith contributed to the $^{56}$Ni production predictions. Erin Kara provided expertise on the QPO analysis and viability of TDE progenitors. Alison Laird contributed to the $^{56}$Ni production predictions. Gavin Lamb provided input on the viable and excluded models discussion. Oliver J. Roberts contributed to the maximum photon energy analysis and identified the third GBM trigger as arising from GRB~250702C. Ryan Seeb contributed to the GBM spectral analysis. V. Ashley Villar contributed to the discussions on the variety of viable supernovae and in particular the viability of ultra-long GRBs being related to type Ixn supernovae.

\section*{Acknowledgments}
We thank \citet{gompertz2025grb250702b}, \citet{carney2025grb250702b}, and \citet{oconnor2025grb250702b} for the sharing of private data and analysis prior to the public release of their papers, enabling far greater conclusions in our work on a tight timeline. We additionally thank B.~O'Connor for extremely helpful comments at an opportune time. We further thank E.~Borowski for helpful information on X-ray binaries.

E.N. acknowledges NASA funding through Cooperative Agreement 80NSSC22K0982. O.J.R., A.G., and S.B. graciously acknowledge NASA funding support from cooperative agreement 80NSSC24M0035. R.S. thanks the NASA OSTEM Internship program for providing support. P.B.'s work was funded by a grant (no. 2020747) from the United States-Israel Binational Science Foundation (BSF), Jerusalem, Israel and by a grant (no. 1649/23) from the Israel Science Foundation. J.D. acknowledges support from NASA under award numbers NAS5-0136, 80NSSC25K7567, and 80NSSC25K7788. A.M.L., C.A.D., A.D.H., and S.E.D.A. acknowledge support from the Science and Technology Funding Council through grants ST/V001035/1 and ST/Y000285/1. The work for J.G., D.L., and P.P. was funded by NASA’s Discovery Program, under the Psyche: Journey to a Metal World mission. Psyche funding is provided to the Johns Hopkins University Applied Physics Laboratory (JHU/APL) through contract 1569206 from NASA Jet Propulsion Laboratory. The work of D.S.S., D.D.F, A.V.R., A.L.L., A.E.T., and M.V.U. was supported by the basic funding programm of the Ioffe Institute no. FFUG-2024-0002; A.E.T. also acknowledges financial support from ASI-INAF Accordo Attuativo HERMES Pathfinder operazioni n. 2022-25-HH.0. A.C.T.'s research was supported by an appointment to the NASA Postdoctoral Program at the NASA Marshall Space Flight Center, administered by Oak Ridge Associated Universities under contract with NASA. G.P.L. acknowledges support from the Royal Society (grant nos. DHF-R1-221175 and DHF-ERE-221005).

\section*{Data Availability}
The \textit{Fermi} and \textit{Swift} data underlying this article are available in the public domain on NASA HEASARC: \url{https://heasarc.gsfc.nasa.gov}. \textit{Psyche}-GRNS data will be available on the Planetary Data System a few months after the event time. Konus-\textit{Wind} and MAXI data can be requested from the respective instrument leads. Data and code generated for this project may be made available upon request to the lead author.



\bibliographystyle{mnras}
\bibliography{bibliography} 



\appendix

\section{GRB~250702B Lightcurve and Background and Source Selections}\label{app:lightcurve}
\begin{figure*}
    \centering
    \includegraphics[width=\textwidth]{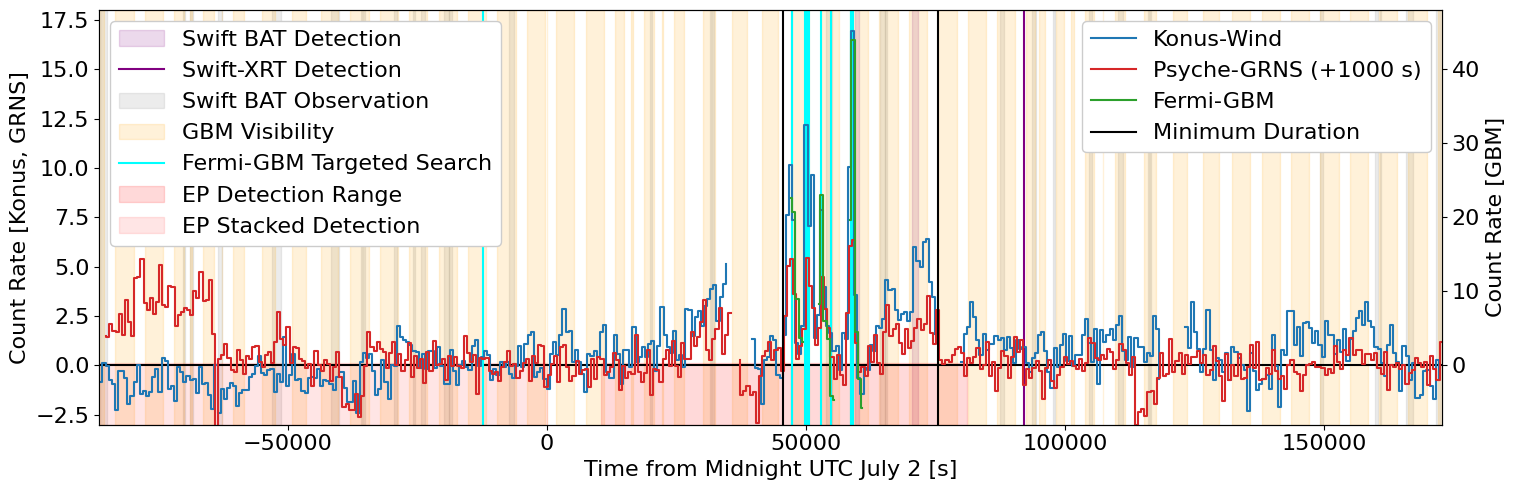}
    \caption{The combined lightcurves and additional detection information utilized to determine the prompt gamma-ray emission times.}
    \label{fig:total_combined_lc}
\end{figure*}

Determining the lightcurve of GRB~250702B requires the development of a self-consistent set of background determinations across instruments as well as intervals of significant emission. We use additional information in order to determine when significant fluctuations are due specifically to GRB~250702B. The evidence in total is summarized in Fig.~\ref{fig:total_combined_lc}. Subsection~\ref{app:bkgd} details the background fitting, subsection~\ref{app:source} the source selections, and subsection~\ref{app:extended-searches} the searches for gamma-ray emission in wider intervals.

\subsection{Background}\label{app:bkgd}
\textit{Swift}-BAT creates sky-images using the balanced mask-weighted technique that automatically subtracts out background. The \textit{Psyche}-GRNS background is an average from the counts between T0-25,000~s to T0+20,000~s and T0+120,000~s to T0+165,000~s. The method for determining Konus-\textit{Wind} and \textit{Fermi}-GBM backgrounds are more complex and detailed below.

\subsubsection{Konus-Wind}
In the waiting mode both Konus-\textit{Wind} (KW) detectors measure count rates in three energy bands (called G1, G2, and G3), with the following energy boundaries: 23--96~keV, 96--398~keV, and 398--1628~keV in the S1 detector, and 18--76~keV, 76--316~keV, and 316--1250~keV in the S2 detector. Long-timescale count rate variations in G1 are mostly related to the galactic X-ray transient activity, while G2 and G3 background variations are typically produced by variations in the solar energetic particle (protons and electrons) flux.

\begin{figure*}
    \centering
    \includegraphics[width=\textwidth]{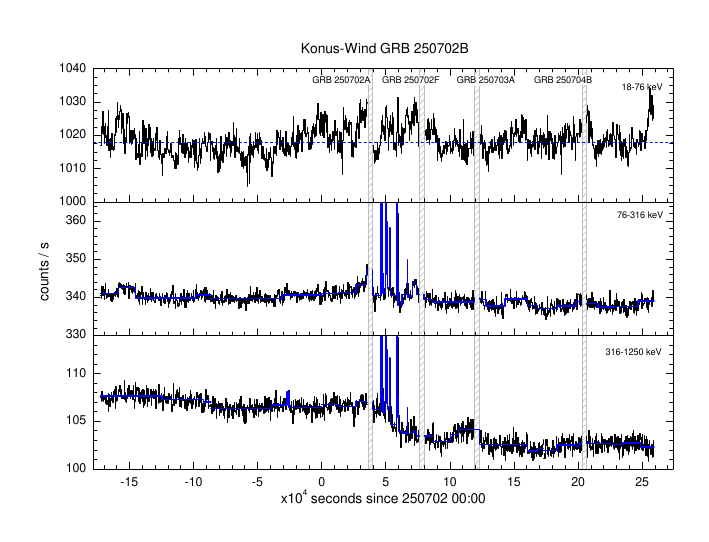}
    \caption{The Konus-\textit{Wind} S2 waiting mode data between T0-172,800~s and T0+25,9200~s. The blue solid lines in the G2 and G3 panels indicate Bayesian block binning, the blue dashed line in the G1 panel is the mean count rate, and the gray hatched intervals are data readout intervals for triggered GRBs, during which Konus did not collect data.}
    \label{fig:kw_bg_lc}
\end{figure*}

We investigate Konus background behavior during the interval between T0-172,800~s and T0+25,9200~s, with the S2 waiting mode data shown in Fig.~\ref{fig:kw_bg_lc}. The average count rates are $\sim$1018~counts/s (G1), $\sim$340~counts/s (G2), $\sim$107~counts/s (G3). Typical long-term variations in background rate are $\sim$10~counts/s (G1), $\sim$5~counts/s (G2), and  $\sim$2~counts/s (G3). In the G2 and G3 bands there is a step-like feature with an amplitude of $\sim$2 and $\sim$4~counts/s, respectively, around T0+55,020~s, which is related to a drop in proton and electron flux seen in 3DP-\textit{Wind}~\citep{Lin_1995SSRv_71_125}.

To estimate background rates for GRB~250702B, we limit our analysis to the interval between T0+39,518~s (end of the data gap after the GRB~250702A readout) and T0+75,847~s (just before GRB~250702F). We select the background interval using Bayesian block decomposition of the lightcurves in each energy band and fit the background using three models: a constant count rate ($a$) in G1, a linear model ($a+bt$) in G2, and a logistic function ($a + b/(1+e^{c(t-t_0)})$ in G3. The model parameters are given in Table~\ref{table:kw-bg-results}.

\begin{table*}
\caption{Konus-\textit{Wind} background fit results, showing the intervals over which the fit is measured and the model parameters used to estimate the background.} 
\centering
\begin{tabular}{ccccc}
    \hline
    Energy band & Intervals - T0 (s) & Model    & Model parameters \\ 
    \hline
     G1         & 59,929--62,690  & Constant & $a=1018.3 \pm 1.6$ (counts/s) \\
     G2         & 39,518--46,074  & Linear   & $a=339.0\pm 0.4$  (counts/s)  \\
                & 59,929--62,690  &          & $b=(-1.2 \pm0.3)\times10^{-4}$  (counts/s$^2$)\\
     G3         & 39,518--46,074  & Logistic & $a=106.36 \pm 0.16$  (counts/s)  \\
                & 47,958--49,842  &          & $b=-2.58 \pm 0.20$ (counts/s) \\
                & 54,297--57,594  &          & $c=(-7 \pm 5)\times10^{-4}$ (1/s) \\
                & 59,929--74,004  &          & $t_0=(54.4\pm 1.1)\times10^{3}$ (s) \\
    \hline
\end{tabular}
\label{table:kw-bg-results}
\end{table*}

\subsubsection{Fermi-GBM}
\label{sec:gbm-background}

Typically, the time-variable background of a GRB detected by \textit{Fermi}-GBM is estimated by fitting a polynomial to the count rate during time intervals before and after the burst \citep[e.g.][]{2021ApJ...913...60P}. This method does not work well for very long-duration events, such as GRB~250702B, because the background may fluctuate more than can be modeled with a simple polynomial, and faint source emission may contaminate the time intervals selected as background.

An alternative background estimation method is to use background rates from orbits preceding and/or following that of the long-duration burst observation \citep{2011arXiv1111.3779F}, and we adopt both this approach and the traditional polynomial method. Charged particles are a significant source of background for \textit{Fermi}-GBM, meaning that the rate and spectrum of the background vary with the geographic position. Further, the directional-dependent response means background also depends on the orientation of the satellite. The background at a given time $t_0$ can be estimated using observations when the spacecraft returns to the same position and orientation as at $t_0$.

The \textit{Fermi} spacecraft returns to the same geographic position every 15 orbits. Because the satellite alternates its pointing every two orbits, this gives two options to estimate the background: 1) using the observations at $t_0\pm30$ orbits, which is at approximately $t_0\pm48$~hours, or 2) averaging the observations at $t_0\pm14$ and $t_0\pm16$ orbits, when the detector orientation is correct, but the geographic positions are slightly offset from that at $t_0\pm15$ orbits.

We estimate the background for the duration of interest from T0+45,600~s to T0+61,800~s. The orientation of \textit{Fermi} was modified due to a Target of Opportunity observation leading up to GRB~250702B. Therefore, it is infeasible to use observations until \textit{Fermi} resumed its nominal observation mode at around T0+27,900~s to estimate the background. Thus, we estimate the background using the observation at T0+30 orbits, between T0+215,685~s and T0+231,885~s. We do not expect any unrelated transients to interfere with the background estimation at energies $\gtrsim$100~keV, since \textit{Fermi}-GBM did not trigger on any astrophysical transients during this time.

\begin{figure}
    \centering
    \includegraphics[width=0.48\textwidth]{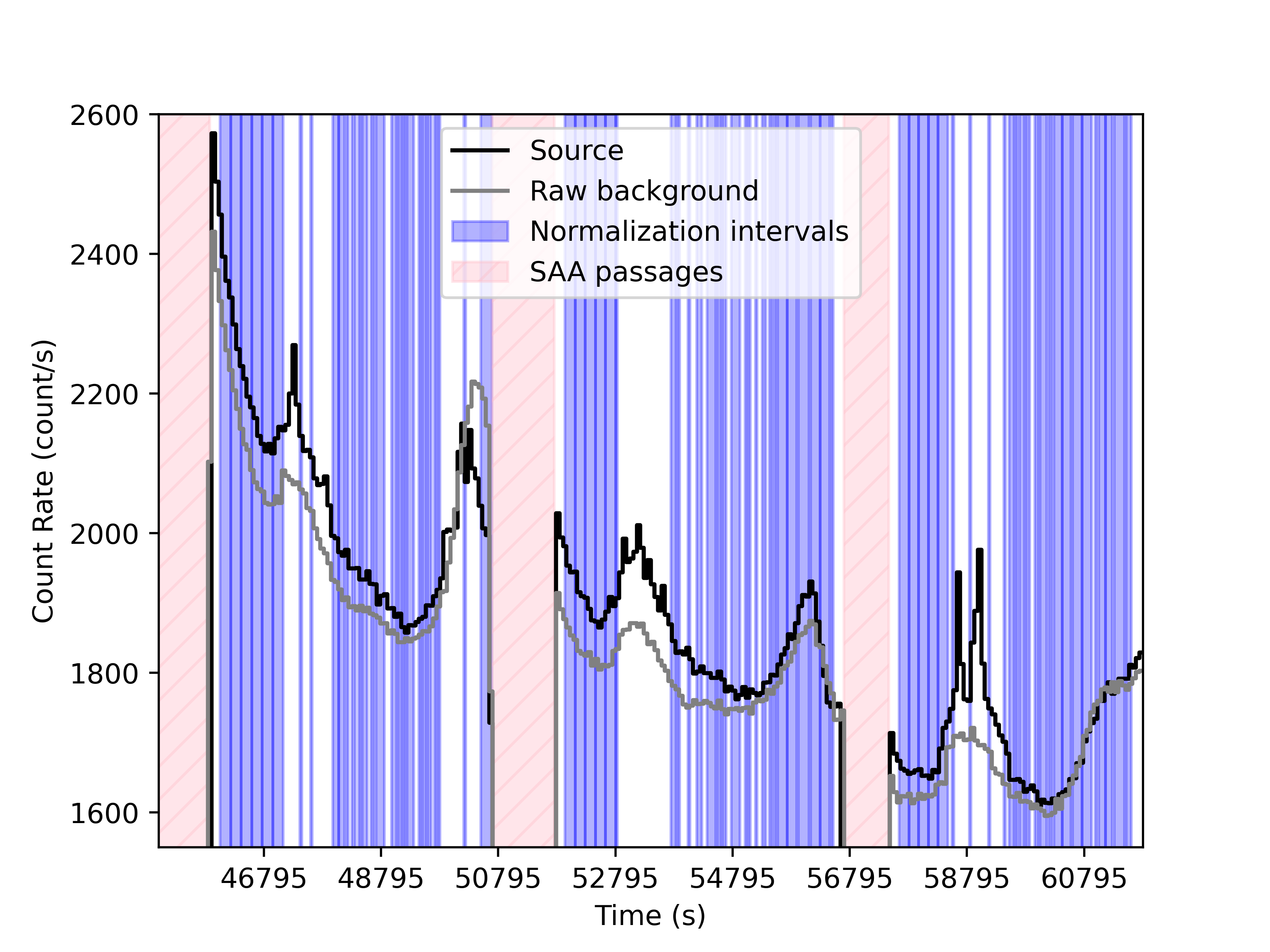}
    \includegraphics[width=0.48\textwidth]{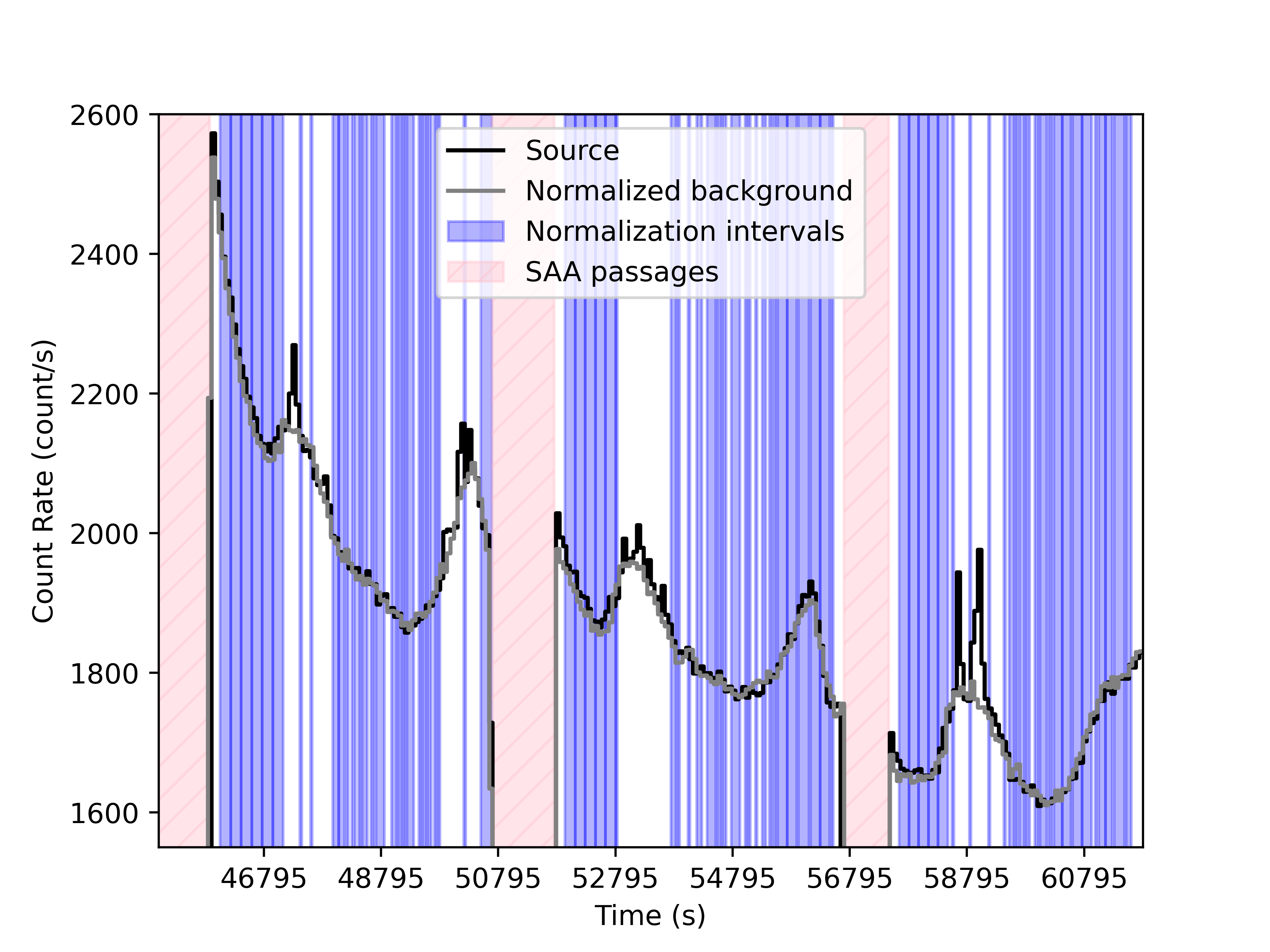}
    \caption{The normalization of \textit{Fermi}-GBM orbital background. \textit{Top}: The raw orbital background overlaid on the source interval lightcurve. The normalization intervals in blue are used to compute normalization factors. \textit{Bottom}: The normalized orbital background overlaid on the source interval lightcurve.}
    \label{fig:gbm-lc}
\end{figure}

The background lightcurve evolution extracted from T0+30 orbits overall matches that of the GRB~250702B interval, although there are offsets in normalization, as seen in Fig.~\ref{fig:gbm-lc}. To resolve this, we select normalization time intervals during which GRB~250702B is occulted by the Earth or Konus detects $<$10 counts/bin in the G1 detector and $<$5 counts/bin in G2 outside of the GBM analysis intervals, in order to minimize source contamination. We add an additional interval between Interval~4 and the SAA passage of \textit{Fermi} because the lightcurve profile differs significantly between the source and orbital background intervals leading up to the SAA. However, there is likely emission (especially soft) from GRB~250702B at this time, which may impact the background estimation and spectral analysis results for Interval~4. The normalization intervals are broken up into time periods of $<$200~s. A normalization factor is calculated for each detector, energy channel, and normalization interval by dividing the source and background counts. Using linear interpolation, normalization factors throughout the GRB~250702B interval at 1~s resolution are computed. Fig.~\ref{fig:gbm-lc} shows a much improved estimate of the overall background rate using this additional scaling step. The background-subtracted lightcurve matches that of Konus-\textit{Wind} and \textit{Psyche}-GRNS, as shown in Fig.~\ref{fig:lc}, validating the background estimation. To estimate the background for spectral analysis, we fit and normalize a polynomial to the orbital background during and around each GBM analysis interval, with the time intervals listed in Table \ref{table:gbm-polynomials-orb}.

As a cross-check, we additionally estimate the background via the usual polynomial fitting method. Using the \textit{Fermi} Gamma-ray Data Tools \citep[GDT-Fermi;][]{GDT-Fermi}, we fit polynomials to the 5~s binned lightcurve for each detector and energy channel in intervals before and after each analysis interval described in Appendix~\ref{app:interval-selections} and summarized in Table~\ref{table:gbm-bright-intervals}. The background is fit with an order 1 polynomial for all intervals except for GBM~Interval~5, where an order 2 polynomial is used. The polynomial fits for the spectral lag measurements are summarized in Table~\ref{table:gbm-polynomials-spectral-lag}. For spectral analysis, we more carefully choose intervals for polynomial fitting to avoid times when there may be dim GRB~250702B emission which may impact the spectral results, through manual inspection of the GBM and Konus source interval and GBM orbital background lightcurves. These are displayed in Table~\ref{table:gbm-polynomials-spectra}.

\begin{table}
\caption{Time intervals to which polynomials are fit to \textit{Fermi}-GBM orbital background data for spectral analysis of each interval in Table~\ref{table:gbm-bright-intervals}. All polynomials are order 1 except for Interval~5, which is order 3.} 
\centering
\begin{tabular}{ccc}
    \hline
    GBM Interval & Background Interval - T0 (s) \\
    \hline
    1 & 47,145--47,455 \\
    2 & 47,530--48,020 \\
    3 & 49,765--50,275 \\
    4 & 50,185--50,445 \\
    5 & 52,780--53,815 \\
    6 & 58,525--58,785 \\
    7 & 58,780--59,185 \\
    \hline
\end{tabular}
\label{table:gbm-polynomials-orb}
\end{table}

\begin{table}
\caption{Time intervals to which polynomials are fit to \textit{Fermi}-GBM data for spectral lag measurements of the corresponding intervals in Table~\ref{table:gbm-bright-intervals-raw}. All polynomials are order 1 except for Interval~5, which is order 2.} 
\centering
\begin{tabular}{ccc}
    \hline
    GBM & Background Interval & Background Interval \\
    Interval & Before - T0 (s) & After - T0 (s) \\
    \hline
    1 & 47,045--47,195 & 47,405--47,545 \\
    2 & 47,430--47,580 & 47,970--48,105 \\
    3 & 49,935--50,085 & 50,210--50,275 \\
    4 & 50,185--50,250 & 50,370--50,525 \\
    5 & 52,379--52,855 & 53,765--54,180 \\
    6 & 58,225--58,575 & 58,735--58,875 \\
    7 & 58,690--58,855 & 59,134--59,379 \\
    \hline
\end{tabular}
\label{table:gbm-polynomials-spectral-lag}
\end{table}

\begin{table}
\caption{Time intervals to which polynomials are fit to \textit{Fermi}-GBM data for spectral analysis of the intervals in Table~\ref{table:gbm-bright-intervals}. The same ranges are used to fit the background for the corresponding intervals in Table~\ref{table:gbm-bright-intervals-raw}. All polynomials are order 1 except for Interval 5, which is order 2.} 
\centering
\begin{tabular}{ccc}
    \hline
    GBM & Background Interval & Background Interval \\
    Interval & Before - T0 (s) & After - T0 (s) \\
    \hline
    1 & 46,991--47,120 & 47,408--47,538 \\
    2 & 47,535--47,614 & 47,968--48,071 \\
    3 & 49,713--49,793 & 50,205--50,259 \\
    4 & 50,205--50,259 & 50,475--50,686 \\
    5 & 52,671--52,825 & 53,738--53,926 \\
    6 & 58,537--58,566 & 58,831--58,860 \\
    7 & 58,830--58,860 & 59,155--59,184 \\
    \hline
\end{tabular}
\label{table:gbm-polynomials-spectra}
\end{table}

\subsection{Source Selections}\label{app:source}
The \textit{Psyche}-GRNS data is predominantly used to confirm the signal variability seen in Konus-\textit{Wind}, allowing for the total duration measurement. Konus-\textit{Wind} analysis intervals are selected using Bayesian block decomposition of the G2 lightcurve. The other instruments have additional source selections, detailed here.

\subsubsection{Swift-BAT}
\textit{Swift}-BAT is a coded mask imager, capable of creating images of the 14--195~keV sky. On July 2, GRB~250702B was within the coded field-of-view of \textit{Swift}-BAT seven times, with five of those occurring after the initial \textit{Fermi}-GBM trigger. None of these observations overlap with the bright intervals listed in Table~\ref{table:gbm-bright-intervals}. There are two detections and two non-detections in the main emission, shown in Fig.~\ref{fig:lc}. One detection confirms the later emission seen in Konus and GRNS as arising from this burst, significantly extending the gamma-ray duration. 

\subsubsection{Fermi-GBM}\label{app:interval-selections}
Fermi-GBM triggered during the emission from GRB~250702B. All four times contained emission from this event, though one trigger was due to an unrelated short GRB as explained in Section~\ref{sec:grb250702c}. The \textit{Fermi}-GBM onboard triggers associated with GRB~250702B and the relevant detectors are shown in Table~\ref{table:gbm-triggers}.

\begin{table}
\caption{\textit{Fermi}-GBM triggers comprising GRB~250702B. The positions are calculated from the \textit{Fermi}-GBM trigger data. The good detectors are those with a viewing angle within $60\degree$ of the source, which are used in analysis.} 
\centering
\begin{tabular}{cccccc}
    \hline
    Name & Trigger Time & Position & Good \\
     & - T0 (s) & (RA, Dec; $\degree$) & Detectors \\
    \hline
    250702548 & 47,342.03 & $(290, 0) \pm 10$ & n8, nb, b1 \\
    250702581 & 50,165.77 & $(286, -9) \pm 8$ & n9, na, nb, b1 \\
    250702682 & 58,893.07 & $(290, -20) \pm 10$ & n8, nb, b1 \\
    \hline
\end{tabular}
\label{table:gbm-triggers}
\end{table}

\begin{figure}
    \centering
    \includegraphics[scale=0.5]{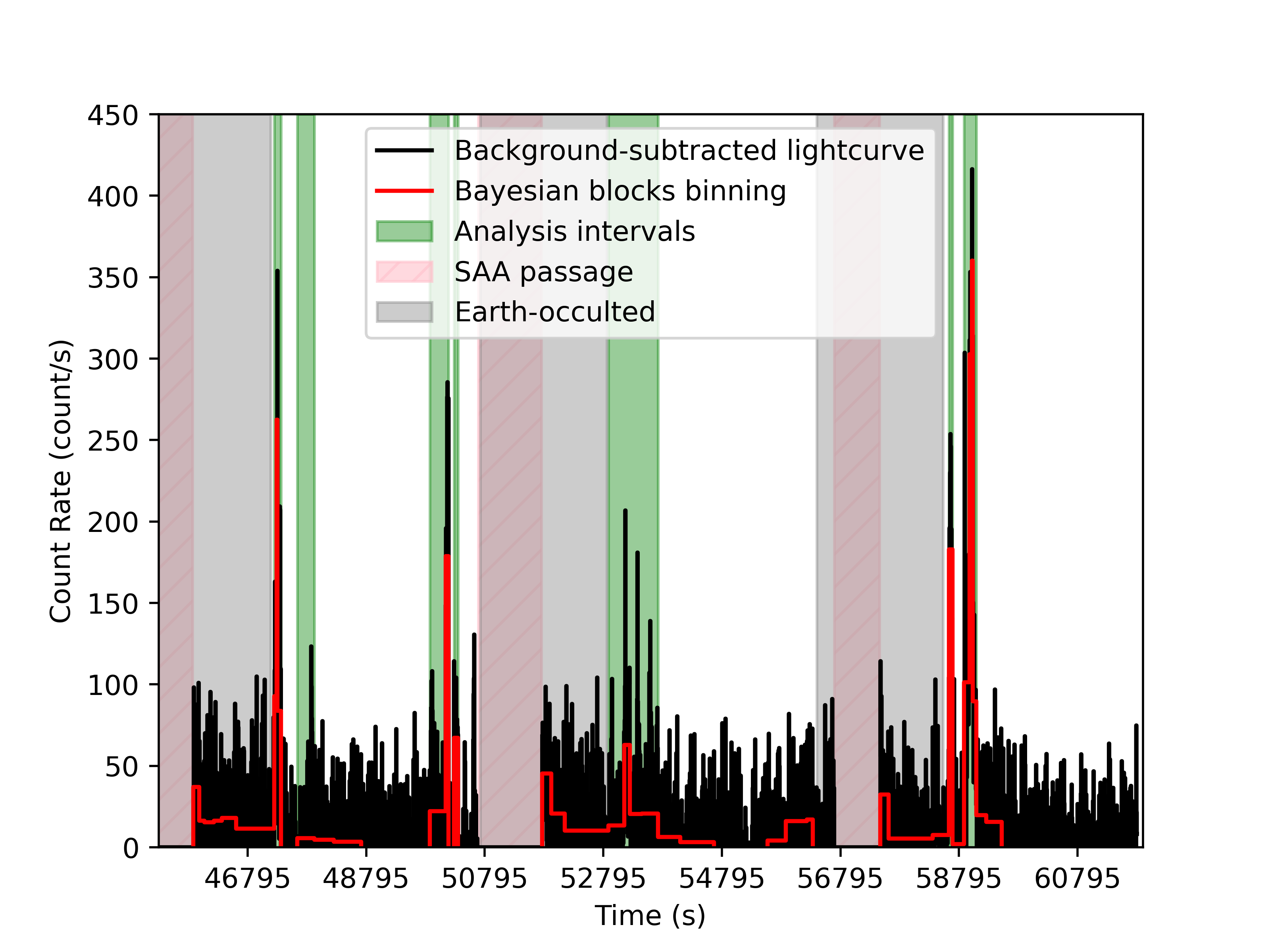}
    \caption{The background-subtracted \textit{Fermi}-GBM lightcurve binned using Bayesian blocks. The analysis intervals which include bins with SNR $>10$ are overlaid.}
    \label{fig:bb-lc}
\end{figure}

The brightest gamma-ray emission occurs between $\sim$T0+46,074~s and $\sim$T0+61,800~s. We bin the \textit{Fermi}-GBM lightcurve for detectors n7, n8, n9, nb, and b1 in this interval using Bayesian blocks, in the 50--500~keV energy range for the NaI detectors and 400--1,000~keV for the BGO detectors. The time periods composed of bins with $\text{SNR}>$10 are summarized in Table~\ref{table:gbm-bright-intervals-raw}. We measure the MVT and spectral lag in each of these intervals as well as perform spectral analysis, finding the peak energy to vary with brightness as is typical for GRBs. In order to constrain spectral curvature in every interval, we merge adjacent bins into a single interval, as displayed in Table~\ref{table:gbm-bright-intervals}. The lightcurve binned using Bayesian blocks is shown in Fig.~\ref{fig:bb-lc}.

\begin{table}
\caption{Intervals with SNR $>$10 in the Bayesian blocks binned lightcurve.} 
\centering
\begin{tabular}{cccc}
    \hline
    GBM Interval & Time Interval & Good \\
    Name & - T0 (s) & Detectors \\
    \hline
    1a & 47,245--47,285 & n8, nb, b1 \\
    1b & 47,285--47,300 & n8, nb, b1 \\
    1c & 47,300--47,355 & n8, nb, b1 \\
    2 & 47,630--47,920 & n8, nb, b1 \\
    3a & 49,865--50,135 & n9, na, nb, b1 \\ 
    3b & 50,135--50,185 & n9, na, nb, b1 \\
    4 & 50,275--50,345 & n9, na, nb, b1 \\
    5a & 52,880--53,150 & n7, n8, nb, b1 \\
    5b & 53,150--53,245 & n7, n8, nb, b1 \\
    5c & 53,245--53,715 & n7, n8, nb, b1 \\
    5d & 53,445--53,715 & n7, n8, nb, b1 \\
    6 & 58,625--58,685 & n8, nb, b1 \\
    7a & 58,880--58,975 & n8, nb, b1 \\
    7b & 58,975--58,990 & n8, nb, b1 \\
    7c & 58,990--59,015 & n8, nb, b1 \\
    7d & 59,015--59,030 & n8, nb, b1 \\
    7e & 59,030--59,085 & n8, nb, b1 \\
    \hline
\end{tabular}
\label{table:gbm-bright-intervals-raw}
\end{table}

\begin{table}
\caption{Intervals determined using the Bayesian blocks binned lightcurve in which we perform detailed \textit{Fermi}-GBM analysis. The good detectors are those with a viewing angle within $60\degree$ of the source, which are used in analysis. These intervals are shaded on the lightcurve in Fig.~\ref{fig:lc}.} 
\centering
\begin{tabular}{ccc}
    \hline
    GBM Interval & Time Interval & Good \\
    Name & - T0 (s) & Detectors \\
    \hline
    1 & 47,245--47,355 & n8, nb, b1 \\
    2 & 47,630--47,920 & n8, nb, b1 \\
    3 & 49,865--50,185 & n9, na, nb, b1 \\
    4 & 50,275--50,345 & n9, na, nb, b1 \\
    5 & 52,880--53,715 & n7, n8, nb, b1 \\
    6 & 58,625--58,685 & n8, nb, b1 \\
    7 & 58,880--59,085 & n8, nb, b1 \\
    \hline
\end{tabular}
\label{table:gbm-bright-intervals}
\end{table}

There are two additional intervals in which the SNR $>$10: T0+51,765--51,915~s, and T0+57,465--57,610~s. These are believed to be due to poor background estimation at times near the SAA passages of \textit{Fermi}, as described in Section~\ref{sec:gbm-background}, and are therefore not included in analysis. The time window containing GRB~250702C, an unrelated short GRB which occurred between T0+53,371.46~s and T0+53,371.97~s, is excised from Interval 5 for analyses.

\subsection{Extended Searches for Gamma-Ray Emission}\label{app:extended-searches}
In order to determine the length of the prompt gamma-ray duration, we perform dedicated searches in \textit{Swift}-BAT, MAXI, and \textit{Fermi}-GBM for emission outside of the main $\sim$25,000~s gamma-ray emission interval. As detailed below, in no instrument do we find significant evidence for gamma-ray emission outside of our 25,000~s interval. 

\subsubsection{Swift-BAT Analysis}
To search for extended hard X-ray to gamma-ray emission, \citet{oconnor2025grb250702b} analyzed the \textit{Swift}-BAT survey data products from the observations with GRB~250702B in the coded field-of-view. Two significant detections of GRB~250702B were made over observations from T0+59,411~s to T0+60,267~s and from T0+70,549~s to T0+71,607~s, as illustrated in Fig.~\ref{fig:lc}. No other significant emission was found in \textit{Swift}-BAT survey data at the position of GRB~250702B from $\sim$1 month prior to July 2 to $\sim$4 days after July 2. 

\subsubsection{MAXI Analysis}
We searched for significant emission in the 2--20~keV energy range in MAXI on July 2nd 2025. We find only three intervals with signals over 2$\sigma$, consistent with initial reporting \citep{2025GCN.40910....1K}. Detections and non-detections are summarized in Table~\ref{table:maxi}.

\begin{table}
\caption{Detections and non-detections around the time of GRB~250702B by MAXI in the 2--20~keV energy range.} 
\centering
\begin{tabular}{ccc}
\hline
Start Time - T0 (s) & Stop Time - T0 (s) & Detection \\
\hline
2,451 & 2,519 & No \\
8,026 & 8,095 & No \\
13,590 & 13,665 & No \\
19,174 & 19,243 & No \\
24,744 & 24,816 & No \\
30,319 & 30,390 & No \\
35,886 & 35,959 & No \\
41,461 & 41,534 & No \\
47,035 & 47,107 & Yes \\
52,614 & 52,688 & Yes \\
58,181 & 58,254 & Yes \\
63,762 & 63,835 & No \\
69,335 & 69,408 & No \\
74,909 & 74,983 & No \\
80,481 & 80,555 & No \\
86,055 & 86,131 & No \\
\hline
\end{tabular}
\label{table:maxi}
\end{table}

\subsubsection{Fermi-GBM Targeted Search Results}
GBM has developed increased sensitivity to transient signals lying below the on-board triggering algorithms by means of subthreshold searches. The Targeted Search was developed for multi-messenger follow-up \citep{Blackburn2015}, and is currently used to identify subthreshold GRB emission triggered by other instruments, such as \textit{Swift}-BAT \citep{Kocevski_2018} and the ECLAIRS telescope onboard the \textit{Space-based multi-band astronomical Variable Objects Monitor} (e.g., \citealt{2024GCN.36856....1B}, \citealt{2024GCN.38184....1R}). The Targeted Search processes continuous time-tagged event data from all 14 detectors coherently around an input time. Three model spectra \citep{Goldstein2016} are folded through the detector responses to produce templates of expected counts, which are then compared to the observed distribution of counts in each energy channel of each detector. The comparison is performed via a log-likelihood ratio (LLR), testing the alternative hypothesis of the presence of a signal with a similar spectrum versus the null hypothesis of only background noise. Treating the LLR as our detection statistic, the model spectrum resulting in the highest LLR is selected as the preferred spectrum, and this procedure is repeated for each bin of data in the search. For more details on the Targeted Search method, see \citet{Blackburn2015,Goldstein2016,goldstein2019}.

To search for gamma-ray emission from GRB~250702B, we run the Targeted Search from T0-86,460~s to T0+345,659~s over periods of time when the source is visible to \textit{Fermi}-GBM, i.e. not in the South Atlantic Anomaly (SAA) and not Earth-occulted. The search is run using overlapping segments of 300~s and processing timescales from 64 ms~to 32.768~s increasing by factors of 2. As performed in standard GBM follow-up, significant events found with the soft spectral template on the 8~s timescale are removed to limit contamination from non-GRB sources \citep{goldstein2019}. 

We find a total of 93 significant candidates, but using a cut on the spatial association probability of $\ge$85\%, we reduce the sample to 38 candidates. We then perform a manual inspection of the lightcurves, localization maps, and spacecraft orbital locations to determine the nature of the candidate. Excluding known GBM triggers, we find 9 candidates consistent with the location and spectral nature of GRB~250702B, detailed in Table~\ref{table:gbm-ts-results}.

\begin{table}
\caption{Candidate subthreshold detections of GRB~250702B by the GBM Targeted Search, where the start times are relative to midnight July 2. LLR is the log-likelihood ratio and $p_{\rm spatial}$ is the spatial p-value.} 
\centering
\begin{tabular}{ccccc}
    \hline
    Start Time - T0 (s) & Timescale & Spectrum & LLR & $p_{\rm spatial}$ \\ 
    \hline
     -12,294.428 & 32.768 & Soft & 14.78 & 97.8 \\
     47,285.332 & 16.384 & Hard & 113.50 & 98.3 \\
     50,182.596 & 4.096 & Normal & 32.21 & 94.0 \\
     50,168.452 & 0.128 & Soft & 14.69 & 88.9 \\
     52,916.956 & 32.768 & Hard & 20.38 & 93.8 \\
     54,900.036 & 32.768 & Soft & 25.07 & 98.3 \\
     58,636.644 & 32.768 & Hard & 98.69 & 99.3 \\
     58,665.316 & 16.384 & Normal & 47.68 & 98.3 \\
     271,823.34 & 32.768 & Soft & 59.88 & 99.2 \\
    \hline
\end{tabular}
\label{table:gbm-ts-results}
\end{table}

There is a possible signal associated with GRB~250702B found on July 1, and another on July 5. In order to estimate the significance of these candidates, we run the identical search settings over $\sim$5~Ms of randomly sampled livetime to generate a background distribution. For every candidate event found with the Targeted Search, a skymap is generated, which includes a systematic localization error. Using these skymaps, we calculate a spatial p-value utilizing GDT-Fermi. This spatial check is important because the source position is near the Galactic center, where we expect an excess of soft and long signals from unrelated sources. We assign a ranking statistic for each candidate as the log-likelihood ratio times the spatial p-value. The more significant of the two is the event on July 5. We find a false alarm rate of $4.5\times10^{-5}$~Hz. However, over a multi-day time range, this event is not significant. Given the lack of signal in BAT, we find no convincing evidence for gamma-ray duration beyond the 25,000~s bright interval on July 2nd.

\subsection{GRB~250702C Dissociation}\label{sec:grb250702c}
During the initial reporting of GRB~250702B there were four separate \textit{Fermi}-GBM triggers whose real-time localizations suggested a common origin \citep{2025GCN.40891....1N}. Follow-up analysis dissociated one trigger, assigning it to a separate burst, GRB~250702C \citep{2025GCN.40931....1N}. The confusion arises because GRB~250702C, a short GRB, lies on top of a long interval of emission from GRB~250702B, corresponding to our GBM Interval 5. By convention, GBM on-board triggers are cataloged according to the specific emission responsible for the trigger, even if there is contemporaneous emission from another source at this time and thus this trigger is GRB~250702C.

\begin{figure*}
    \centering
    \includegraphics[width=0.49\textwidth]{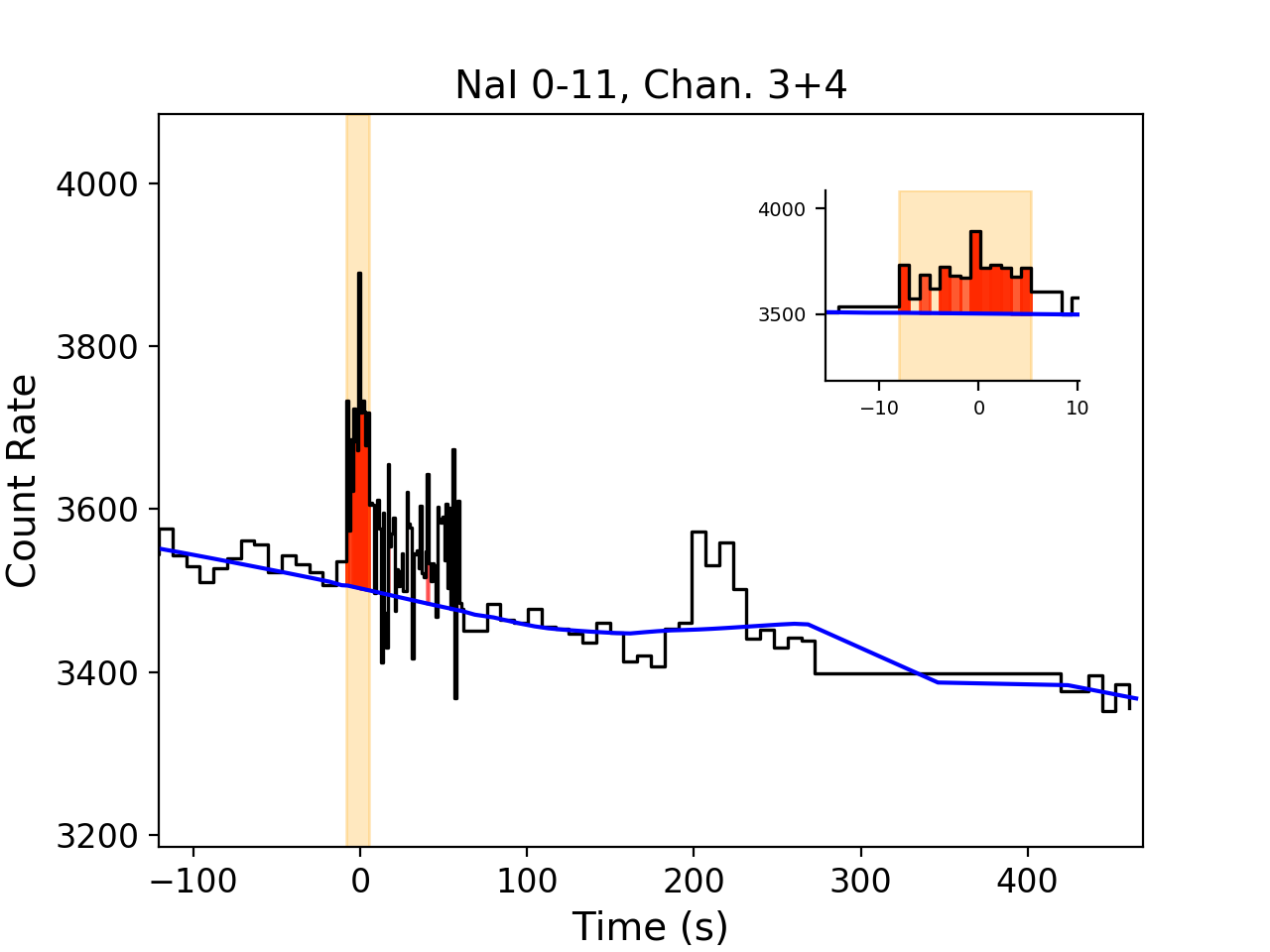}
    \includegraphics[width=0.49\textwidth]{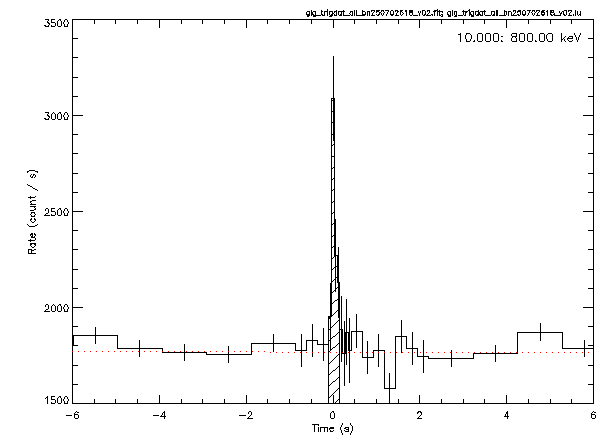}
    \includegraphics[width=0.49\textwidth]{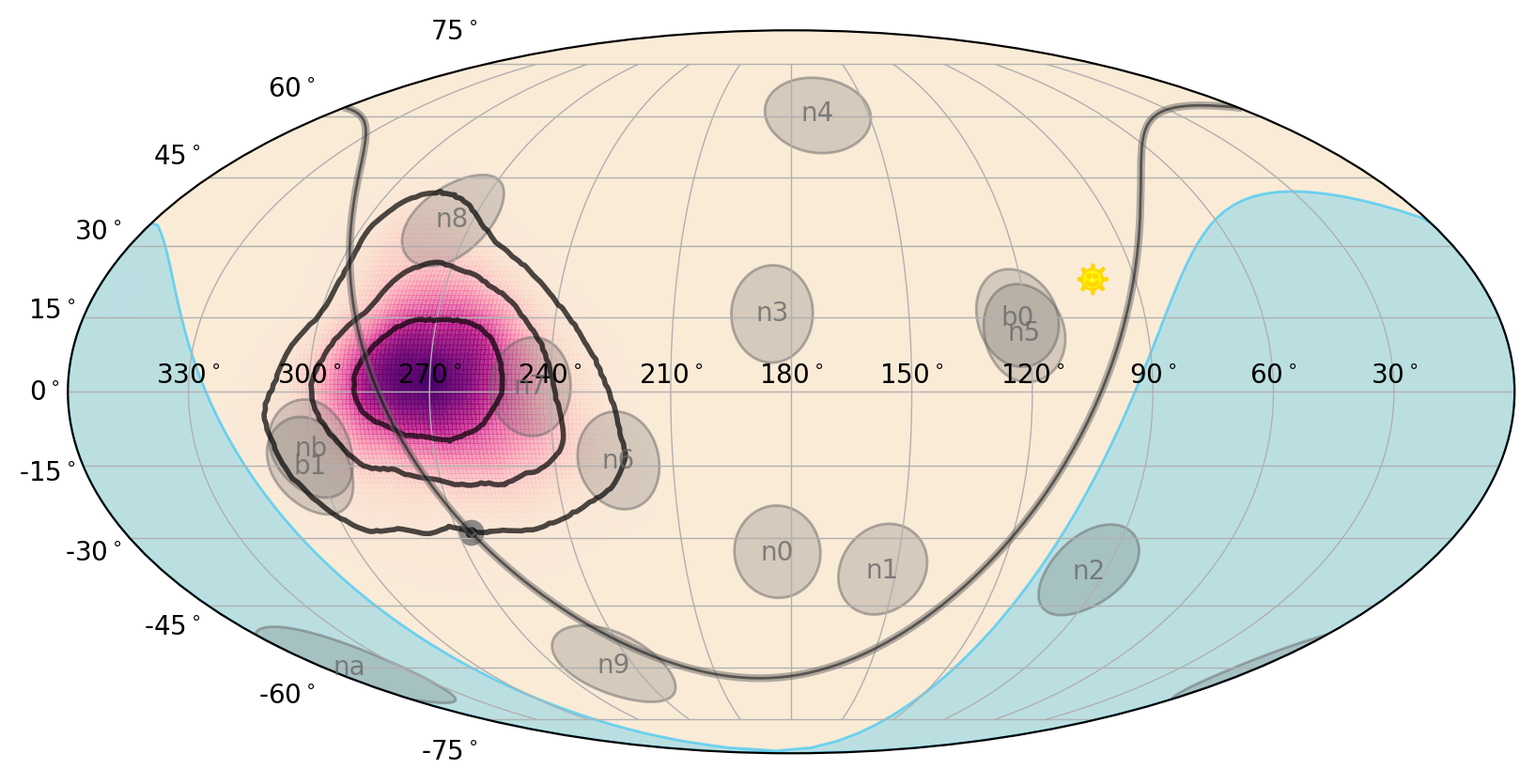}
    \includegraphics[width=0.49\textwidth]{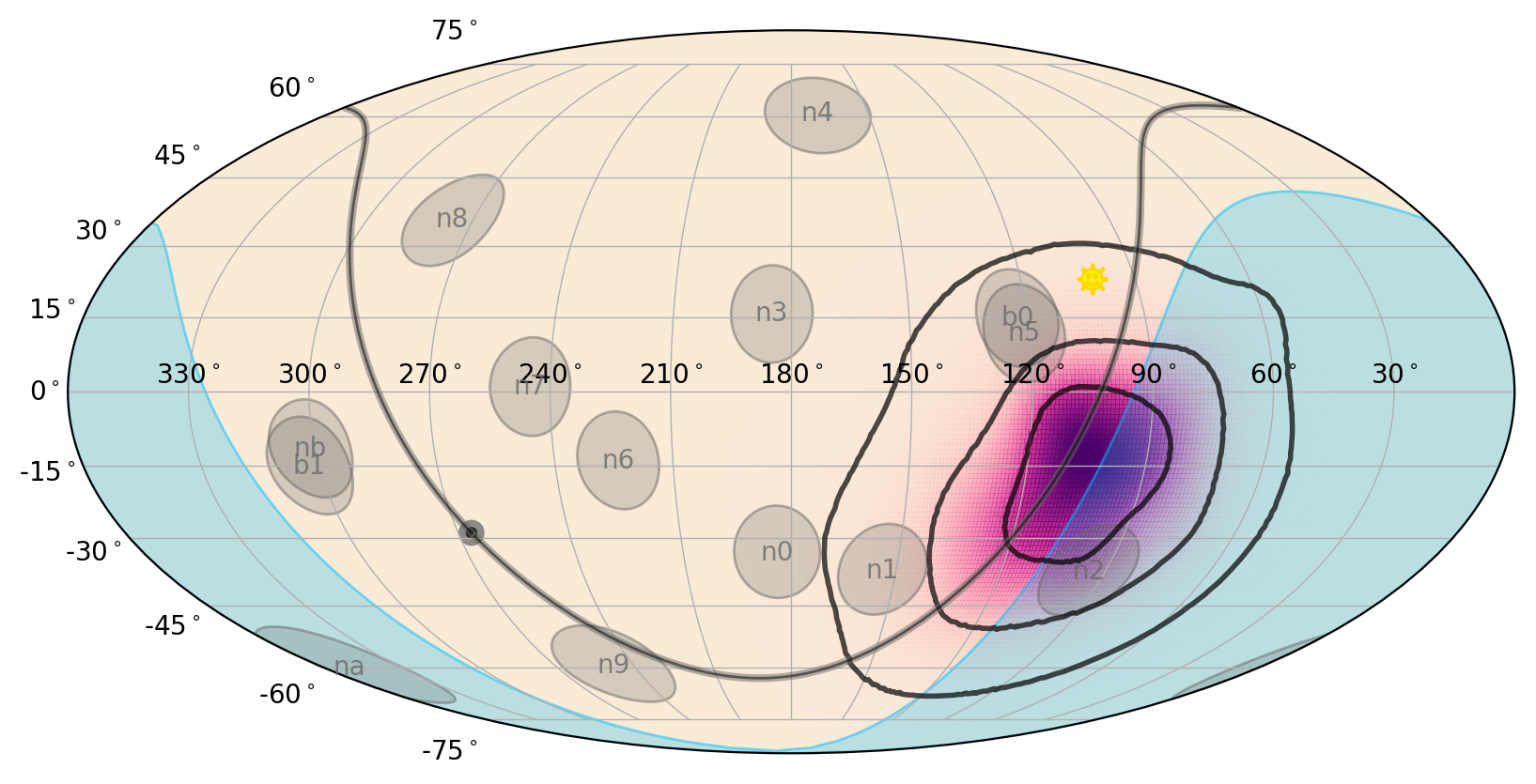}
    \caption{\textit{Left}: The automated \textit{Fermi}-GBM lightcurve selection and localization of the trigger now assigned to GRB~250702C. \textit{Right}: The manual lightcurve selection on the short burst GRB~250702C. The longer emission from GRB~250702B dominated the photon counts and pulled the localization to this source, despite the trigger burst originating from a different localization.}
    \label{fig:GRB250702C}
\end{figure*}

The confusion arose because the automated \textit{Fermi}-GBM localization software \citep{goldstein2020evaluation} selects the dominant emission around the trigger time, which in this case was actually due to GRB~250702B. Manual analysis at fine time resolution around the specific trigger time identified a very short and significant pulse right at trigger time, as displayed in Fig.~\ref{fig:GRB250702C}. Localization shows it originated from a different position on the sky. Thus, it is a separate burst that happened to occur on top of emission from GRB~250702B.

\section{Additional Analysis and Inference Details}
Section~\ref{sec:analysis} contains the key results from the gamma-ray analysis. However, the details of these analyses are sometimes contained here.

\subsection{Minimum Variability Timescale}\label{app:mvt}
\begin{figure*}
    \centering
    \includegraphics[width=0.48\textwidth]{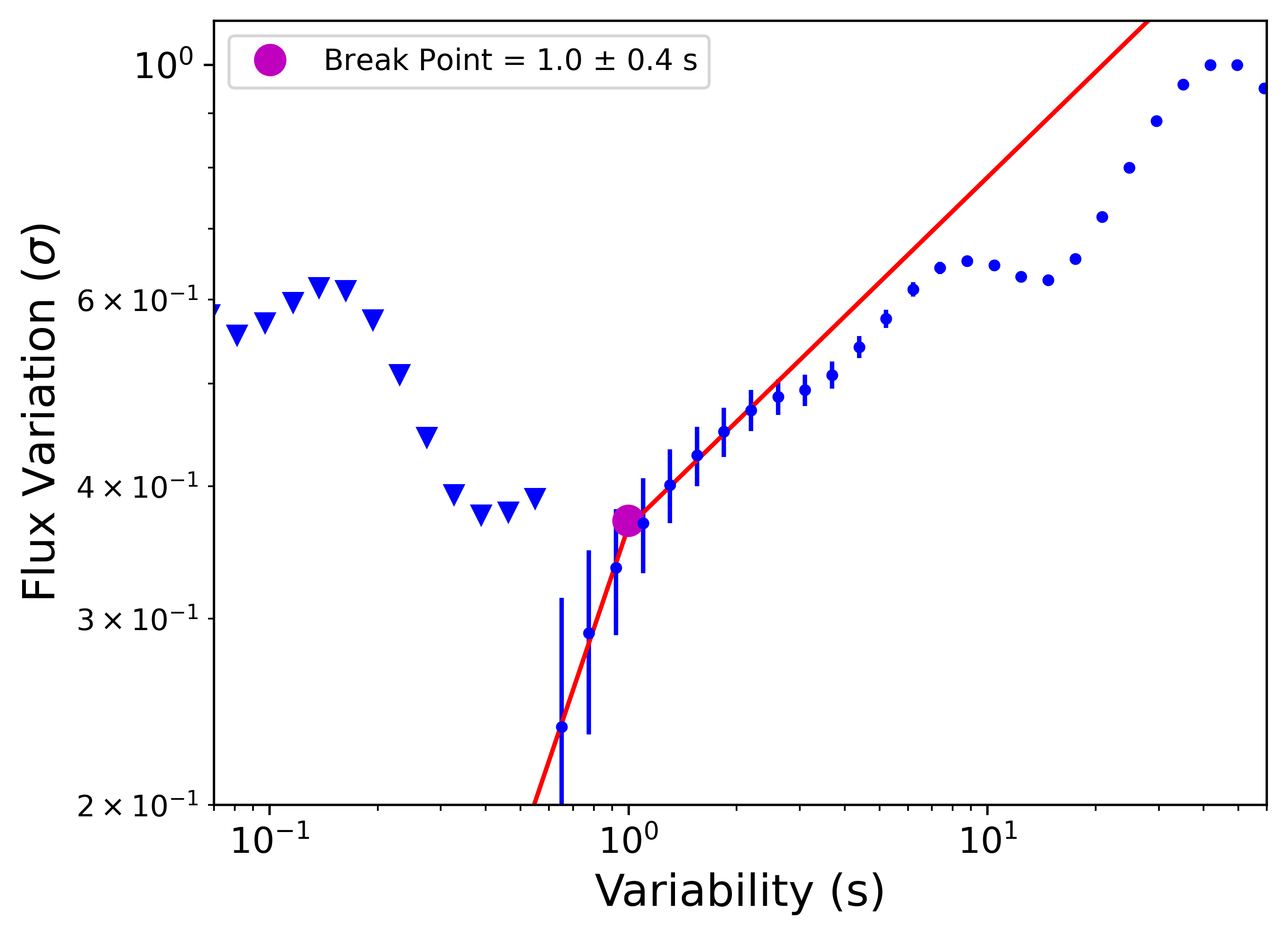}
    \includegraphics[width=0.49\textwidth]{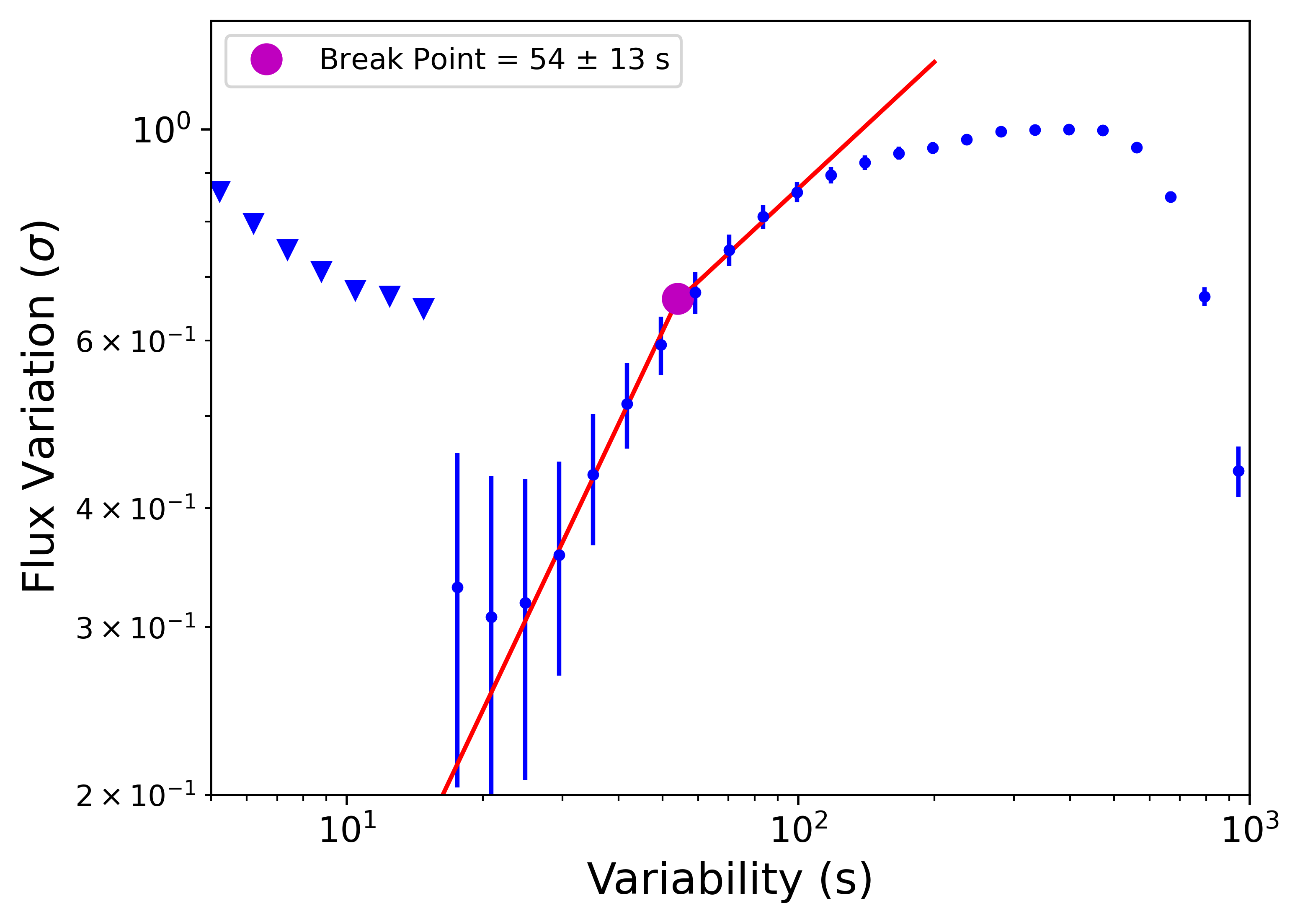}
    \caption{The flux variation as a function of variability timescale for T0+50,068–50,359~s (left), the time interval which yields the smallest MVT of GRB~250702B, and Swift~J1644+57 (right). A broken power law is fit to these data, with the break point corresponding to the MVT.}
    \label{fig:mvt}
\end{figure*}

Here, we display plots showing the measurement of key MVT values. The lowest MVT measured for GRB~250702B and the MVT measurement of Swift~J1644+57 are shown in Fig.~\ref{fig:mvt}. We show the measured MVT from the \textit{Swift}-BAT GUANO intervals of GRB~250702B in Table~\ref{table:bat-mvt}.

\begin{table}
\caption{\textit{Swift}-BAT MVT results for GRB~250702B.} 
\centering
\begin{tabular}{cc}
    \hline
    Interval - T0 (s) & MVT (s) \\ [0.5ex]
    \hline
    47,323--47,355 & $<5.48$ \\
    53,355--53,555 & $<40.27$ \\
    58,880--59,075 & $6.8\pm0.9$ \\
    \hline
\end{tabular}
\label{table:bat-mvt}
\end{table}

\subsection{Spectral Analysis}\label{app:spectral-analysis}
We here detail the spectral analysis of GRB~250702B in \textit{Fermi}-GBM and Konus-\textit{Wind}. For comparison, the Konus intervals map onto the GBM intervals in the following way: Konus~1~$\rightarrow$~GBM~1~and~2, Konus~2 $\rightarrow$~GBM~3~and~4, Konus~3~$\rightarrow$~GBM~5, and Konus~4~$\rightarrow$~GBM~6~and~7, as visualized in Fig.~\ref{fig:lc}.

\subsubsection{Fermi-GBM}\label{app:gbm-spectral-analysis}
We perform time-integrated spectral analysis for each interval in Table~\ref{table:gbm-bright-intervals}, using the detectors with observing angles within $60\degree$ of the GRB~250702B position and data binned at 5~s resolution. Both the orbital background method and polynomial fitting are used to estimate the background, and we compare the results from these two methods. Most spectral fits with the polynomial background use the standard energy selection of 8--900~keV for the NaI detectors and 280~keV--40~MeV for the BGO detectors. However, the polynomial background estimation in Interval~3 is problematic $\lesssim$75~keV, and we therefore perform spectral analysis using 75--900~keV for the NaI detectors. The orbital background is not reliable at low energies, which may be due to interfering point sources and/or weak soft emission from GRB~250702B outside of the known intervals. We thus use energy ranges of 75--900~keV in Intervals~2~and~3, 100--900~keV in Interval~4, and 50--900~keV in the other intervals for the NaI detectors, and 280~keV--40~MeV for the BGO detectors.

A forward-folding analysis is performed using GDT-Fermi, in which a spectral model is convolved with the detector response matrix. The detector response represents the relationship between the incident energies of photons and observed counts in the detector energy channels of \textit{Fermi}-GBM for a particular observation. The GRB~250702B position determined by \textit{Swift}-XRT \citep{2025GCN.40919....1K} is used when computing the detector response matrices. The result of the convolution, which is the expected energy distribution of source counts based on the model, is combined with the background estimate. This is compared with the measured data using the PG statistic \citep[pgstat;][]{1999ascl.soft10005A}, which assumes a Poisson signal and Gaussian background.

The following spectral models are fit where $N(E)$ is the photon spectrum and $E$ is the photon energy: 
\begin{enumerate}
    \item power law (PL): 
        \begin{equation}
            N(E) = A \Big(\frac{E}{E_{\text{piv}}}\Big)^\alpha
        \end{equation}
        where the free parameters are amplitude $A$ and photon index $\alpha$. The pivot energy $E_{\rm piv}$ is fixed to 100~keV, following \citet{2021ApJ...913...60P}.
    \item blackbody (BB):
        \begin{equation}
            N(E) = A \frac{E^2}{e^{\frac{E}{kT}} - 1}
        \end{equation}
        where the free parameters are amplitude $A$ and blackbody temperature $kT$.
    \item cutoff power law (CPL), also known as Comptonized:
        \begin{equation}
            N(E) = A \Big(\frac{E}{E_{\text{piv}}}\Big)^\alpha e^{(2+\alpha) \frac{E}{E_p}}
        \end{equation}
        where the free parameters are amplitude $A$, photon index $\alpha$, peak energy of the $\nu F_\nu$ spectrum $E_p$. The pivot energy $E_{piv}$ is fixed to 100~keV.
    \item Band function \citep{1993ApJ...413..281B}:
        \begin{equation}
            N(E) = A 
            \begin{cases}
                \Big(\frac{E}{E_{\text{piv}}}\Big)^\alpha e^{-(2+\alpha)\frac{E}{E_p}} & \text{if } E \leq \xi E_p \\
                \Big(\frac{E}{E_{\text{piv}}}\Big)^\beta \Big(\xi \frac{E_p}{E_\text{piv}}\Big)^{\alpha-\beta} e^{\beta-\alpha} & \text{if } E > \xi E_p \\
            \end{cases}
        \end{equation}
        where $\xi = \frac{\alpha-\beta}{2+\alpha}$. The free parameters are amplitude $A$, low-energy photon index $\alpha$, high-energy photon index $\beta$, and peak energy of the $\nu F_\nu$ spectrum $E_p$. The pivot energy $E_{\rm piv}$ is fixed to 100~keV.
    \item smoothly broken power law \citep[SBPL;][]{2006ApJS..166..298K}:
        \begin{equation}
            N(E) = A \Big(\frac{E}{E_\text{piv}}\Big)^b 10^{a-a_\text{piv}}
        \end{equation}
        where $a = m n \ln\Big(\frac{e^q + e^{-q}}{2}\Big)$, $a_\text{piv} = m n \ln\Big(\frac{e^{q_\text{piv}} + e^{-q_\text{piv}}}{2}\Big)$, $q = \log(\frac{E}{E_b})/n$, $q_\text{piv} = \log(\frac{E_\text{piv}}{E_b})/n$, $m = (\beta-\alpha)/2$, and $b = (\alpha+\beta)/2$. The free parameters are amplitude $A$, low-energy photon index $\alpha$, high-energy photon index $\beta$, and break energy $E_b$. The smoothness parameter $n$ is fixed to 0.3, following \citet{2021ApJ...913...60P}.
    \item double smoothly broken power law (2SBPL; \citet{2018AA...613A..16R}):
        \begin{multline}
            N(E) = A E_b^{\alpha_1} \Bigg( \bigg( \Big( \frac{E}{E_b} \Big)^{-\alpha_1 n_1} + \Big( \frac{E}{E_b} \Big)^{-\alpha_2 n_1} \bigg)^{\frac{n_2}{n_1}} \\
             + \Big( \frac{E}{E_j}  \Big)^{-\beta n_2} \bigg( \Big( \frac{E_j}{E_b} \Big)^{-\alpha_1 n_1} + \Big( \frac{E_j}{E_b} \Big)^{-\alpha_2 n_1} \bigg)^{\frac{n_2}{n_1}} \Bigg)^{-\frac{1}{n_2}}
        \end{multline}
        where $E_j = \big(-(2+\alpha_2)/(2+\beta)\big)^{1/n_2(\beta-\alpha_2)} E_p$. The free parameters are amplitude $A$, low-energy photon index $\alpha_1$, photon index between the break and peak energies $\alpha_2$, high-energy photon index $\beta$, peak energy $E_p$, and break energy $E_b$. The break energy smoothness parameter $n_1$ is fixed to 5.38, and the peak energy smoothness parameter $n_2$ is fixed to 2.69, following \citet{2018AA...613A..16R}.
\end{enumerate}
We also test the following models: Band function with an exponential cutoff \citep{2018ApJ...864..163V}, Band function with a SBPL \citep{2018ApJ...864..163V}, PL+BB, CPL+BB, Band+BB \citep{2011ApJ...727L..33G}, and SBPL+BB. These do not yield solid constraints and are never statistically preferred, so they are excluded from the results. 

\begin{figure*}
    \centering
    \includegraphics[width=0.9\textwidth]{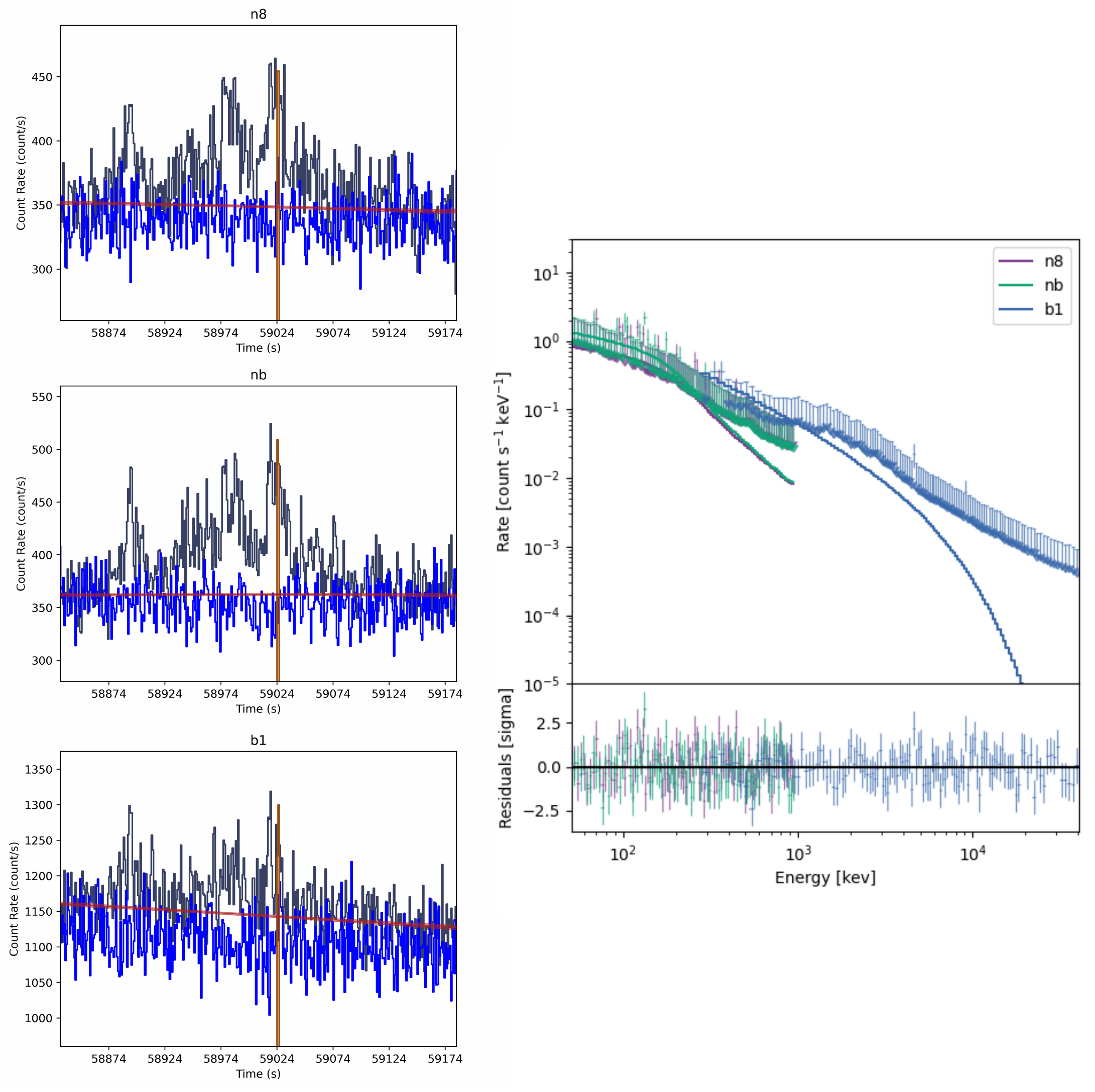}
    \caption{Spectral analysis results for the peak flux interval between T0+59,024.082 and T0+59,025.106~s using the orbital background estimation method. We use an energy range of 50--900~keV for the NaI detectors and 280~keV--40~MeV for the BGO detectors. \textit{Left}: Lightcurves for each detector showing the signal during the GRB~250702B interval (black) with the peak flux interval shaded (light red), the raw orbital background (blue), and the polynomial fit to the orbital background and then normalized (red). \textit{Right}: Count spectrum showing the CPL model fit to data.}
    \label{fig:peak-flux-spectrum-orb}
\end{figure*}

\begin{figure*}
    \centering
    \includegraphics[width=0.9\textwidth]{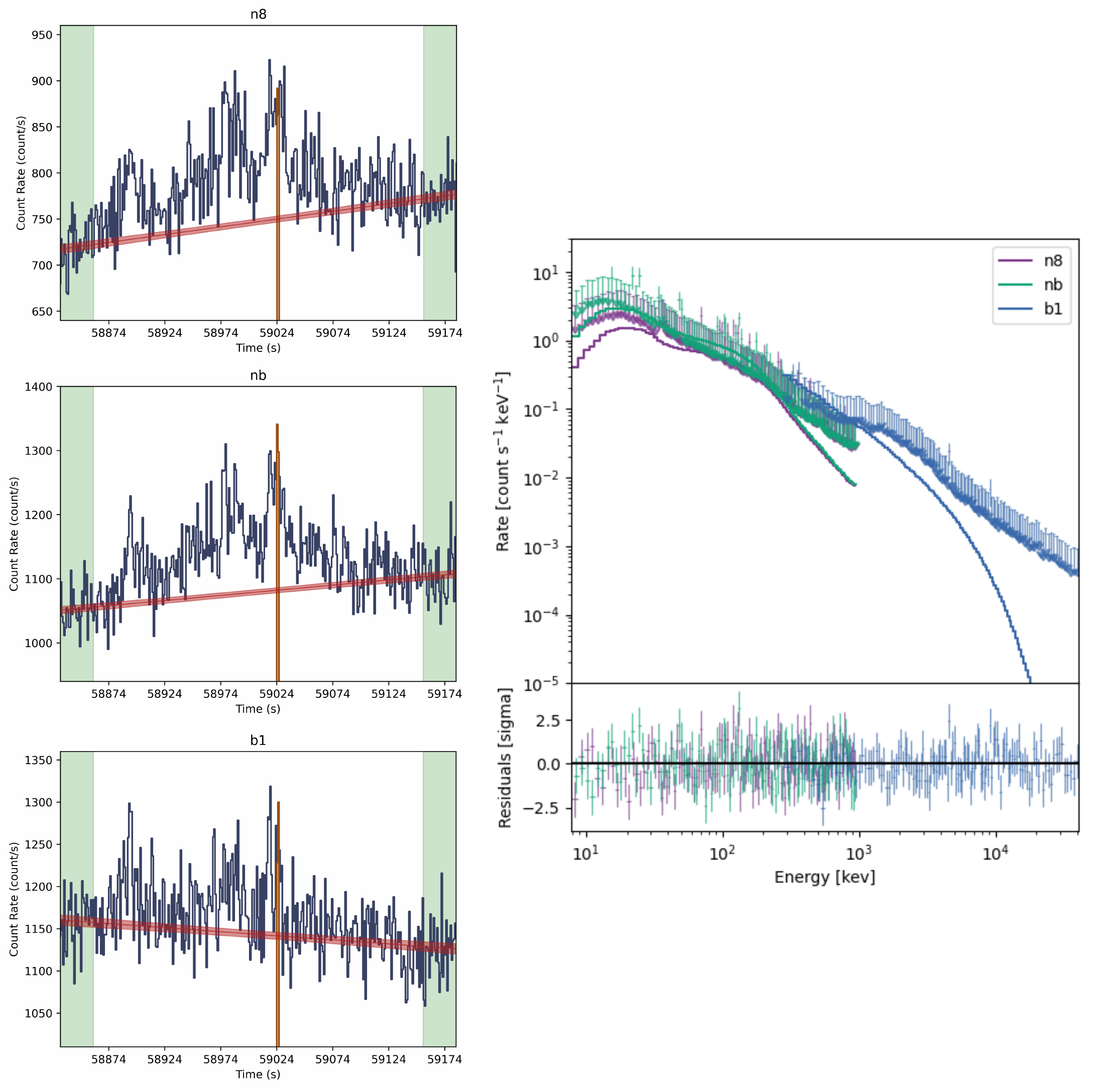}
    \caption{Spectral analysis results for the peak flux interval between T0+59,024.082 and T0+59,025.106~s using the polynomial background estimation method. We use an energy range of 8--900~keV for the NaI detectors and 280~keV--40~MeV for the BGO detectors. \textit{Left}: Lightcurves for each detector showing the signal during the GRB~250702B interval (black) with the peak flux interval shaded (light red), the intervals in which the background polynomial is fit (green), and the polynomial fit  (red). \textit{Right}: Count spectrum showing the CPL model fit to data.}
    \label{fig:peak-flux-spectrum-poly}
\end{figure*}

The results for all successful time-integrated fits of each analysis interval are in Tables~\ref{table:interval-1-spectra}--\ref{table:interval-7-spectra}, with the best fit models summarized in Table~\ref{table:spectra-summary}. We also perform spectral analysis for two peak flux intervals: T0+59,024.082--59,025.106~s and T0+50,163.206--50,164.230~s using the Interval~3 and 7 backgrounds, respectively. The results are summarized in Table~\ref{table:peak-flux-spectra-3} and Table~\ref{table:peak-flux-spectra-7}. The T0+50,163.206--50,164.230~s peak flux spectral results are plotted in Fig.~\ref{fig:peak-flux-spectrum-orb} and Fig.~\ref{fig:peak-flux-spectrum-poly}. Contrary to the usual GBM standard, we report fluences and fluxes in the 1--10,000~keV bolometric range, which cannot be directly compared with the values in the GBM catalog. Due to difficulties estimating the background in Intervals~3~and~4, we caution against drawing strong inferences from those measurements, including the Interval~3 peak flux results. The Interval~2~and~5 spectra are also unreliable because they consist of dim emission over a long time period.

\begin{table*}
\caption{Spectral analysis results for Interval~1 using \textit{Fermi}-GBM detectors n8, nb, and b1. The fluence is given in the 1--10,000~keV range. Uncertainties are given at the 90\% confidence level. For the NaI detectors, we use an energy range of 50--900~keV with the orbital background and 8--900~keV with the polynomial background. CPL is the preferred model with both the orbital and polynomial background.} 
\centering
\begin{tabular}{ccccccccccc}
    \hline
    Background & Model & $\alpha$ & $\beta$ & $E_p$ & $E_b$/kT & Stat/ & Fluence \\ 
     & & & & (keV) & (keV) & DOF & (erg/cm$^2$) \\
    \hline
    \rule{0pt}{2.5ex}
    Orbital & PL & $-1.40${\raisebox{0.5ex}{\tiny$^{+0.03}_{-0.03}$}} & & & & 451/303 & $1.06${\raisebox{0.5ex}{\tiny$^{+0.07}_{-0.06}$}}$\times10^{-4}$ \\ [0.5ex]
     & BB & & & & $582${\raisebox{0.5ex}{\tiny$^{+11}_{-12}$}} & 521/303 & $2.144${\raisebox{0.5ex}{\tiny$^{+0.07}_{-0.08}$}}$\times10^{-4}$ \\ [0.5ex]
     & \textbf{CPL} & $-0.71${\raisebox{0.5ex}{\tiny$^{+0.04}_{-0.05}$}} & & $3400${\raisebox{0.5ex}{\tiny$^{+400}_{-300}$}} & & 327/302 & $1.69${\raisebox{0.5ex}{\tiny$^{+0.05}_{-0.06}$}}$\times10^{-4}$ \\ [0.5ex]
     Polynomial & PL & $-1.40${\raisebox{0.5ex}{\tiny$^{+0.04}_{-0.04}$}} & & & & 301/361 & $6.6${\raisebox{0.5ex}{\tiny$^{+0.5}_{-0.4}$}}$\times10^{-5}$ \\ [0.5ex]
      & \textbf{CPL} & $-1.22${\raisebox{0.5ex}{\tiny$^{+0.08}_{-0.08}$}} & & $3500${\raisebox{0.5ex}{\tiny$^{+200}_{-2000}$}} & & 286/360 & $6.1${\raisebox{0.5ex}{\tiny$^{+0.6}_{-0.5}$}}$\times10^{-5}$ \\ [0.5ex]
      & SBPL & $-1.24${\raisebox{0.5ex}{\tiny$^{+0.08}_{-0.08}$}} & $-1.9${\raisebox{0.5ex}{\tiny$^{+0.2}_{-0.6}$}} & & $900${\raisebox{0.5ex}{\tiny$^{+1000}_{-500}$}} & 283/359 & $5.9${\raisebox{0.5ex}{\tiny$^{+0.5}_{-0.4}$}}$\times10^{-5}$ \\ [0.5ex]
    \hline
\end{tabular}
\label{table:interval-1-spectra}
\end{table*}

\begin{table*}
\caption{Spectral analysis results for Interval~2 using \textit{Fermi}-GBM detectors n8, nb, and b1. The fluence is given in the 1--10,000~keV range. Uncertainties are given at the 90\% confidence level. For the NaI detectors, we use an energy range of 75--900~keV with the orbital background and 8--900~keV with the polynomial background. BB is the preferred model with the orbital background and CPL is the preferred model with the polynomial background.} 
\centering
\begin{tabular}{ccccccccccc}
    \hline
    Background & Model & $\alpha$ & $\beta$ & $E_p$ & $E_b$/kT & Stat/ & Fluence \\ 
     & & & & (keV) & (keV) & DOF & (erg/cm$^2$) \\
    \hline
    \rule{0pt}{2.5ex}
    Orbital & PL & $-1.35${\raisebox{0.5ex}{\tiny$^{+0.03}_{-0.03}$}} & & & & 836/284 & $1.25${\raisebox{0.5ex}{\tiny$^{+0.08}_{-0.09}$}}$\times10^{-4}$ \\
     & \textbf{BB} & & & & $694${\raisebox{0.5ex}{\tiny$^{+12}_{-13}$}} & 442/284 & $3.89${\raisebox{0.5ex}{\tiny$^{+0.14}_{-0.13}$}}$\times10^{-4}$ \\
     & CPL & $1.44${\raisebox{0.5ex}{\tiny$^{+0.03}_{-0.03}$}} & & $2670${\raisebox{0.5ex}{\tiny$^{+60}_{-70}$}} & & 445/283 & $3.90${\raisebox{0.5ex}{\tiny$^{+0.13}_{-0.14}$}}$\times10^{-4}$ \\
    Polynomial & PL & $-1.22${\raisebox{0.5ex}{\tiny$^{+0.11}_{-0.2}$}} & & & & 435/361 & $2.3${\raisebox{0.5ex}{\tiny$^{+0.6}_{-0.9}$}}$\times10^{-5}$ \\
     & \textbf{CPL} & $2.7${\raisebox{0.5ex}{\tiny$^{+0.2}_{-0.3}$}} & & $2200${\raisebox{0.5ex}{\tiny$^{+200}_{-300}$}} & & 425/360 & $5.9${\raisebox{0.5ex}{\tiny$^{+0.7}_{-4}$}}$\times10^{-5}$ \\
    \hline
\end{tabular}
\label{table:interval-2-spectra}
\end{table*}

\begin{table*}
\caption{Spectral analysis results for Interval~3 using \textit{Fermi}-GBM detectors n9, na, nb, and b1. The fluence is given in the 1--10,000~keV range. Uncertainties are given at the 90\% confidence level. For the NaI detectors, we use an energy range of 75--900~keV with the orbital background and 75--900~keV with the polynomial background. CPL is the preferred model with the orbital background and PL is the preferred model with the polynomial background.} 
\centering
\begin{tabular}{ccccccccccc}
    \hline
    Background & Model & $\alpha$/ & $\beta$ & $E_p$ & $E_b$/kT & Stat/ & Fluence \\ 
     & & $\alpha_1$; $\alpha_2$ & & (keV) & (keV) & DOF & (erg/cm$^2$) \\
    \hline
    \rule{0pt}{2.5ex}
    Orbital & PL & $-1.37${\raisebox{0.5ex}{\tiny$^{+0.03}_{-0.03}$}} & & & & 698/366 & $1.42${\raisebox{0.5ex}{\tiny$^{+0.08}_{-0.08}$}}$\times10^{-4}$ \\ [0.5ex]
     & \textbf{CPL} & $-0.10${\raisebox{0.5ex}{\tiny$^{+0.04}_{-0.04}$}} & & $2700${\raisebox{0.5ex}{\tiny$^{+800}_{-800}$}} & & 449/365 & $2.62${\raisebox{0.5ex}{\tiny$^{+0.12}_{-0.12}$}}$\times10^{-4}$ \\ [0.5ex]
     & Band & $-0.10${\raisebox{0.5ex}{\tiny$^{+0.04}_{-0.05}$}} & $-5${\raisebox{0.5ex}{\tiny$^{+2}_{-5}$}} & $2700${\raisebox{0.5ex}{\tiny$^{+200}_{-200}$}} & & 449/364 & $2.63${\raisebox{0.5ex}{\tiny$^{+0.2}_{-0.13}$}}$\times10^{-5}$ \\ [0.5ex]
     & 2SBPL & $-0.453${\raisebox{0.5ex}{\tiny$^{+0.013}_{-0.014}$}}; $-0.68${\raisebox{0.5ex}{\tiny$^{+0.11}_{-0.12}$}} & $-6${\raisebox{0.5ex}{\tiny$^{+2}_{-3}$}} & $3000${\raisebox{0.5ex}{\tiny$^{+300}_{-300}$}} & $400${\raisebox{0.5ex}{\tiny$^{+300}_{-200}$}} & 430/362 & $2.8${\raisebox{0.5ex}{\tiny$^{+0.2}_{-0.2}$}}$\times10^{-4}$ \\ [0.5ex]
    Polynomial & \textbf{PL} & $-1.50${\raisebox{0.5ex}{\tiny$^{+0.05}_{-0.06}$}} & & & & 585/366 & $9.8${\raisebox{0.5ex}{\tiny$^{+0.9}_{-0.7}$}}$\times10^{-5}$ \\ [0.5ex]
     & BB & & & & $79${\raisebox{0.5ex}{\tiny$^{+2}_{-2}$}} & 630/366 & $1.99${\raisebox{0.5ex}{\tiny$^{+0.09}_{-0.10}$}}$\times10^{-5}$ \\ [0.5ex]
     & SBPL & $-1.34${\raisebox{0.5ex}{\tiny$^{+0.12}_{-0.14}$}} & $-1.60${\raisebox{0.5ex}{\tiny$^{+0.12}_{-0.2}$}} & & $400${\raisebox{0.5ex}{\tiny$^{+1000}_{-300}$}} & 583/364 & $9.4${\raisebox{0.5ex}{\tiny$^{+1.1}_{-0.9}$}}$\times10^{-5}$ \\ [0.5ex]
    \hline
\end{tabular}
\label{table:interval-3-spectra}
\end{table*}

\begin{table*}
\caption{Spectral analysis results for Interval~4 using \textit{Fermi}-GBM detectors n9, na, nb, and b1. The fluence is given in the 1--10,000~keV range. Uncertainties are given at the 90\% confidence level. For the NaI detectors, we use an energy range of 100--900~keV with the orbital background and 8--900~keV with the polynomial background. CPL is the preferred model with both the orbital and polynomial background.} 
\centering
\begin{tabular}{ccccccccccc}
    \hline
    Background & Model & $\alpha$ & $\beta$ & $E_p$ & $E_b$/kT & Stat/ & Fluence \\ 
     & & & & (keV) & (keV) & DOF & (erg/cm$^2$) \\
    \hline
    \rule{0pt}{2.5ex}
    Orbital & PL & $-1.41${\raisebox{0.5ex}{\tiny$^{+0.04}_{-0.05}$}} & & & & 305/343 & $4.47${\raisebox{0.5ex}{\tiny$^{+0.03}_{-0.03}$}}$\times10^{-5}$ \\ [0.5ex]
     & BB & & & & $610${\raisebox{0.5ex}{\tiny$^{+20}_{-20}$}} & 299/343 & $8.2${\raisebox{0.5ex}{\tiny$^{+0.4}_{-0.5}$}}$\times10^{-5}$ \\ [0.5ex]
     & \textbf{CPL} & $-0.48${\raisebox{0.5ex}{\tiny$^{+0.06}_{-0.07}$}} & & $3500${\raisebox{0.5ex}{\tiny$^{+500}_{-500}$}} & & 265/342 & $7.2${\raisebox{0.5ex}{\tiny$^{+0.5}_{-0.5}$}}$\times10^{-5}$ \\ [0.5ex]
     & Band & $-0.47${\raisebox{0.5ex}{\tiny$^{+0.07}_{-0.08}$}} & $-3.6${\raisebox{0.5ex}{\tiny$^{+1.3}_{-6}$}} & $3400${\raisebox{0.5ex}{\tiny$^{+500}_{-500}$}} & & 264/341 & $7.2${\raisebox{0.5ex}{\tiny$^{+0.6}_{-0.6}$}}$\times10^{-5}$ \\ [0.5ex]
     & SBPL & $-0.72${\raisebox{0.5ex}{\tiny$^{+0.05}_{-0.06}$}} & $-9.8${\raisebox{0.5ex}{\tiny$^{+3}_{-0.2}$}} & & $6300${\raisebox{0.5ex}{\tiny$^{+1000}_{-1100}$}} & 261/341 & $7.4${\raisebox{0.5ex}{\tiny$^{+0.5}_{-0.6}$}}$\times10^{-5}$ \\ [0.5ex]
    Polynomial & PL & $-1.42${\raisebox{0.5ex}{\tiny$^{+0.05}_{-0.06}$}} & & & & 398/481 & $3.0${\raisebox{0.5ex}{\tiny$^{+0.3}_{-0.3}$}}$\times10^{-5}$ \\ [0.5ex]
     & \textbf{CPL} & $-1.0${\raisebox{0.5ex}{\tiny$^{+0.2}_{-0.2}$}} & & $530${\raisebox{0.5ex}{\tiny$^{+200}_{-120}$}} & & 386/480 & $9.7${\raisebox{0.5ex}{\tiny$^{+1.4}_{-1.4}$}}$\times10^{-6}$ \\ [0.5ex]
     & SBPL & $-1.0${\raisebox{0.5ex}{\tiny$^{+0.3}_{-0.2}$}} & $-1.8${\raisebox{0.5ex}{\tiny$^{+0.2}_{-0.3}$}} & & $150${\raisebox{0.5ex}{\tiny$^{+70}_{-60}$}} & 383/479 & $1.9${\raisebox{0.5ex}{\tiny$^{+1.3}_{-0.7}$}}$\times10^{-5}$ \\ [0.5ex]
    \hline
\end{tabular}
\label{table:interval-4-spectra}
\end{table*}

\begin{table*}
\caption{Spectral analysis results for Interval~5 using \textit{Fermi}-GBM detectors n7, n8, nb, and b1. The fluence is given in the 1--10,000~keV range. Uncertainties are given at the 90\% confidence level. For the NaI detectors, we use an energy range of 50--900~keV with the orbital background and 8--900~keV with the polynomial background. CPL is the preferred model with both the orbital and polynomial background.} 
\centering
\begin{tabular}{ccccccccccc}
    \hline
    Background & Model & $\alpha$ & $\beta$ & $E_p$ & kT & Stat/ & Fluence \\ 
     & & & & (keV) & (keV) & DOF & (erg/cm$^2$) \\
    \hline
    \rule{0pt}{2.5ex}
    Orbital & PL & $-1.286${\raisebox{0.5ex}{\tiny$^{+0.013}_{-0.014}$}} & & & & 2331/395 & $4.86${\raisebox{0.5ex}{\tiny$^{+0.13}_{-0.13}$}}$\times10^{-4}$ \\ [0.5ex]
     & BB & & & & $713${\raisebox{0.5ex}{\tiny$^{+6}_{-6}$}} & 1155/395 & $1.30${\raisebox{0.5ex}{\tiny$^{+0.02}_{-0.02}$}}$\times10^{-3}$ \\ [0.5ex]
     & \textbf{CPL} & $0.40${\raisebox{0.5ex}{\tiny$^{+0.02}_{-0.02}$}} & & $3070${\raisebox{0.5ex}{\tiny$^{+60}_{-60}$}} & & 1117/394 & $1.21${\raisebox{0.5ex}{\tiny$^{+0.02}_{-0.02}$}}$\times10^{-3}$ \\ [0.5ex]
    Polynomial & PL & $-1.17${\raisebox{0.5ex}{\tiny$^{+0.02}_{-0.02}$}} & & & & 9296/482 & $3.37${\raisebox{0.5ex}{\tiny$^{+0.12}_{-0.13}$}}$\times10^{-4}$ \\ [0.5ex]
     & \textbf{CPL} & $-0.71${\raisebox{0.5ex}{\tiny$^{+0.03}_{-0.04}$}} & & $4700${\raisebox{0.5ex}{\tiny$^{+600}_{-600}$}} & & 9109/481 & $5.1${\raisebox{0.5ex}{\tiny$^{+0.2}_{-0.2}$}}$\times10^{-4}$ \\ [0.5ex]
    \hline
\end{tabular}
\label{table:interval-5-spectra}
\end{table*}

\begin{table*}
\caption{Spectral analysis results for Interval~6 using \textit{Fermi}-GBM detectors n8, nb, and b1. The fluence is given in the 1--10,000~keV range. Uncertainties are given at the 90\% confidence level. For the NaI detectors, we use an energy range of 50--900~keV with the orbital background and 8--900~keV with the polynomial background. CPL is the preferred model with both the orbital and polynomial background.} 
\centering
\begin{tabular}{ccccccccccc}
    \hline
    Background & Model & $\alpha$/ & $\beta$ & $E_p$ & $E_b$/kT & Stat/ & Fluence \\ 
     & & $\alpha_1$; $\alpha_2$ & & (keV) & (keV) & DOF & (erg/cm$^2$) \\
    \hline
    \rule{0pt}{2.5ex}
    Orbital & PL & $-1.51${\raisebox{0.5ex}{\tiny$^{+0.03}_{-0.04}$}} & & & & 278/303 & $6.3${\raisebox{0.5ex}{\tiny$^{+0.6}_{-0.5}$}}$\times10^{-5}$ \\ [0.5ex]
     & BB & & & & $65.5${\raisebox{0.5ex}{\tiny$^{+1.3}_{-1.3}$}} & 458/303 & $1.16${\raisebox{0.5ex}{\tiny$^{+0.04}_{-0.04}$}}$\times10^{-5}$ \\ [0.5ex]
     & \textbf{CPL} & $-1.14${\raisebox{0.5ex}{\tiny$^{+0.06}_{-0.06}$}} & & $3400${\raisebox{0.5ex}{\tiny$^{+900}_{-700}$}} & & 226/302 & $7.2${\raisebox{0.5ex}{\tiny$^{+0.3}_{-0.4}$}}$\times10^{-5}$ \\ [0.5ex]
    Polynomial & PL & $-1.37${\raisebox{0.5ex}{\tiny$^{+0.03}_{-0.03}$}} & & & & 360/361 & $5.9${\raisebox{0.5ex}{\tiny$^{+0.3}_{-0.3}$}}$\times10^{-5}$ \\ [0.5ex]
     & \textbf{CPL} & $-1.08${\raisebox{0.5ex}{\tiny$^{+0.06}_{-0.07}$}} & & $2900${\raisebox{0.5ex}{\tiny$^{+900}_{-700}$}} & & 291/360 & $5.7${\raisebox{0.5ex}{\tiny$^{+0.3}_{-0.3}$}}$\times10^{-5}$ \\ [0.5ex]
     & 2SBPL & $-0.69${\raisebox{0.5ex}{\tiny$^{+0.02}_{-0.02}$}}; $-1.27${\raisebox{0.5ex}{\tiny$^{+0.05}_{-0.06}$}} & $-6${\raisebox{0.5ex}{\tiny$^{+3}_{-4}$}} & $3600${\raisebox{0.5ex}{\tiny$^{+1300}_{-1300}$}} & $44${\raisebox{0.5ex}{\tiny$^{+8}_{-7}$}} & 286/357 & $6.0${\raisebox{0.5ex}{\tiny$^{+0.7}_{-0.9}$}}$\times10^{-5}$ \\ [0.5ex]
    \hline
\end{tabular}
\label{table:interval-6-spectra}
\end{table*}

\begin{table*}
\caption{Spectral analysis results for Interval~7 using \textit{Fermi}-GBM detectors n8, nb, and b1. The fluence is given in the 1--10,000~keV range. Uncertainties are given at the 90\% confidence level. For the NaI detectors, we use an energy range of 50--900~keV with the orbital background and 8--900~keV with the polynomial background. CPL is the preferred model with the orbital background. SBPL is the preferred model with the polynomial background.} 
\centering
\begin{tabular}{ccccccccccc}
    \hline
    Background & Model & $\alpha$ & $\beta$ & $E_p$ & $E_b$/kT & Stat/ & Flux \\ 
     & & & & (keV) & (keV) & DOF & (erg/cm$^2$/s) \\
    \hline
    \rule{0pt}{2.5ex}
    Orbital & PL & $-1.50${\raisebox{0.5ex}{\tiny$^{+0.02}_{-0.03}$}} & & & & 345/303 & $1.6${\raisebox{0.5ex}{\tiny$^{+0.2}_{-0.2}$}}$\times10^{-4}$ \\ [0.5ex]
     & BB & & & & $56.4${\raisebox{0.5ex}{\tiny$^{+0.7}_{-0.7}$}} & 697/303 & $2.66${\raisebox{0.5ex}{\tiny$^{+0.06}_{-0.07}$}}$\times10^{-5}$ \\ [0.5ex]
     & \textbf{CPL} & $-1.24${\raisebox{0.5ex}{\tiny$^{+0.04}_{-0.04}$}} & & $4300${\raisebox{0.5ex}{\tiny$^{+1000}_{-800}$}} & & 288/302 & $1.79${\raisebox{0.5ex}{\tiny$^{+0.07}_{-0.07}$}}$\times10^{-4}$ \\ [0.5ex]
     & Band & $-1.24${\raisebox{0.5ex}{\tiny$^{+0.04}_{-0.05}$}} & $-3.6${\raisebox{0.5ex}{\tiny$^{+1.4}_{-6}$}} & $4300${\raisebox{0.5ex}{\tiny$^{+1100}_{-900}$}} & & 288/301 & $1.79${\raisebox{0.5ex}{\tiny$^{+0.06}_{-0.06}$}}$\times10^{-4}$ \\ [0.5ex]
    Polynomial & PL & $-1.36${\raisebox{0.5ex}{\tiny$^{+0.02}_{-0.02}$}} & & & & 747/361 & $1.52${\raisebox{0.5ex}{\tiny$^{0.06}_{-0.06}$}}$\times10^{-4}$ \\ [0.5ex]
     & CPL & $-1.13${\raisebox{0.5ex}{\tiny$^{+0.04}_{-0.04}$}} & & $3600${\raisebox{0.5ex}{\tiny$^{+1000}_{-900}$}} & & 657/360 & $1.48${\raisebox{0.5ex}{\tiny$^{+0.08}_{-0.08}$}}$\times10^{-4}$ \\ [0.5ex]
     & \textbf{SBPL} & $-0.5${\raisebox{0.5ex}{\tiny$^{+0.2}_{-0.2}$}} & $-1.60${\raisebox{0.5ex}{\tiny$^{+0.05}_{-0.06}$}} & & $85${\raisebox{0.5ex}{\tiny$^{+20}_{-14}$}} & 620/359 & $1.19${\raisebox{0.5ex}{\tiny$^{+0.08}_{-0.08}$}}$\times10^{-4}$ \\ [0.5ex]
    \hline
\end{tabular}
\label{table:interval-7-spectra}
\end{table*}

\begin{table*}
\caption{Spectral analysis results for the peak flux interval between T0+50,163.206 and T0+50,164.230~s using \textit{Fermi}-GBM detectors n9, na, nb, and b1, and the Interval~3 orbital and polynomial backgrounds. The flux is given in the 1--10,000~keV range. Uncertainties are given at the 90\% confidence level. For the NaI detectors, we use an energy range of 75--900~keV with the orbital background and 75--900~keV with the polynomial background. CPL is the preferred model with both the orbital and polynomial background.} 
\centering
\begin{tabular}{ccccccccccc}
    \hline
    Background & Model & $\alpha$/ & $\beta$ & $E_p$ & $E_b$/kT & Stat/ & Flux \\ 
     & & $\alpha_1$; $\alpha_2$ & & (keV) & (keV) & DOF & (erg/cm$^2$/s) \\
    \hline
    \rule{0pt}{2.5ex}
     Orbital & PL & $-1.39${\raisebox{0.5ex}{\tiny$^{+0.07}_{-0.08}$}} & & & & 197/366 & $2.6${\raisebox{0.5ex}{\tiny$^{+0.4}_{-0.4}$}}$\times10^{-6}$ \\ [0.5ex]
     & BB & & & & $204${\raisebox{0.5ex}{\tiny$^{+10}_{-11}$}} & 194/366 & $1.26${\raisebox{0.5ex}{\tiny$^{+0.10}_{-0.12}$}}$\times10^{-6}$ \\ [0.5ex]
     & \textbf{CPL} & $-0.69${\raisebox{0.5ex}{\tiny$^{+0.10}_{-0.12}$}} & & $3800${\raisebox{0.5ex}{\tiny$^{+1200}_{-1100}$}} & & 179/365 & $3.8${\raisebox{0.5ex}{\tiny$^{+0.3}_{-0.5}$}}$\times10^{-6}$ \\ [0.5ex]
     & SBPL & $-0.93${\raisebox{0.5ex}{\tiny$^{+0.09}_{-0.11}$}} & $-7${\raisebox{0.5ex}{\tiny$^{+3}_{-3}$}} & & $7000${\raisebox{0.5ex}{\tiny$^{+3000}_{-3000}$}} & 180/364 & $4.1${\raisebox{0.5ex}{\tiny$^{+0.4}_{-0.5}$}}$\times10^{-6}$ \\ [0.5ex]
     & 2SBPL & $2.65${\raisebox{0.5ex}{\tiny$^{+0.04}_{-0.05}$}}; $-1.12${\raisebox{0.5ex}{\tiny$^{+0.12}_{-0.2}$}} & $-5${\raisebox{0.5ex}{\tiny$^{+3}_{-5}$}} & $5000${\raisebox{0.5ex}{\tiny$^{+3000}_{-2000}$}} & $133${\raisebox{0.5ex}{\tiny$^{+8}_{-9}$}} & 180/362 & $3.9${\raisebox{0.5ex}{\tiny$^{+0.5}_{-0.5}$}}$\times10^{-6}$ \\ [0.5ex]
     Polynomial & PL & $-1.42${\raisebox{0.5ex}{\tiny$^{+0.07}_{-0.09}$}} & & & & 193/366 & $2.4${\raisebox{0.5ex}{\tiny$^{+0.3}_{-0.4}$}}$\times10^{-6}$ \\ [0.5ex]
     & BB & & & & $140${\raisebox{0.5ex}{\tiny$^{+7}_{-8}$}} & 199/366 & $7.9${\raisebox{0.5ex}{\tiny$^{+0.7}_{-0.8}$}}$\times10^{-7}$ \\ [0.5ex]
     & \textbf{CPL} & $-0.93${\raisebox{0.5ex}{\tiny$^{+0.11}_{-0.14}$}} & & $5000${\raisebox{0.5ex}{\tiny$^{+2000}_{-2000}$}} & & 184/365 & $3.3${\raisebox{0.5ex}{\tiny$^{+0.3}_{-0.6}$}}$\times10^{-6}$ \\ [0.5ex]
    \hline
\end{tabular}
\label{table:peak-flux-spectra-3}
\end{table*}

\begin{table*}
\caption{Spectral analysis results for the peak flux interval between T0+59,024.082 and T0+59,025.106~s using \textit{Fermi}-GBM detectors n8, nb, and b1, and the Interval~7 orbital and polynomial backgrounds. The flux is given in the 1--10,000~keV range. Uncertainties are given at the 90\% confidence level. For the NaI detectors, we use an energy range of 50--900~keV with the orbital background and 8--900~keV with the polynomial background. CPL is the preferred model with both the orbital and polynomial background.} 
\centering
\begin{tabular}{ccccccccccc}
    \hline
    Background & Model & $\alpha$ & $\beta$ & $E_p$ & $E_b$/kT & Stat/ & Fluence \\ 
     & & & & (keV) & (keV) & DOF & (erg/cm$^2$) \\
    \hline
    \rule{0pt}{2.5ex}
    Orbital & PL & $-1.43${\raisebox{0.5ex}{\tiny$^{+0.08}_{-0.1}$}} & & & & 143/303 & $2.5${\raisebox{0.5ex}{\tiny$^{+0.5}_{-0.4}$}}$\times10^{-6}$ \\ [0.5ex]
     & BB & & & & $89${\raisebox{0.5ex}{\tiny$^{+5}_{-5}$}} & 149/303 & $5.8${\raisebox{0.5ex}{\tiny$^{+0.5}_{-0.6}$}}$\times10^{-7}$ \\ [0.5ex]
     & \textbf{CPL} & $-1.01${\raisebox{0.5ex}{\tiny$^{+0.13}_{-0.2}$}} & & $3000${\raisebox{0.5ex}{\tiny$^{+2000}_{-2000}$}} & & 133/302 & $2.8${\raisebox{0.5ex}{\tiny$^{+0.3}_{-0.4}$}}$\times10^{-6}$ \\ [0.5ex]
     & Band & $-0.98${\raisebox{0.5ex}{\tiny$^{+0.2}_{-0.2}$}} & $-2.8${\raisebox{0.5ex}{\tiny$^{+0.8}_{-7}$}} & $3000${\raisebox{0.5ex}{\tiny$^{+2000}_{-2000}$}} & & 133/301 & $2.7${\raisebox{0.5ex}{\tiny$^{+0.3}_{-0.4}$}}$\times10^{-6}$ \\ [0.5ex]
     & SBPL & $-1.06${\raisebox{0.5ex}{\tiny$^{+0.13}_{-0.2}$}} & $-2.5${\raisebox{0.5ex}{\tiny$^{+0.6}_{-3}$}} & & $2000${\raisebox{0.5ex}{\tiny$^{+2000}_{-1000}$}} & 133/301 & $2.7${\raisebox{0.5ex}{\tiny$^{+0.3}_{-0.3}$}}$\times10^{-6}$ \\ [0.5ex]
    Polynomial & PL & $-1.33${\raisebox{0.5ex}{\tiny$^{+0.06}_{-0.08}$}} & & & & 186/361 & $2.4${\raisebox{0.5ex}{\tiny$^{+0.3}_{-0.3}$}}$\times10^{-6}$ \\ [0.5ex]
     & \textbf{CPL} & $-1.00${\raisebox{0.5ex}{\tiny$^{+0.14}_{-0.2}$}} & & $3000${\raisebox{0.5ex}{\tiny$^{+2000}_{-2000}$}} & & 172/360 & $2.5${\raisebox{0.5ex}{\tiny$^{+0.3}_{-0.3}$}}$\times10^{-6}$ \\ [0.5ex]
     & SBPL & $-0.8${\raisebox{0.5ex}{\tiny$^{+0.2}_{-0.2}$}} & $-1.8${\raisebox{0.5ex}{\tiny$^{+0.2}_{-0.5}$}} & & $400${\raisebox{0.5ex}{\tiny$^{+200}_{-200}$}} & 171/359 & $2.0${\raisebox{0.5ex}{\tiny$^{+2}_{-0.6}$}}$\times10^{-6}$ \\ [0.5ex]
    \hline
\end{tabular}
\label{table:peak-flux-spectra-7}
\end{table*}

\citet{oganesyan2025transient} analyze \textit{Fermi}-GBM data to study GRB~250702B. Their spectral analysis also contains both orbital background and polynomial background approaches. They probe the three brightest intervals in GBM corresponding to the triggers and report a best-fit spectrum of a power-law in all cases. The orbital background utilized for their third fit interval uses background intervals during which \textit{Fermi}-GBM was in a Target of Opportunity and then while \textit{Fermi} was in a modified observing profile related to the stuck solar panel, thus the assumptions which underlie the orbital background tool are not necessarily met \citep{2011arXiv1111.3779F}. In their second fit interval, the change in the fit statistic between the PL and CPL model is significant according to the general GBM team criterion \cite[e.g.][]{2021ApJ...913...60P}. The reason for the lack of preference for turnover in their first interval is not immediately obvious.

\subsubsection{Konus-Wind}\label{app:konus-spectral-analysis}
We explore the spectral evolution of the burst using three-channel spectra, constructed from the counts in the G1, G2 and G3 energy bands. We select spectrum accumulation intervals corresponding to each separate emission episode determined using Bayesian blocks and peak flux intervals selected within each of the intervals. Because the emission is weak during Interval~3, there is no peak flux spectrum reported. The total time-integrated spectrum is constructed as a sum of counts in all episodes.

Details of the Konus-\textit{Wind} three-channel spectral analysis can be found in~\citep{tsvetkova2021konus}. We perform the spectral analysis in \texttt{XSPEC}~v.12.15.0 \citep{1999ascl.soft10005A}, using PL and CPL models parameterized by the peak energy of the $\nu F(\nu)$ spectrum and with the energy flux as the model normalization. Since a CPL fit to a three-channel spectrum has zero degrees of freedom (and, in the case of convergence, $\chi^2 = 0$), we do not report the statistic for such fits. We calculate the confidence intervals of the parameters using the command \texttt{steppar} in \texttt{XSPEC}.

\subsection{Quasi-periodic Oscillations}\label{app:qpo}
For the cross-spectrum analysis we combine the G2 and G3 energy ranges and consider two independent observations from two Konus detectors, S1 and S2. The lightcurves are extracted with a binning of $\Delta t = 2.944$~s, and we compute power spectra using the \texttt{stingray} package over a continuous time interval 30~ks long, between T0+45,600~s and T0+75,600~s. We calculate the cross power spectrum (CPS) between the two detectors using the combined G2+G3 lightcurves, employing the \texttt{Crossspectrum} class in \texttt{stingray} with Leahy normalization. The spectra are logarithmically rebinned in frequency space to reduce statistical scatter and enhance the visibility of broad features. An important advantage of the CPS over individual power spectra is that white-noise contributions, being uncorrelated between detectors, do not contribute to the cross spectrum. Possible detector-related systematics are likewise suppressed, removing the need for explicit modeling of these components.

To estimate the uncertainties on the CPS, we adopt a Monte Carlo approach. Specifically, we generate 1,000 realizations of the lightcurves by varying the count rates in each time bin independently according to their measured variances. For each simulated lightcurve, we recompute the CPS, and from this ensemble we derive the mean and standard deviation of the power at each frequency. This procedure yields the statistical uncertainties on the measured CPS that serve as inputs to the subsequent fitting analysis.

The CPS is modeled using the sum of a power law and a Lorentzian component of the form  
\begin{equation}
L(\nu) = \frac{A \gamma^2}{\gamma^2 + (\nu - \nu_0)^2}
\end{equation}
where $\nu_0$ is the centroid frequency, $\gamma$ is the full width at half maximum, and $A$ is the amplitude. 
In particular, noting a potentially interesting excess around 3~mHz, we perform a fit and summarize the best-fit values in Table~\ref{tab:fits}.

\begin{figure*}
    \centering
    \includegraphics[width=0.8\textwidth]{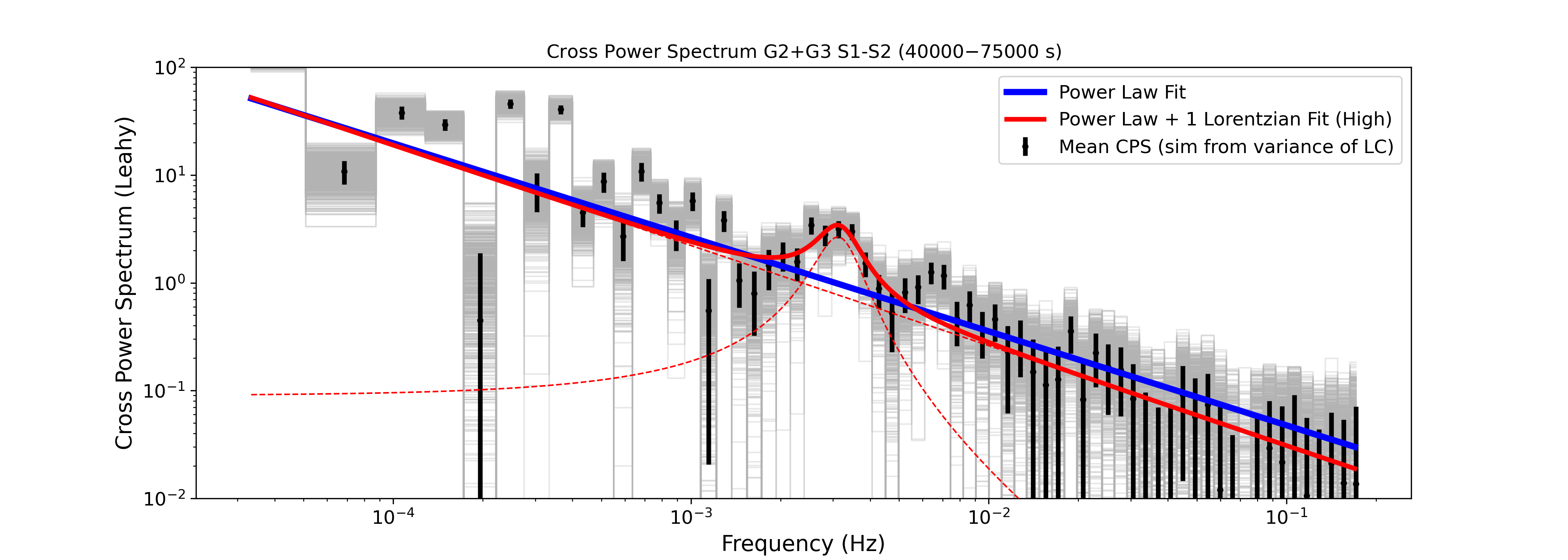}
    \includegraphics[width=0.8\textwidth]{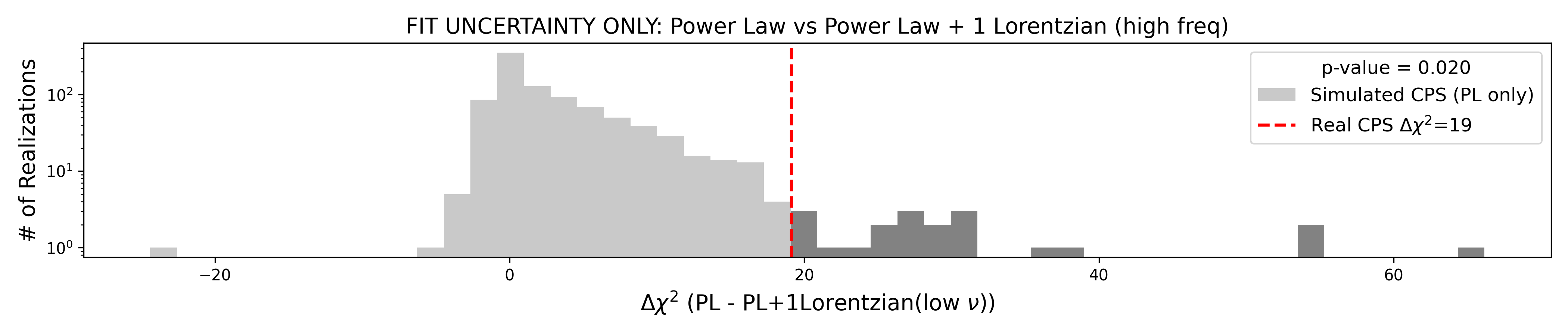}
    \caption{\textit{Top}: Cross power spectrum (CPS) of G2+G3 S1–S2 in the interval T0+45,600–75,600~s. The gray lines show individual simulated CPS realizations derived by varying the individual time bins of the lightcurve according to the variance of the count rates, while the black points with error bars mark the mean CPS and their Gaussian standard deviations. The solid blue line represents the best-fit power-law model. The red solid line corresponds to the power-law plus one Lorentzian (at high frequency) model. The dotted red lines show the individual power-law and Lorentzian components. \textit{Bottom}: Distribution of simulated $\Delta \chi^2$ values obtained under the null hypothesis (power-law model only), compared with the observed data. Comparison between the power-law and power-law+Lorentzian models. The light gray histograms show the distribution from Monte Carlo simulations, while the dark gray regions indicate realizations with $\Delta \chi^2$ greater than or equal to the observed value. The dashed red line marks the $\Delta \chi^2$ measured from the real data, with the corresponding p-values reported in the legend.
    }
    \label{fig:cps_g2d1xg2d2_base}
\end{figure*}

We compare the goodness of fit of a simple power-law model with a model including one Lorentzian component on top of the power law. To determine the significance of the improvement of the models over the null hypothesis of a power-law-only behavior, we simulate 1,000 CPS assuming the power-law-only best fit model and for each simulated CPS we perform the fit of the power-law-only (null-hypothesis) and the fit assuming the power-law+Lorentzian model (alternate hypothesis). As shown in Fig.~\ref{fig:cps_g2d1xg2d2_base}, the observed $\Delta \chi^2$ values fall within the distribution expected from simulations under the null hypothesis, with p-values on the order of 10 percent at most. No statistically significant features are found, and the inclusion of additional Lorentzian components does not improve the fit beyond the level expected from random fluctuations. We therefore conclude that there is no compelling evidence for QPOs in the CPS during this interval.

\begin{table*}
\caption{Best fit parameters for power-law and power-law+Lorentzian models fit to the CPS. $K$ is the power-law normalization and $\Gamma$ is the power-law index.}
\centering
\begin{tabular}{cccccccc}
\hline
Model & $K$ & $\Gamma$ & $\nu_0$ (mHz) & $\gamma$ (mHz) & $A$ & $\chi^2$/DOF \\
\hline
PL                               & $0.007 \pm 0.003$ & $0.87 \pm 0.06$ & -- & -- & -- & 462/64 \\
PL + 1 $L(\nu)$ (high $\nu$)     & $0.004 \pm 0.003$ & $0.92 \pm 0.08$ & $3.1 \pm 0.3$ & $1.1 \pm 0.7$ & $2.6 \pm 1.2$ & 405/61 \\
\hline
\end{tabular}
\label{tab:fits}
\end{table*}

For the power spectrum analysis, we use the same time period (T0+45,600--75,600~s) and binning ($\Delta t = 2.944$~s) as in the CPS analysis. For each light curve of interest, we create a power spectral density (PSD) with the (rms/mean)$^2$ normalization using the \texttt{pyLag} spectral-timing python package \citep{Wilkins2019}. In this approach, we utilize the unbinned PSD to search for potential QPOs, as has been commonly adopted in the search for QPOs around supermassive black holes \citep[e.g.][]{Vaughan2005,Vaughan2010,Gierlinski2008,Alston2014,Masterson2025}. Each point in an unbinned PSD is distributed following a $\chi^2$ distribution with two degrees of freedom (denoted $\chi^2_2$; see \citealt{Vaughan2005,Vaughan2010} for more details). This distribution has a large variance compared to a Gaussian distribution, and hence, large fluctuations are relatively common and must be handled with statistical care. 

To search for QPOs, we follow a maximum likelihood approach, fitting the unbinned PSD using the Whittle likelihood given by 
\begin{equation}
    \ln\mathcal{L} = -\sum_j \left(\frac{I_j}{S_j} + \ln S_j\right),
\end{equation}
where $j$ denotes a sum over all frequencies $f_j$, $I_j$ is the observed power at frequency $f_j$, and $S_j$ is the model power at frequency $f_j$. For QPOs which appear as an excess in a single frequency bin, the statistical significance of a single outlier bin can be estimated by measuring $T_R = \max_j R_j$, where $R_j = 2I_j / S_j$ \citep{Gierlinski2008,Alston2014}. For QPOs which span multiple frequency bins, we must adopt a different approach that accounts for the spread of the power across many channels \citep[see e.g.,][for further discussion]{Masterson2025}. 

The Konus-\textit{Wind} PSD for GRB250702B does not show any obvious single frequency channel outliers (i.e., $T_R \gtrsim 20$), but does show what looks like a potential broad feature around 3 mHz. Hence, we adopt the following approach to estimating the significance of this broad feature. We fit the data with two models: (1) a power-law plus constant model, where the constant accounts for the white noise at high frequencies, and (2) power-law plus constant plus Lorentzian, where the Lorentzian is to model a potential QPO. To compare these two models, we compute the likelihood ratio test (LRT) statistic, given by $T_\mathrm{LRT} = -2\left(\ln\mathcal{L}_{H_0} - \ln\mathcal{L}_{H_1}\right)$ \citep[see Section 8 of][for more details]{Vaughan2010}, where $H_0$ indicates the power-law only model (the null hypothesis) and $H_1$ indicates the power-law plus Lorentzian model. Similar to the CPS analysis, we test the statistical significance of this additional model component by simulating 1,000 realizations of the underlying broadband noise and asking how often an additional Lorentzian yields a comparable $T_\mathrm{LRT}$. To simulate these realizations, we adopt the methodology outlined in \cite{Timmer1995}. In brief, this method takes an underlying PSD model and draws a normally distributed value for the real and imaginary part of the Fourier transform at each frequency that are proportional to the square root of the model power at that frequency. This method thus randomizes both the phase and the amplitude of the PSD, correctly accounting for the $\chi^2_2$ distribution of power that arises from summing in quadrature two normally distributed variables (the real and imaginary part of the Fourier transform). For each realization, we fit the simulated unbinned PSD with both models and record the $T_\mathrm{LRT}$. We then adopt the ratio of simulations with $T_\mathrm{LRT}$ greater than the observed value to the total number of simulations as the resulting p-value for adding the Lorentzian component to the model.

\begin{figure*}
    \centering
    \includegraphics[width=\linewidth]{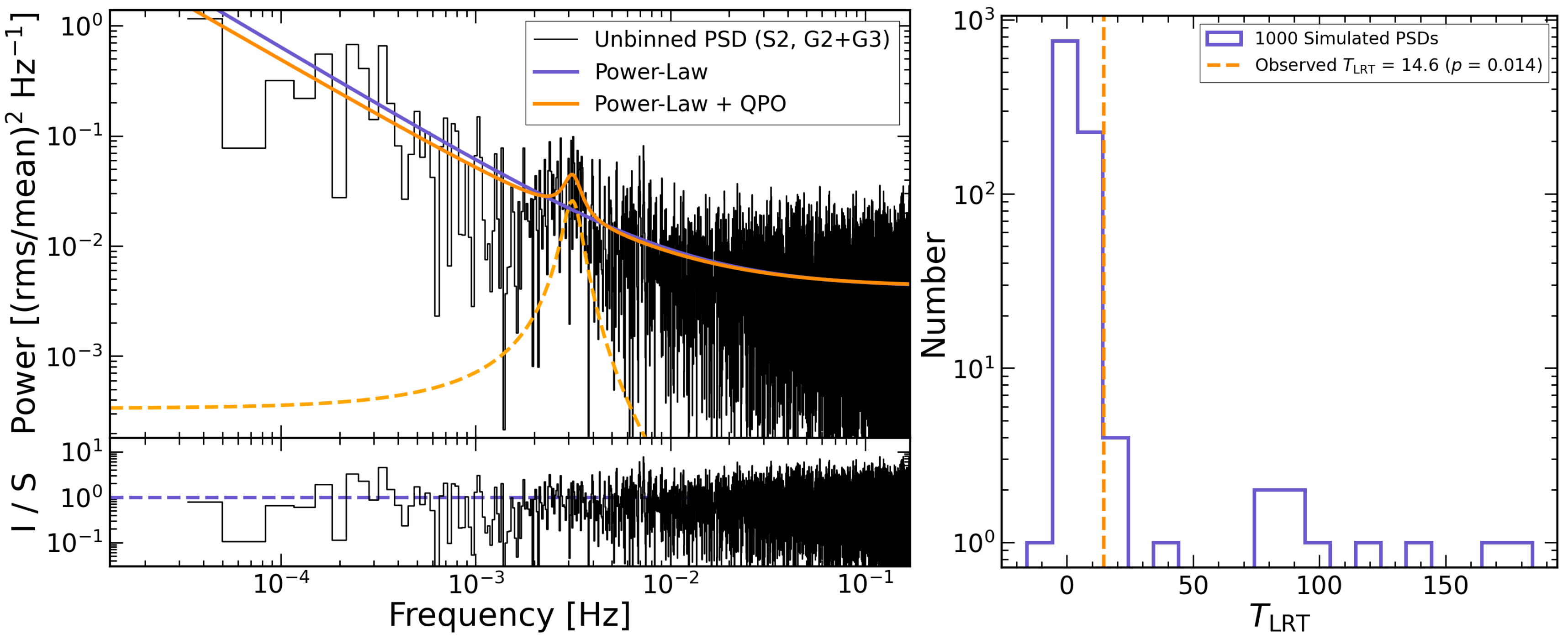}
    \caption{Example of PSD analysis for the S2, G2+G3 data. \textit{Left}: Unbinned PSD (black) from the S2, G2+G3 data, fit with both a power-law plus constant model (purple) and a power-law plus constant plus Lorentzian model (orange). The bottom panel shows the data divided by the best-fitting power-law plus constant plus Lorentzian model. \textit{Right}: The resulting $T_\mathrm{LRT}$ from adding a Lorentzian to the PSD model for 1000 simulated light curves with the same underlying power-law PSD. The dashed orange line shows the observed value of $T_\mathrm{LRT} = 14.6$, which indicates a significance of $p = 0.014$.}
    \label{fig:qpo_psd}
\end{figure*}

The result of our PSD analysis is comparable that of the CPS; in short, we do not find any evidence for statistically significant ($>$3$\sigma$) evidence for a QPO in the GRB250702B. We test many different combinations of Konus-\textit{Wind} data, using the S2 detector and G2, G3, and G2+G3 energy ranges. For each dataset, the best-fitting Lorentzian sits at around $\sim$3~mHz, similar to what is seen in the CPS. In Fig.~\ref{fig:qpo_psd}, we show an example of the PSD and best-fit models for the S2, G2+G3 data, along with the distribution of simulated $T_\mathrm{LRT}$ values. The resulting p-values for each energy range are: 0.030 (S2, G2), 0.5 (S2, G3), and 0.014 (S2, G2+G3). Although the statistical significance can depend on the choice of broadband noise model, we expect that any additional complexities in the model (e.g., a break in the power law) will only decrease the significance of a QPO-like feature. Hence, we find no strong evidence for a QPO in the Konus-\textit{Wind} data of GRB~250702B.

\subsection{Bulk Lorentz Factor}\label{app:gamma}
In order for the source to be optically thin to $\gamma\gamma$-annihilation, in which the optical depth at a given energy $E_p<E<E_{\rm cut}$ is $\tau_{\gamma\gamma}(\Gamma_0,E)<1$, where $E_{\rm cut}$ is the cutoff energy, it must be moving ultrarelativistically with $\Gamma_0\gg1$ \citep{Baring-Harding-97,Lithwick-Sari-01,Granot+08,Hascoet+12}. 

In this case, the spectrum above $E_{\rm cut}$ will be optically thick to $\gamma\gamma$-annihilation and will show either an exponential or a power-law flux suppression \citep[][]{Granot+08}. High-energy spectral cutoffs have been observed in several GRBs, e.g., GRB~090926A \citep{Ackermann+11,2017A&A...606A..93Y}, and GRBs~100724B and 160509A 
\citep[][also see \citealt{Tang+15} and \citealt{2023ApJ...956..101S} for additional sources]{2018ApJ...864..163V}, and were used to obtain an estimate of $\Gamma_0$. When there is no clear high-energy spectral break and the emission is only seen up to a certain maximum energy, $E_{\rm max}$, instead, then the same arguments can be used to place a lower limit on $\Gamma_0>\Gamma_{0,\rm min}$. Many works employ the one-zone assumption in which both the annihilating photons are co-spatial \cite[e.g.][]{Lithwick-Sari-01,Abdo+09}. Such treatments yield an estimate of $\Gamma_0$ larger by a factor of $\sim2$ than those that either consider two distinct emission zones \citep[e.g.][]{Zou+11} or account for the spatial, temporal, and directional dependence of the annihilating 
photons (\citealt{Granot+08,Hascoet+12}). 

Here, we consider the analytic model of \citet{Granot+08} that assumes an ultrarelativistic thin spherical shell and calculates $\tau_{\gamma\gamma}$ along the trajectories of all observed photons to yield
\begin{multline}\label{eq:Gamma_0}
    \Gamma_0 = 100\left[\frac{396.9}{C_2(1+z)^{\Gamma_{\rm ph}}}\left(\frac{L_0}{10^{52}\,{\rm erg\,s}^{-1}}\right)
    \left(\frac{5.11\,{\rm GeV}}{E_{\rm cut}}\right)^{1+\Gamma_{\rm ph}} \right. \\
    \left. \left(\frac{-\Gamma_{\rm ph}}{2}\right)^{-5/3} \frac{33.4\,{\rm ms}}{t_v}\right]^{1/(2-2\Gamma_{\rm ph})}
\end{multline}
where $L_0 = 4\pi d_L^2(1+z)^{-\Gamma_{\rm ph}-2}F_0$, $t_v$ is the variability timescale in milliseconds, $\Gamma_{\rm ph}$ is the photon index of the power law spectrum at $E>E_{\rm pk}$, $F_0$ is the (unabsorbed) energy flux ($\nu F_\nu$) obtained at 511~keV from the power law component, and $d_L$ is the luminosity distance. The parameter $C_2\approx1$ is constrained from observations of spectral cutoffs in other GRBs \citep{2018ApJ...864..163V}. \citet{Gill-Granot-18} confirmed the results of this model by performing numerical simulations, where they showed that it yields an accurate estimate of the bulk Lorentz factor from observations of spectral cutoffs if the emission region remains optically thin to Thomson scattering due to the produced $e^\pm$ pairs. 

The estimate of $\Gamma_0$ in equation~\ref{eq:Gamma_0} should be compared with $\Gamma_{0,\max}=(1+z)E_{\rm cut}/m_ec^2$, which is the maximum bulk Lorentz factor for a given $E_{\rm cut}$ and for which the cutoff energy in the comoving frame is at the self-annihilation threshold of $E_{\rm cut}'=(1+z)E_{\rm cut}/\Gamma_{0,\max}=m_ec^2$. When the emission region is Thomson thick due to the created pairs, $E_{\rm cut}'$ can be slightly lower than $m_ec^2$ due to Compton downscattering \citep{Gill-Granot-18}. When this is not the case, the true bulk Lorentz factor is obtained from the minimum of the two estimates. 

For GBM~Interval~7 of GRB~250702B, the MVT is $\sim$1.4~s and the maximum observed photon energy is $E_{\max}\sim5$~MeV, with both parameters listed in the observer frame. Here, we consider two different spectral model fits, namely PL and CPL, that both yield an acceptable fit. The CPL model shows a spectral peak at $E_p\sim4$~MeV, which we interpret as the high-energy spectral cutoff energy ($E_{\rm cut}=E_{\rm peak}$) with a hard photon index $\alpha=-1.24$ below this energy. For a photon flux of $2.9\times10^{-4}$~ph/s/cm$^{2}$/keV at $E=511$~keV, we obtain an estimate of the true bulk Lorentz factor of $\Gamma_0\simeq81$ for the source redshift of 1.036. Alternatively, when considering the PL model that extends up to a maximum observed photon energy $E_{\max}\sim5$~MeV with photon index $\Gamma_{\rm ph}=-1.50$, we constrain the bulk Lorentz factor from below and find $\Gamma_{0,\min}\simeq56$ for a photon flux of $2.2\times10^{-4}$~ph/s/cm$^{2}$/keV. In this case, since there is no spectral information at $E>E_{\max}$, we consider $E_{\rm max}$ as a lower limit on the photon energy of a possible high-energy spectral cutoff at some energy $E_{\rm cut}>E_{\rm max}$. For the same $E_{\rm cut}$ ($E_{\max}$) in the CPL (PL) model, we find the maximum corresponding bulk Lorentz factor of $\Gamma_{0,\max}\simeq 17$ (20). 

When $\Gamma_{0,\max}$ is smaller than $\Gamma_0$ or $\Gamma_{0,\min}$, it implies a large Thomson optical depth due to the produced $e^\pm$-pairs, with $\tau_{T,\pm}\simeq1.8\times10^4$ ($2.8\times10^3$) for the cutoff power-law (power-law) model. \citet{Gill-Granot-18} showed that in this case the spectrum above the peak energy of the Band spectrum would become hard with photon index $\Gamma_{\rm ph}\gtrsim-2$, as obtained in this GRB, which would manifest as a flat or slightly rising spectrum in $\nu F_\nu$ above $E_p$ with a sharp spectral cutoff at $E_{\rm cut}'<m_ec^2$ in the comoving frame. In this situation, $\Gamma_{0,\max}$ would under-predict the true bulk Lorentz factor by up to a factor of ten, and the true Lorentz factor is much closer to $\Gamma_0$ or $\Gamma_{0,\min}$ when calculated using equation~\ref{eq:Gamma_0}, which we adopt here. 

The coasting bulk Lorentz factor of the ejecta can also be inferred from the afterglow when the lightcurve shows a clear deceleration peak, e.g., in the optical and/or X-rays, which are often located between the injection and cooling break frequencies, $\nu_m$ and $\nu_c$, respectively, of the slow-cooling synchrotron spectrum. Then, depending on the power-law index, $p$, of the energy distribution of the radiating electrons, $N_e(\gamma)\propto \gamma^{-p}$ for $\gamma>\gamma_m$, and the radial profile of the external medium density, $n_{\rm ext} = n_0(R/R_0)^{-k}$, the afterglow flux density declines at $T>T_{\rm dec}$ as a power law with $d\ln F_\nu/d\ln T = [k(3p-5)-12(p-1)]/4(4-k)$. 

In the thin-shell approximation, $T_{\rm dec} = (1+z)[(3-k)E_{\rm k,iso}/2^{5-k}\pi Ac^{5-k}\Gamma_0^{2(4-k)}]^{1/(3-k)}$ marks the time at which the flow decelerates by sweeping up mass $M_{\rm sw}(R) = 4\pi \rho(R)R^3/(3-k) = M_0/\Gamma_0=E_{\rm k,iso}/\Gamma_0^2c^2$ in the external medium, where $\rho(R)=m_pn_{\rm ext}(R)=AR^{-k}$ and $M_0$ is the mass of the ejecta. No afterglow lightcurve peak in the optical and X-rays was seen in GRB~250702B and all of the observations were obtained post-deceleration, so only a lower limit can be placed on $\Gamma_0$. In fact, the afterglow modeling in \citet{levan2025day} suggests that the steeply decaying X-ray and optical afterglow observations were made after an early jet break at $T_{\rm jet}\lesssim0.5$~days and where the jet propagates inside a wind ($k=2$) medium. Therefore, $T_{\rm dec}<T_{\rm jet}$, which yields $\Gamma_0 \gtrsim 23\,(1+z)^{1/4}E_{\rm k,iso,54}^{1/4}A_*^{-1/4}T_{\rm jet,4}^{-1/4}$ for $k=2$ and $A=3\times10^{35}A_*\,$~cm$^{-1}$. This lower limit is consistent with the estimate of the bulk Lorentz factor obtained above.

These values are broadly consistent with measurements from multiwavelength afterglow modeling. \citet{levan2025day} infer a bulk Lorentz factor at $\sim$0.5~day of $\sim$40, which would be lower than the initial value due to deceleration. \citet{oconnor2025grb250702b} measure a comparable $\Gamma_0$. With the additional restriction of $\Gamma_0\theta_j\geq1$, with $\theta_j$ the half-jet opening angle, to prevent the jet from spreading laterally in the coasting phase they measure $\Gamma_0\gtrsim 100$. Thus, all approaches require an ultrarelativistic jet. 

\section{Detailed Discussion on Excluded Progenitors}\label{app:progenitor}
The following subsections discuss each physical progenitor in detail, and why it is excluded. Beyond the progenitor systems discussed below, there are two possible options to explain the extreme duration as arising not from intrinsic accretion timescales but from propagation effects. One is the option of a lensed event, where the duration arises from seeing temporally-delayed lenses. While this could, in principle, explain the gamma-ray signature, it is not compatible with the $\sim$day earlier signal seen only in X-rays. The other option is dust echoes of the initial signal from surrounding material in the host galaxy. However, this should be negligible above $\sim$100~keV, which is incompatible with our observation of multiple MeV photons in late pulses, as noted in \citet{levan2025day}.

\subsection{X-Ray Binaries (and other Galactic sources)}
Early on in the study of this event, the alignment with the Galactic plane suggested a possible source within the Milky Way. Many Galactic sources which involve compact objects produce flares which are recovered in the gamma-ray burst monitors. As one particular example, we highlight X-ray binaries. These are binary systems where a compact object accretes material from a donor star. This can launch jets which have been observed with $\sim$week durations in the GRB monitors \citep[e.g.,][]{jenke2016fermi}. However, it is rare for X-ray binaries (and other Galactic sources) to exceed being detected to a few hundred keV, and none have been seen to 5~MeV. To this we add the 0.05\% chance alignment with the extragalactic putative host galaxy \citep{levan2025day}.

\subsection{Magnetar Giant Flares}
Magnetars are neutron stars with the most extreme persistent magnetic fields in the cosmos \citep{thompson1995soft}. These sources produce flares, the most extreme of which are giant flares, observed up to MeV energies and seen in the Milky Way and other nearby galaxies \citep{2021ApJ...907L..28B,negro2024role}. However, their emission size is comparable to the size of a neutron star, with durations of the prompt spike lasting fractions of a second. The duration is inconsistent by orders of magnitude.

\subsection{Neutron Star Mergers}
Neutron star mergers refer to both binary neutron star and neutron star-black hole mergers. In either scenario, the accretion timescale is on the order of 0.1~s, suggesting prompt durations under 10~s in duration \citep{fryer2019understanding}. However, short gamma-ray bursts with extended emission have properties expected from a compact merger origin, and have durations up to $\sim$100~s \citep{norris06}. Further, recent works have identified long-duration gamma-ray bursts which must arise from merger events due to offsets from their host galaxies \citep{rastinejad2022kilonova,gompertz2023case,levan2024heavy} and may arise from neutron star mergers. These bursts also reach $\sim$100~s in duration. Thus, neutron star mergers are inconsistent in duration by orders of magnitude.

\subsection{White Dwarf Mergers}\label{sec:whitedwarf}
White dwarfs merging with other stellar remnants (white dwarf, neutron star, black hole) have been discussed as possible GRB progenitors. We first note that it is not guaranteed that these mergers will produce jets. White dwarf-white dwarf binaries can produce a wide variety explosions including thermonuclear (type Ia) supernovae and short-duration GRBs assuming the collapse of white dwarfs with strong magnetic fields~\citep{2023mgm..conf..217R,2025ApJ...978L..38C}. White dwarf-neutron star and white dwarf-black hole mergers have been proposed for a wide range of peculiar supernova explosions~\citep{2012MNRAS.419..827M,2023ApJ...956...71K}. None of these signatures have ever been associated with GRBs.

However, if we assume such mergers produce ultrarelativistic jets, stellar remnant merger options involving one white dwarf are the most natural to explain ultra-long GRBs.  This is because, compared to neutron stars and black holes, white dwarfs have large radii:  $\sim$3$\times 10^8$~cm for a white dwarf near the Chandrasekhar limit and $10^9$~cm for a 0.5~$M_{\odot}$ white dwarf.  The corresponding orbital separation when the white dwarf overfills it Roche limit is (1--5)$\times 10^9$~cm.  The specific angular momentum from these mergers is 7$\times 10^{17}$--3$\times 10^{18}$~cm/s, higher than most collapsar progenitors.  Simulations of these mergers showed high accretion rates for 15--150~s \citep{1999ApJ...520..650F}, but the disk formed in these mergers could continue to accrete for up to 10,000s for an $\alpha$-disk model where the viscous $\alpha \approx 0.01$.  The duration does not increase dramatically with increasing compact remnant mass so we would expect the same durations for IMBHs.  The accretion rate drops considerably after the first 100~s make a strong long-duration jet unlikely. Given the $\sim$day lead up to the strongest signal in GRB~250702B, these models are excluded.

\subsection{Relativistic Tidal Disruption Event}\label{app:tde}
Tidal disruption events occur when a star encounters a massive black hole and is disrupted when the gravitational gradient of the black hole overcomes the binding energy of the star. These events are commonly detected and studied in optical and X-ray wavelengths. A rare subset of TDEs occur when the accretion onto the black hole generates a relativistic jet. Four events are known, of which three were discovered by the \textit{Swift}-BAT and seen in hard X-rays for a few days \citep{sakamoto2011grb,krimm2011swift,krimm2011swift2,andreoni2022very}. For these reasons, relativistic TDEs are a viable model for ultra-long GRBs, and are thus an attractive model to consider for GRB~250702B.

We find TDEs involving a massive black hole to be incompatible with the gamma-ray results for GRB~250702B. First, we note only one relativistic TDE has ever reached the on-board trigger threshold of a GRB monitor. Further, none have been observed above a few hundred keV, let alone several MeV. And of the relativistic TDEs seen in gamma-rays and followed in X-rays, their durations are more than an order of magnitude longer than the emission in GRB~250702B. Near-infrared observations shows GRB~250702B to be non-nuclear in the host galaxy, while nearly all TDEs arise from nuclear positions \citep{levan2025day}.

For better comparison, we repeat our MVT analysis procedure using a \textit{Swift}-BAT mask-weighted light curve in 1~ms bins for Swift~J1644+57 \citep{2011Natur.476..421B}, the relativistic TDE with the shortest known variation timescale, finding a value of $\sim$54~s. With a redshift of 0.33 \citep{2011Sci...333..199L}, this corresponds to a rest-frame MVT of $\sim$40~s. Taking the Schwarzschild crossing time limit, this corresponds to a black hole mass of a few million solar masses, matching the results of X-ray reverberation mapping \citep{kara2016relativistic}. Further, the physical scales involved in tidal disruption events are expected to be orders of magnitude larger than the Schwarzschild light crossing time on both analytic grounds and results from simulations \citep{rees1988tidal,ayal2000tidal}. This led some to invoke a white dwarf merging with an IMBH for Swift~J1644+57 \citep{krolik2011swift}.

\subsection{Intermediate Mass Black Hole Tidal Disruption Event}\label{app:imbh}
In principle, an IMBH may resolve the issues with a relativistic TDE. Indeed \citet{levan2025day} invoke this as a viable model, highlighting it also explains the non-nuclear origin of the transient with respect to the host galaxy. However, there are multiple difficulties with this model which are discussed in the text.

\subsection{Typical (Carbon-Oxygen) Collapsar}\label{app:collapsar}
Most long GRBs arise from a collapsar origin. Theoretically, this is a rapidly rotating compact star which results in strong accretion after core-collapse \citep{1999ApJ...524..262M}. This powers bipolar ultrarelativistic jets which produce the prompt and afterglow signatures. The shockwave generated as the jet propagates through the star in concert with the winds from the accretion disk explode the star, powering a supernova, which is also regularly observed. Observationally, effectively all supernovae seen following long GRBs are type Ic broad-line \citep{cano2017observer}. The concordance model from theory and simulation is collapsar GRBs arise from the deaths of rapidly rotating stars which have been stripped of both hydrogen and helium, i.e., carbon-oxygen (CO) stars.

This origin is well-matched to the properties of GRB~250702B. It explains the relativistic jets and the follow-up data showing both a consistency with an external shock as well as the transient occurring within the stellar field of the host galaxy \citep{levan2025day}. The long GRB population has typical MVT values (or upper limits) of $\sim$0.1-10~s and peak energies from a few tens to hundreds of keV \citep{veres2023extreme,2021ApJ...913...60P}. The values of GRB~250702B are well within the known populations for these parameters. 

However, the prompt duration is far too long. The duration of the prompt GRB emission is determined by how long internal dissipation occurs in the jet. This is then tied to the accretion timescale onto the central engine. As discussed in Section~\ref{sec:COM}, the total accretion timescale is a combination of the free-fall timescale for the material to hang up in a disk and then the disk accretion timescale.  For systems spun up by tides from tight binaries, the angular momentum (shown in Fig.~\ref{fig:angularmom}) is only sufficient to form a fairly compact disk (1,000--10,000\,km).  With these models, it is difficult to produce accretion timescales that exceed a few thousand seconds (Fig.~\ref{fig:helium-star-power} shows a CO binary where the orbital separation is at the Roche limit).  The corresponding luminosity for our 32~M$_\odot$ progenitor in a tight binary is shown in Fig.~\ref{fig:helium-star-power}.  \citet{fryer2025explaining} explore the maximal duration viable through stars spun up near breakup by a tight binary for a broad number of stellar systems finding maximal accretion timescales of $\sim$1,000~s. Even allowing for extension of this timescale due to different engine models or due to observing the emitting region (not the accretion itself), this is incompatible with the total $\sim$100,000~s duration of GRB~250702B.

\subsection{Helium Collapsars}
A collapsar from a helium star progenitor may extend the duration because the larger radius leads to a longer freefall time. \citet{fryer2025explaining} also studied tidally spun-up helium star binaries.  In general, the angular momentum at any given mass coordinate for a helium-star binary is lower than that of a CO binary.  In addition, only the tightest of helium-star binaries have sufficient angular momentum to form a disk.  Figures~\ref{fig:angularmom} and \ref{fig:tacc} show the angular momentum and accretion timescales for helium star binaries with orbital separations at the Roche limit.  The corresponding jet power is shown in Fig.~\ref{fig:helium-star-power}.  The lower angular momentum means that the jet power peaks early and decays earlier.  But the helium-star merger has additional material (in the helium envelope) that can produce long-term accretion that extends beyond that of a typical CO Collapsar.  At these late times, the luminosity has already dropped considerably, over two orders of magnitude below peak. Thus, helium collapsars cannot explain the gamma-ray duration.

\subsection{Supergiant Collapsars}
Blue supergiants were invoked to explain the extreme duration of GRB~111209A \citep{gendre2013ultra,perna2018ultra,ioka2016ultra}. Now more than a decade later, no viable mechanism to achieve sufficient rotation in the progenitor star has been identified. Even if we assume mild coupling between burning layers, the maximum accretion timescale will be determined by the free-fall time onto the disk.  For example, using equation~\ref{eq:freefall}, the accretion timescale for a $10^{12} {\rm \, cm}$, 40\,M$_\odot$ blue supergiant would be 15,000~s, insufficient for the duration of GRB~250702B. Further, such events would likely have the strongest accretion at early times, in contrast to what is seen for GRB~250702B. Thus, this model is excluded. 

\subsection{Collapsar with Magnetar Central Engine}
Magnetar central engines are sometimes invoked to explain specific observational signatures seen in GRBs, such as plateaus following short GRBs \citep{gompertz2014magnetar}. However, they are disfavored as viable engines for collapsar GRBs for both observational and theoretical reasons. Observationally, collapsar GRBs arise predominantly from low metallicity galaxies \citep{fruchter2006long}. This was confirmation of a prediction from early theoretical papers \citep[e.g.][]{1999ApJ...524..262M} where the low metallicity is a requirement to form the black hole to power the jet. Higher metallicity star forming regions produce fewer large stars, and the large stars they do produce still end as neutron stars because of significant mass loss through winds. Magnetars are thus not important contributors to the long GRB population \citep{fryer2025explaining}. Further, in core collapse supernovae, even if the stellar remnant forms an extreme magnetic field, it will be buried for decades to a millennia \citep{1989ApJ...346..847C,1999A&A...345..847G}, precluding tapping of the magnetic fields to drive the relativistic jets. Even if we ignore these problems and treat ultra-longs as somehow arising from magnetar central engines, the kinetic energy extracted from the rotation of the magnetar would peak at early times and decay as a power law in time. Thus, the delayed peak in GRB~250702B is incompatible with a magnetar central engine.

\subsection{Binary Helium Star Merger}
The merger of two evolved cores (e.g. two helium cores) has also also been invoked as a means to produce a spun up core prior to collapse.  Detailed studies of these mergers found that, although the merged system had more angular momentum than typical single-star models, they did not produce systems spinning more rapidly than tidally-spun up models~\citep{2005ApJ...623..302F}.  Unless the collapse occurs shortly after the merger, these systems will likely have durations that are shorter than tidally spun-up systems.  Even if the collapse occurs right after the merger, the peak accretion rates and jet powers will occur well below a few thousand seconds. These events are thus also inconsistent with the observed duration in GRB~250702B. 


\bsp	
\label{lastpage}
\end{document}